\documentclass[10pt]{iopart}
\expandafter\let\csname equation*\endcsname\relax
\expandafter\let\csname endequation*\endcsname\relax

\usepackage{amsmath, amssymb, graphicx, color, url, bm}
\usepackage[breaklinks = true, colorlinks = true, linkcolor = blue, urlcolor = blue, citecolor = blue]{hyperref}
\usepackage[numbers,sort&compress]{natbib}

\begin{document}

\title[Dark energy two decades after]{Dark energy two decades after: Observables, probes, consistency tests}

\author{Dragan Huterer$^1$ and Daniel L Shafer$^2$}

\address{$^1$ Department of Physics, University of Michigan,
450 Church Street, Ann Arbor, MI 48109, USA}
\ead{huterer@umich.edu}

\address{$^2$ Department of Physics and Astronomy, Johns Hopkins University,
3400 North Charles Street, Baltimore, MD 21218, USA}
\ead{dshafer2@jhu.edu}

\begin{abstract}
The discovery of the accelerating universe in the late 1990s was a watershed
moment in modern cosmology, as it indicated the presence of a fundamentally
new, dominant contribution to the energy budget of the universe. Evidence for
dark energy, the new component that causes the acceleration, has since become
extremely strong, owing to an impressive variety of increasingly precise
measurements of the expansion history and the growth of structure in the
universe. Still, one of the central challenges of modern cosmology is to shed
light on the physical mechanism behind the accelerating universe. In this
review, we briefly summarize the developments that led to the discovery of
dark energy. Next, we discuss the parametric descriptions of dark energy and
the cosmological tests that allow us to better understand its nature. We then
review the cosmological probes of dark energy. For each probe, we briefly
discuss the physics behind it and its prospects for measuring dark energy
properties. We end with a summary of the current status of dark energy
research.
\end{abstract}

\noindent\textit{Keywords}: dark energy, observational cosmology, large-scale structure \pacs{98.80.Es, 95.36.+x} \submitto{\RPP}

\maketitle
\ioptwocol

\section{Introduction}

The discovery of the accelerating universe in the late 1990s
\cite{riess98,perlmutter99} was a watershed moment in modern cosmology. It
unambiguously indicated the presence of a qualitatively new component in the
universe, one that dominates the energy density today, or of a modification of the laws of gravity. Dark energy quickly became a centerpiece of the new standard cosmological model, which also features baryonic matter, dark matter, and radiation (photons and relativistic neutrinos). Dark energy naturally resolved some tensions in cosmological parameter measurements of the 1980s and early 1990s, explaining in particular the fact that the geometry of the universe was consistent with the flatness predicted by inflation, while the matter density was apparently much less than the critical value necessary to close the universe.

The simplest and best-known candidate for dark energy is the energy of the vacuum, represented in Einstein's equations by the cosmological-constant term. Vacuum energy density, unchanging in time and spatially smooth, is currently in good agreement with existing data. Yet, there exists a rich set of other dark energy models, including evolving scalar fields, modifications to general relativity, and other physically-motivated possibilities. This has spawned an active research area focused on describing and modeling dark energy and its effects on the expansion rate and the growth of density fluctuations, and this remains a vibrant area of cosmology today.

Over the past two decades, cosmologists have been investigating how best to measure the properties of dark energy. They have studied exactly what each cosmological probe can say about this new component, devised novel cosmological tests for the purpose, and planned observational surveys with the principal goal of precision dark energy measurements. Both ground-based and space-based surveys have been planned, and there are even ideas for laboratory tests of the physical phenomena that play a role in some dark energy models. Current measurements have already sharply improved constraints on dark energy; as a simple example, the statistical evidence for its existence, assuming a cosmological constant but not a flat universe, is nominally over 66$\sigma$\footnote{To obtain this number, we maximized the likelihood over all parameters, first with the dark energy density a free parameter and then with it fixed to zero, using the same current data as in figure~\ref{fig:BAO_CMB_SN}. We quote the number of standard deviations of a (one-dimensional) Gaussian distribution corresponding to this likelihood ratio.}. Future observations are expected to do much better still, especially for models that allow a time-evolving dark energy equation of state. They will allow us to map the expansion and growth history of the universe at the percent level, beginning deep in the matter-dominated era, into the period when dark energy dominates, and up to the present day.

\begin{figure*}[t]
\centering
\includegraphics[width=0.8\textwidth]{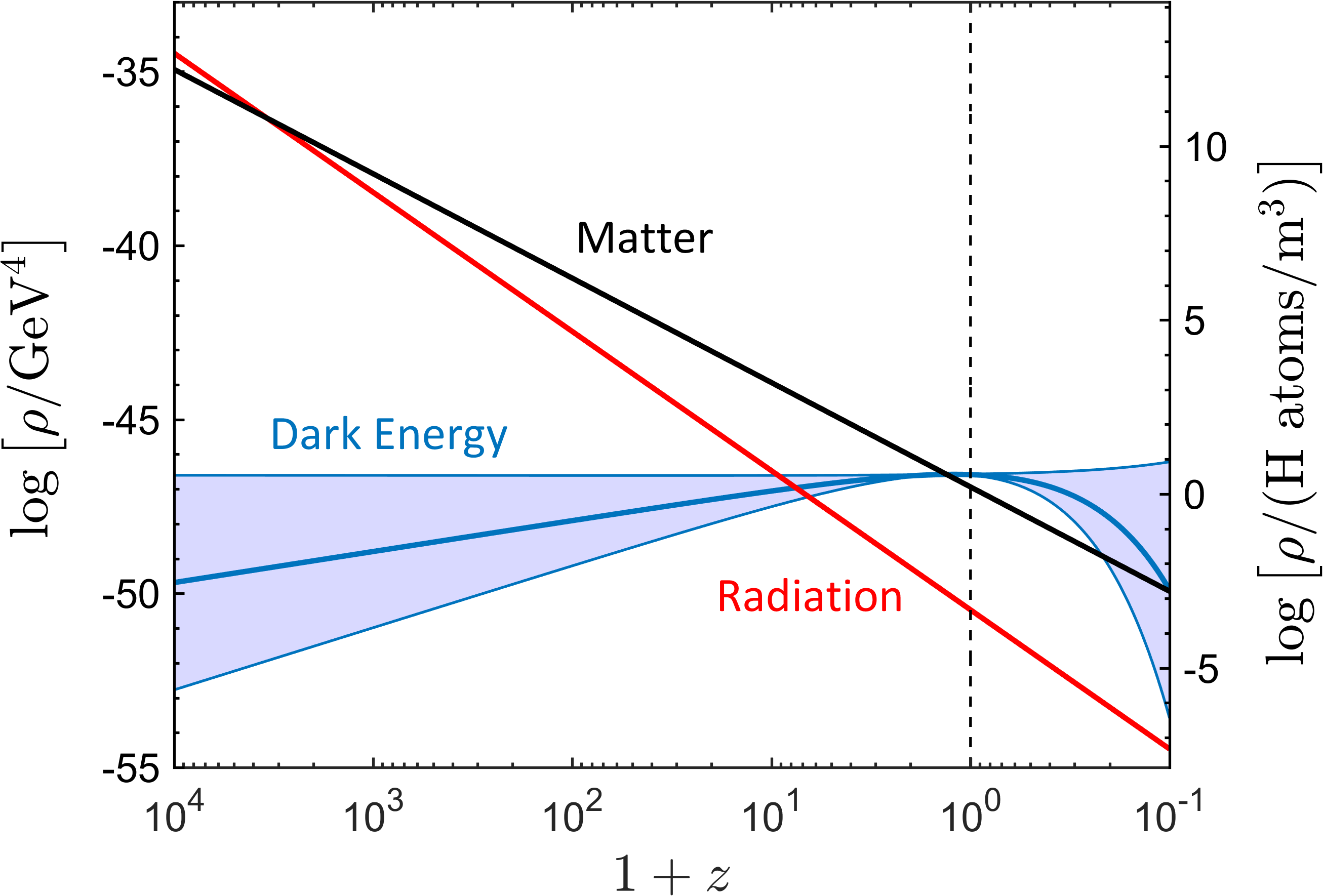}
\caption{Energy density of species in the universe as a function of $(1 + z)$, where $z$ is the redshift. The dashed vertical line indicates the present time ($z = 0$), with the past to the left and future to the right. Note that matter ($\propto (1 + z)^3$) and radiation ($\propto (1 + z)^4$) energy densities scale much faster with the expanding universe than the dark energy density, which is exactly constant for a cosmological constant $\Lambda$. The shaded region for dark energy indicates the energy densities allowed at 1$\sigma$ (68.3\% confidence) by combined constraints from current data assuming the equation of state is allowed to vary as $w(z) = w_0 + w_a \, z/(1 +z)$.}
\label{fig:de_density}
\end{figure*}

Despite the tremendous observational progress in measuring dark energy properties, no fundamentally new insights into the physics behind this mysterious component have resulted. Remarkably, while the error bars have shrunk dramatically, current constraints are still roughly consistent with the specific model that was originally quoted as the best fit in the late 1990s ---  a component contributing about 70\% to the current energy budget with an equation-of-state ratio $w \simeq -1$. This has led some in the particle physics and cosmology community to suspect that dark energy really is just the cosmological constant $\Lambda$ and that its unnaturally-small value is the product of a multiverse, such as would arise from the framework of eternal inflation or from the landscape picture of string theory, which generically features an enormous number of vacua, each with a different value for $\Lambda$. In this picture, we live in a vacuum which is able to support stars, galaxies, and life, making our tiny $\Lambda$ a necessity rather than an accident or a signature of new physics. As such reasoning may be untestable and therefore arguably unscientific, many remain hopeful that cosmic acceleration can be explained by testable physical theory that does not invoke the anthropic principle. For now, improved measurements provide by far the best opportunity to better understand the physics behind the accelerating universe.

\begin{figure*}[ht]
\centering
\includegraphics[width=0.47\textwidth]{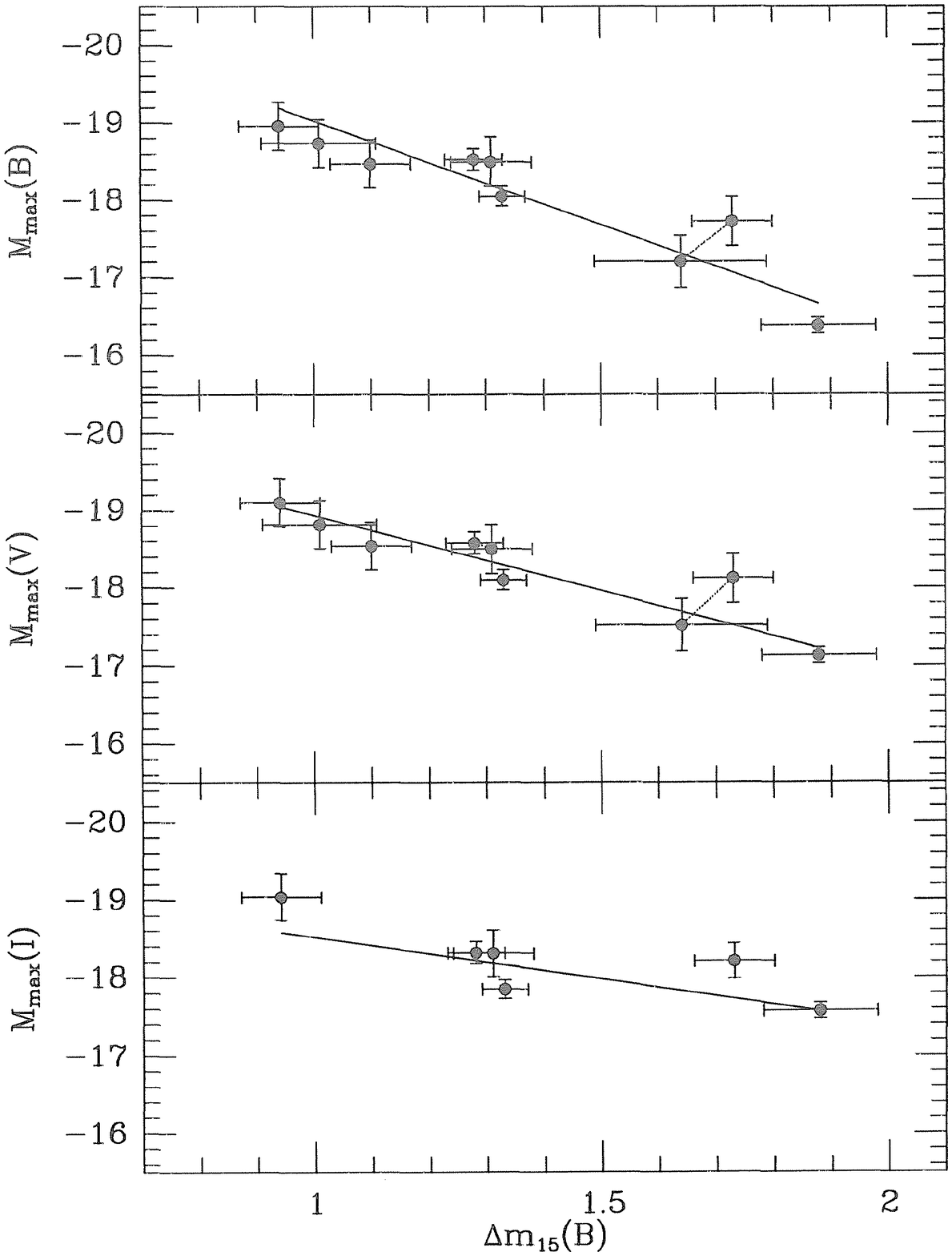}
\includegraphics[width=0.5\textwidth]{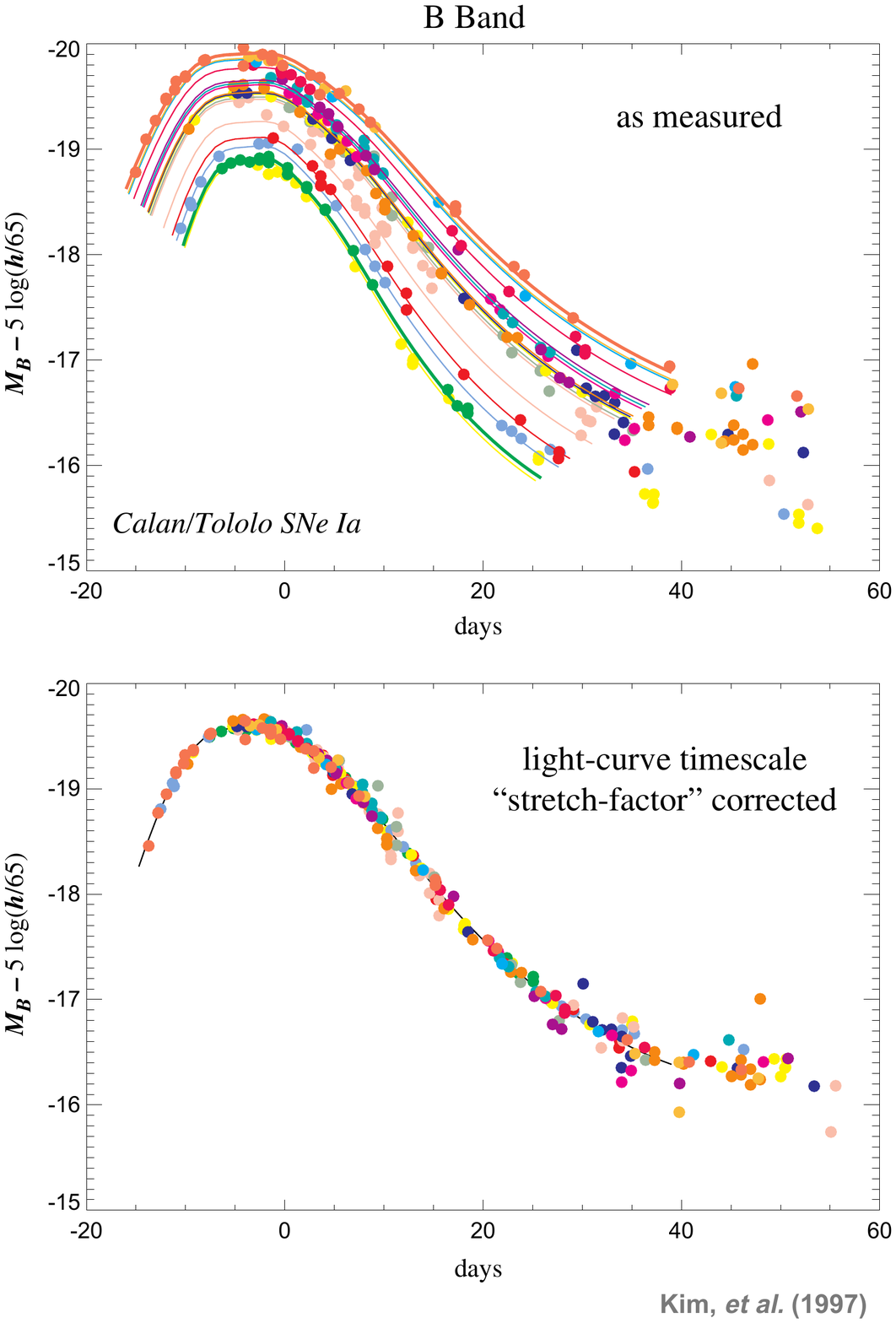}
\caption{Key properties of type Ia supernovae that enabled them to become a powerful tool to discover the acceleration of the universe. \textit{Left panel}: The Phillips relation, reproduced from his 1993 paper \cite{Phillips_93}. The (apparent) magnitude of SNe Ia is correlated with $\Delta m_{15}$, the decay of the light curve 15~days after the maximum. \textit{Right panel}: Light curves for a sample of SNe Ia before (top) and after (bottom) correction for stretch (essentially, the Phillips relation); reproduced from \cite{Kim:1997rm}.}
\label{fig:Phillips_rel}
\end{figure*}

Figure~\ref{fig:de_density} shows the energy density of species in the
universe as a function of $(1 + z)$, which is equivalent to the inverse of the
scale factor $a$. The dashed vertical line indicates the present time ($z =
0$), with the past to the left and the future to the right. Notice that
radiation, which scales as $(1 + z)^4$, dominates the early universe. Matter
scales as $(1 + z)^3$ and overtakes radiation at $z \simeq 3400$,
corresponding to $t \simeq 50,000~\text{yr}$ after the big bang. Dark energy
shows a very different behavior; vacuum energy density is precisely constant
in time, and even dynamical dark energy, when constrained to fit current data,
allows only a modest variation in density with time. The shaded region in
figure~\ref{fig:de_density} indicates the region allowed at 1$\sigma$ (68.3\%
confidence) by combined constraints from current data (see figure~\ref{fig:BAO_CMB_SN}) assuming the equation of state is allowed to vary as $w(a) =
w_0 + w_a \, (1 - a)$.

Our goal is to broadly review cosmic acceleration for physicists and astronomers who have a basic familiarity with cosmology but may not be experts in the field. This review complements other excellent, and often more specialized, reviews of the subject that focus on dark energy theory \cite{Copeland_review,Padmanabhan_review,Li:2011sd}, cosmology \cite{Peebles_Ratra_03}, the physics of cosmic acceleration \cite{Uzan_06}, probes of dark energy \cite{Hut_Tur_00,Weinberg:2012es}, dark energy reconstruction \cite{Sahni_review}, dynamics of dark energy models \cite{Linder_review}, the cosmological constant \cite{Carroll_LivRevRel,WeinbergRMP}, and dark energy aimed at astronomers \cite{Frieman:2008sn}. A parallel review of dark energy theory is presented in this volume by P.\ Brax.

The rest of this review is organized as follows. In section~\ref{sec:history},
we provide a brief history of the discovery of dark energy and how it changed
our understanding of the universe. In section~\ref{sec:basic}, we outline the
mathematical formalism that underpins modern cosmology. In
section~\ref{sec:parameters}, we review empirical parametrizations of dark
energy and other ways to quantify our constraints on geometry and growth of
structure, as well as modified gravity descriptions. We review the principal
cosmological probes of dark energy in section~\ref{sec:principal_probes} and
discuss complementary probes in section~\ref{sec:other_probes}. In
section~\ref{sec:summary}, we summarize key points regarding the observational
progress on dark energy.

\section{The road to dark energy} \label{sec:history}

In the early 1980s, inflationary theory shook the world of cosmology by
explaining several long-standing conundrums
\cite{Guth:1980zm,Linde:1981mu,Albrecht:1982wi}. The principal inflationary
feature is a mechanism to accelerate the expansion rate so that the universe appears precisely flat at late times. As one of inflation's cornerstone predictions, flatness became the favored possibility among cosmologists. At the same time, various direct measurements of mass in the universe were typically producing answers that were far short of the amount necessary to close the universe.

Notably, the baryon-to-matter ratio measured in galaxy clusters, combined with the baryon density inferred from big bang nucleosynthesis, effectively ruled out the flat, matter-dominated universe, implying instead a low matter density $\Omega_m \sim 0.3$ \cite{1991MNRAS.253P..29F,1991ApJ...379...52W,White:1993wm}. Around the same time, measurements of galaxy clustering --- both the amplitude and shape of the correlation function --- indicated strong preference for a low-matter-density universe and further pointed to the concordance cosmology with the cosmological constant contributing to make the spatial geometry flat \cite{Maddox:1990hb,Efstathiou:1990xe}. The relatively high values of the measured Hubble constant at the time ($H_0 \simeq 80~\text{km/s/Mpc}$ \cite{Freedman:1994fc}), combined with the lower limit on the age of the universe inferred from the ages of globular clusters ($t_0 > 11.2$~Gyr at 95\% confidence \cite{Krauss_Chaboyer}), also disfavored a high-matter-density universe. Finally, the discovery of massive clusters of galaxies at high redshift $z \sim 1$ \cite{Donahue:1997sp,Bahcall:1998ur} independently created trouble for the flat, matter-dominated universe.

\begin{figure*}[ht]
\centering
\includegraphics[width=0.7\textwidth]{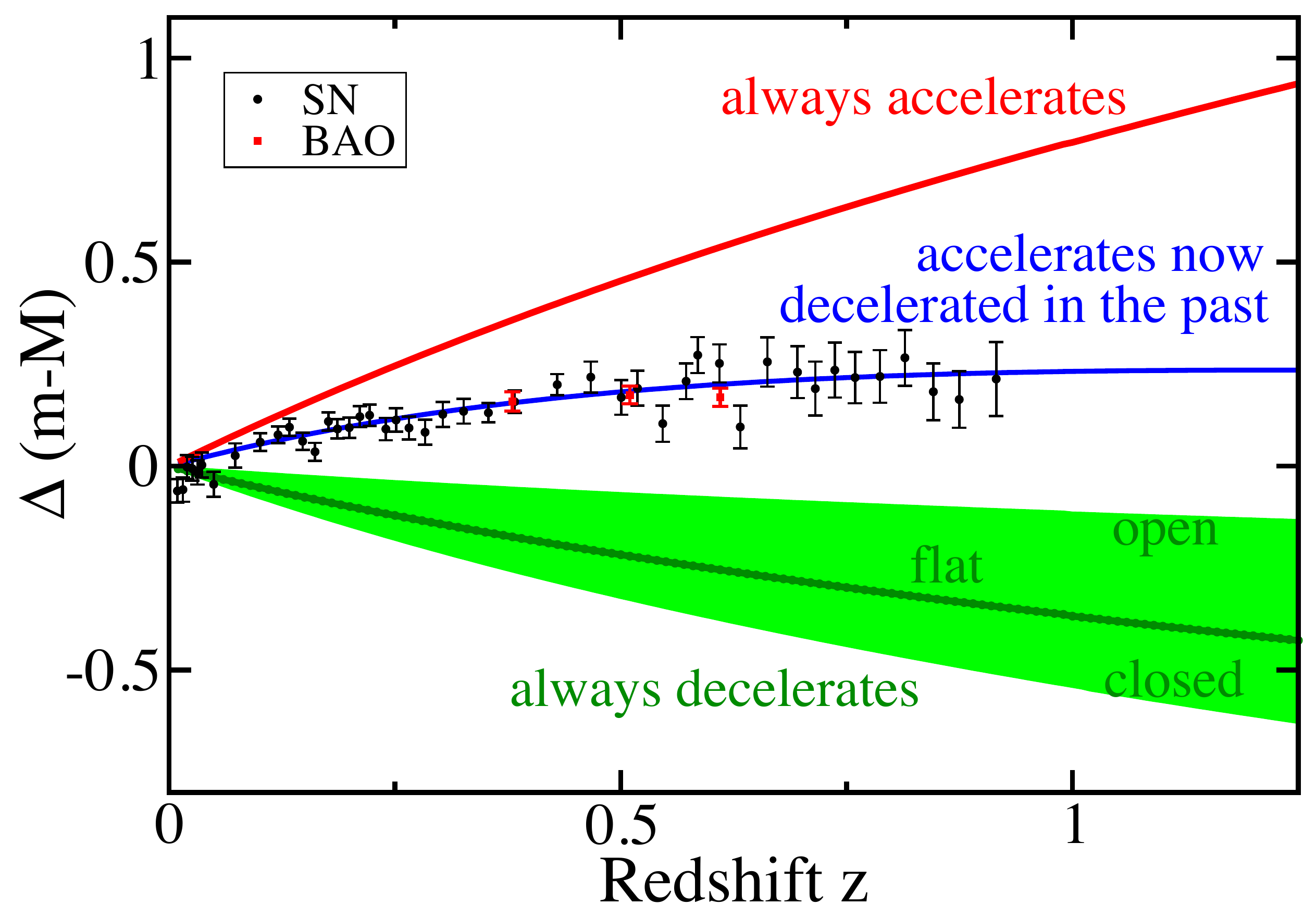}
\caption{Evidence for the transition from deceleration in the past to acceleration today. The blue line indicates a model that fits the data well; it features acceleration at relatively late epochs in the history of the universe, beginning a few billion years ago but still billions of years after the big bang. For comparison, we also show a range of matter-only models in green, corresponding to $0.3 \leq \Omega_m \leq 1.5$ and thus spanning the open, flat, and closed geometries without dark energy. Finally, the red curve indicates a model that \emph{always} exhibits acceleration and that also does not fit the data. The black data points are binned distance moduli from the Supercal compilation \cite{Supercal} of 870~SNe, while the three red data points represent the distances inferred from the most recent BAO measurements (BOSS DR12 \cite{Alam:2016hwk}).}
\label{fig:deceleration}
\end{figure*}

While generally in agreement with measurements, such a low-density universe \emph{still} conflicted with the ages of globular clusters, even setting aside inflationary prejudice for a flat universe. Also, the results were not unambiguous: throughout the 1980s and early 1990s, there was claimed evidence for a much higher matter density from measurements of galaxy fluxes \cite{Loh:1986wg} and peculiar velocities (e.g.\ \cite{1993ApJ...405..437N,Bernardeau:1994vz,Dekel:1993hq}), along with theoretical forays that have since been disfavored, such as inflation models that result in an open universe \cite{Bucher:1994gb,Ratra:1994vw} and extremely low values of the Hubble constant \cite{Bartlett:1994je}. Even the early type Ia supernova studies yielded inconclusive results \cite{Perlmutter_97}.

Attempts to square the theoretical preference for a flat universe with uncertain measurements of the matter density included a proposal for the existence of a nonzero cosmological constant $\Lambda$. This term, corresponding to the energy density of the vacuum, would need to have a tiny value by particle physics standards in order to be comparable to the energy density of matter today. Once considered by Einstein to be the mechanism that guarantees a static universe \cite{Einstein:1917ce}, it was soon disfavored when it became clear that such a static universe is unstable to small perturbations, and it was abandoned once it became established that the universe is actually expanding. Entertained as a possibility in 1980s \cite{peebles84,turner84}, the cosmological constant was back in full force in the 1990s \cite{Efstathiou:1990xe,kofman93,Krauss:1995yb,Ostriker:1995su,Frieman_PNGB,Stompor:1995jd,Coble,Liddle:1995pd}. Nevertheless, it was far from clear that anything other than matter, plus a small amount of radiation, comprises the energy density in the universe today.

A breakthrough came in late 1990s. Two teams of supernova observers, the Supernova Cosmology Project (led by Saul Perlmutter) and the High-Z Supernova Search Team (led by Brian Schmidt) developed an efficient approach to use the world's most powerful telescopes working in concert to discover and follow up supernovae. These teams identified procedures to guarantee finding batches of SNe in each run (for a popular review of this, see \cite{Perlmutter_Schmidt}).

Another breakthrough came in 1993 by Mark Phillips, an astronomer working in Chile \cite{Phillips_93}. He noticed that the SN Ia luminosity (or absolute magnitude) is correlated with the decay time of the SN light curve. Phillips considered the quantity $\Delta m_{15}$, the decay of the light from the SN 15 days after the maximum. He found that $\Delta m_{15}$ is strongly correlated with the SN intrinsic brightness (estimated using other methods). The \textit{Phillips relation} (left panel of figure~\ref{fig:Phillips_rel}) is roughly the statement that ``broader is brighter.'' That is, SNe with broader light curves tend to have a larger intrinsic luminosity. This broadness can be quantified by a ``stretch factor'' that scales the width of the light curve \cite{perlmutter99}. By applying a correction based on stretch (right panel of figure~\ref{fig:Phillips_rel}), astronomers found that the intrinsic dispersion of the SN Ia luminosity, initially $\sim$0.3--0.5~mag, can be reduced to $\sim$0.2~mag after correction for stretch. Note that this corresponds to an error in distance of $\delta d_L / d_L = [\ln(10)/5] \, \delta m \sim 10\%$. The Phillips relation was the second key ingredient that enabled SNe Ia to achieve the precision needed to reliably probe the contents of the universe.

A third important ingredient was the ability to separate intrinsic variation in individual SN luminosities from extinction due to intervening dust along the line of sight, which leads to reddening. This separation requires SN Ia color measurements, achieved by observing and fitting SN Ia light curves in multiple wavebands (e.g.\ \cite{Riess:1996pa}).

The final, though chronologically the first, key step for the discovery of dark energy was the development and application of charge-coupled devices (CCDs) in observational astronomy, and they equipped cameras of increasingly
large size \cite{BTC,Rockosi2002,Miyazaki:2002wa,LSST_camera,panstarrs_camera,MMT,Flaugher:2015pxc}.

Some of the early SN Ia results came in the period 1995--1997 but were based on only a few high-redshift SNe and therefore had large uncertainties (e.g.\ \cite{Perlmutter_97}).

\begin{figure*}[ht]
\centering
\includegraphics[width=0.8\textwidth]{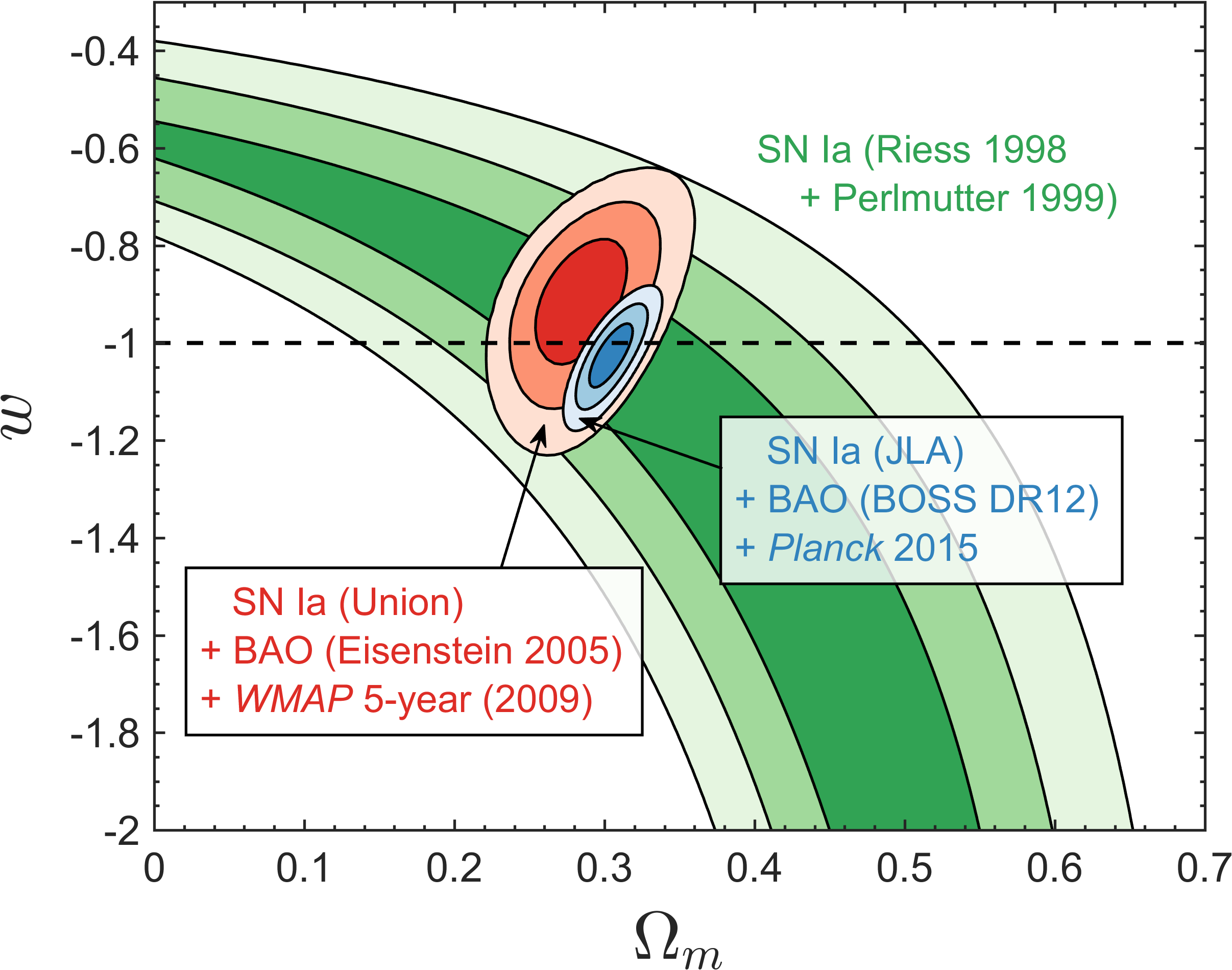}
\caption{History of constraints on key dark energy parameters $\Omega_m$ and a constant equation of state $w$, assuming a flat universe such that $\Omega_\text{de} = 1 - \Omega_m$. The three sets of contours show the status of measurements around the time of dark energy discovery (circa 1998; green), roughly a decade later following precise measurements of CMB anisotropies and the detection of the BAO feature (circa 2008; red), and in the present day, nearly two decades after discovery (circa 2016; blue). Note that, to estimate the 1998 constraints, we analyze a combined set of SNe from the two independent analyses, discarding duplicates but first comparing the very similar low-redshift samples to infer the (arbitrary) magnitude offset between the two.}
\label{fig:om_w_history}
\end{figure*}

\subsection{The discovery of dark energy}

The definitive results, based on 16 \cite{riess98} and 42 \cite{perlmutter99} high-redshift supernovae, followed soon thereafter. The results of the two teams agreed and indicated that the distant SNe are dimmer than would be expected in a matter-only universe. In other words, they were farther away than expected, suggesting that the expansion rate of the universe is increasing, contrary to the expectation for a matter-dominated universe with \emph{any} amount of matter. Over the following decade, larger and better SN samples \cite{Knop:2003iy,Astier:2005qq,WoodVasey:2007jb,Miknaitis:2007jd,Kowalski:2008ez} confirmed and strengthened the original findings, while discoveries of very-high-redshift ($z > 1$) objects played an important role by providing evidence for the expected earlier epoch of deceleration \cite{Riess:2001gk,Riess:2004nr,Riess:2006fw}.

This accelerated expansion of the universe requires the presence of a new component with strongly negative pressure. To see this, consider the \textit{acceleration equation}, which governs the behavior of an expanding universe (see section~\ref{sec:basic} for a more complete introduction to basic cosmology):
\begin{equation*}
\frac{\ddot{a}}{a} = -\frac{4\pi G}{3} \left(\rho + 3p \right) = -\frac{4\pi G}{3} \left(\rho_m + \rho_\text{de} + 3 p_\text{de} \right) \ ,
\label{eq:Friedmann_2}
\end{equation*}
where $\rho$ and $p$ are the energy density and pressure of all components in the universe, including matter and a new component we call dark energy (radiation contributes negligibly at redshifts much less than $\sim$1000, and the pressure of cold dark matter can also be ignored). Accelerated expansion of the universe is equivalent to $\ddot{a} > 0$, and this can happen only when the pressure of the new component is
strongly negative. In terms of the dark energy equation of state $w \equiv
p_\text{de}/\rho_\text{de}$, acceleration only occurs when $w < -1/3 \, (1 +
\rho_m/\rho_\text{de})$; therefore, regardless of matter density, acceleration
never occurs when $w > -1/3$.

Stronger evidence for dark energy has followed in parallel with drastically
improved constraints on other cosmological parameters, particularly by the
cosmic microwave background (CMB) anisotropy measurements and measurements of
the baryon acoustic oscillation (BAO) feature in the clustering of galaxies
(both of which will be discussed at length in
section~\ref{sec:principal_probes}). In figure~\ref{fig:deceleration}, we show the
Hubble diagram (plot of magnitude vs.\ redshift) for modern SN Ia data from
the ``Supercal'' compilation \cite{Supercal}, binned in redshift, along with
recent BAO measurements that also measure distance vs.\ redshift \cite{Alam:2016hwk}, and finally the theory expectation for the
currently favored $\Lambda$-cold-dark-matter ($\Lambda$CDM) model, a $\Lambda$-only model, and matter-only models
without dark energy spanning the open, closed, and flat geometry. In figure~\ref{fig:om_w_history}, we show the evolution of constraints in the plane of
matter density relative to critical $\Omega_m$ and dark energy equation of state $w$, beginning around the time of dark energy discovery (circa 1998), then about a
decade later (circa 2008), and finally in the present day (circa 2016), nearly two decades later.

The discovery of the accelerating universe via SN Ia observations was a
dramatic event that, almost overnight, overturned the previously favored
matter-only universe and pointed to a new cosmological standard model dominated by a 
negative-pressure component. This component that causes the expansion of the universe to accelerate
was soon named ``dark energy'' by cosmologist Michael Turner \cite{Huterer:1998qv}. The
physical nature of dark energy is currently unknown, and the search for it is
the subject of worldwide research that encompasses theory, observation, and
perhaps even laboratory experiments. The physics behind dark energy has connections to
fundamental physics, to astrophysical observations, and to the ultimate fate
of the universe.

\section{Modern cosmology: the basics} \label{sec:basic}

We will begin with a brief overview of the physical foundations of modern cosmology.

The cosmological principle states that, on large enough scales, the universe
is homogeneous (the same everywhere) and isotropic (no special direction). It
is an assumption but also a testable hypothesis, and indeed there is excellent
observational evidence that the universe satisfies the cosmological principle
on its largest spatial scales (e.g.\ \cite{Scrimgeour:2012wt,Laurent:2016eqo}). Under these
assumptions, the metric can be written in the Robertson-Walker (RW) form
\[
ds^2 = dt^2 - a^2(t) \left[\frac{dr^2}{1 - kr^2} + r^2 \left(d\theta^2 + \sin^2\theta \, d\phi^2 \right) \right] ,
\]
where $r$, $\theta$, and $\phi$ are comoving spatial coordinates, $t$ is time,
and the expansion is described by the cosmic scale factor $a(t)$, where the
present value is $a(t_0) = 1$ by convention. The quantity $k$ is the intrinsic
curvature of three-dimensional space; $k = 0$ corresponds to a spatially flat
universe with Euclidean geometry, while $k > 0$ corresponds to positive
curvature (spherical geometry) and $k < 0$ to negative curvature (hyperbolic
geometry).

The scale factor $a(t)$ is a function of the energy densities and pressures of the components that fill the universe. Its evolution is governed by the Friedmann equations, which are derived from Einstein's equations of general relativity using the RW metric:
\begin{align}
H^2 \equiv \left(\frac{\dot a}{a} \right)^2 &= \frac{8\pi G \, \rho}{3} - \frac{k}{a^2} + \frac{\Lambda}{3} \ , \label{eq:FriedmannI} \\
\frac{\ddot{a}}{a} &= -\frac{4\pi G}{3} \left(\rho + 3p \right)  + \frac{\Lambda}{3} \ , \label{eq:FriedmannII}
\end{align}
where $H$ is the Hubble parameter, $\Lambda$ is the cosmological constant term, $\rho$ is the total energy density, and $p$ is the pressure. Note that the cosmological constant $\Lambda$ can be subsumed into the energy density $\rho$, but separating out $\Lambda$ reflects how it was incorporated historically, before the discovery of the accelerating universe.

We can define the critical density $\rho_\text{crit} \equiv 3 H^2/(8 \pi G)$ as the density that leads to a flat universe with $k = 0$. Then the effect of dark energy on the expansion rate can be described by its present-day energy density relative to critical $\Omega_\text{de}$ and its equation of state $w$, which is the ratio of pressure to energy density:
\begin{equation}
\Omega_\text{de} \equiv \frac{\rho_{\text{de}, 0}}{\rho_{\text{crit}, 0}} \ ; \qquad w \equiv \frac{p_\text{de}}{\rho_\text{de}} \ .
\label{eq:Ode}
\end{equation}
The simplest possibility is that the equation of state is constant in time. This is in fact the case for (cold, nonrelativistic) matter ($w_\text{matter} = 0$) and radiation ($w_\text{rad} = 1/3$). However, it is also possible that $w$ evolves with time (or redshift). The continuity equation,
\begin{equation}
\dot{\rho} = - 3 H (\rho + p) \ ,
\end{equation}
is not an independent result but can be derived from \eqref{eq:FriedmannI} and \eqref{eq:FriedmannII}. An expression of conservation of energy, it can used to solve for the dark energy density as a function of redshift for an arbitrary equation of state $w(z)$:
\begin{align}
\rho_\text{de}(z) &= \rho_{\text{de}, 0} \, \exp \left[3 \int_0^z \frac{1 + w(z')}{1 + z'} dz' \right] \label{eq:rhode} \\
&= \rho_{\text{de}, 0} \, (1 + z)^{3(1 + w)} , \nonumber
\end{align}
where the second equality is the simplified result for constant $w$.

The expansion rate of the universe $H \equiv \dot{a}/a$ from
\eqref{eq:FriedmannI} can then be written as
(again for $w = \text{constant}$)
\begin{align}
H^2 = H_0^2 &\left[\Omega_m (1 + z)^3 + \Omega_r (1 + z)^4 \vphantom{\sum} \right. \label{eq:Hz} \\
&+ \left. \Omega_\text{de} (1 + z)^{3(1 + w)} + \Omega_k (1 + z)^2 \right] \ , \nonumber
\end{align}
where $H_0$ is the present value of the Hubble parameter (the Hubble constant), $\Omega_m$ and $\Omega_r$ are the matter and radiation energy densities relative to critical, and the dimensionless curvature ``energy density'' $\Omega_k$ is defined such that $\sum_i \Omega_i = 1$. Since $\Omega_r \simeq 8 \times 10^{-5}$, we can typically ignore the radiation contribution for low-redshift ($z \lesssim 10$) measurements; however, near the epoch of recombination ($z \sim 1000$), radiation contributes significantly, and at earlier times ($z \gtrsim 3300$), it dominates.

\subsection{Distances and geometry}

Observational cosmology is complicated by the fact that we live in an
expanding universe where distances must be defined carefully. Astronomical
observations, including those that provide clues about nature of dark energy,
fundamentally rely on two basic techniques, measuring fluxes from objects and
measuring angles on the sky. It is therefore useful to define two types of
distance, the luminosity distance and the angular diameter distance. The
luminosity distance $d_L$ is the distance at which an object with a certain
luminosity produces a certain flux ($f = L/(4\pi d_L^2)$), while the angular
diameter distance $d_A$ is the distance at which a certain (transverse)
physical separation $x_\text{trans}$ produces a certain angle on the sky
($\theta = x_\text{trans}/d_A$). For a (homogeneous and isotropic) Friedmann-Robertson-Walker universe,
the two are closely related and given in terms of the comoving distance
$r(z)$:
\begin{equation}
d_L(z) = (1 + z) \, r(z) \ ; \qquad d_A(z) = \frac{1}{1 + z} \, r(z) \ .
\label{eq:distances}
\end{equation}
The comoving distance can be written compactly as
\begin{equation}
r(z) = \lim_{\Omega_k' \to \Omega_k} \frac{c}{H_0 \sqrt{\Omega_k'}} \, \sinh \left[\sqrt{\Omega_k'}{\int_0^z \frac{H_0}{H(z')} \, dz'} \right] ,
\label{eq:rz}
\end{equation}
which is valid for all $\Omega_k$ (positive, negative, zero) and where $H(z)$ is the Hubble parameter (e.g.\ \eqref{eq:Hz}). 

Having specified the effect of dark energy on the expansion rate and the distances, its effect on any quantity that fundamentally only depends on the expansion rate can be computed. For example, number counts of galaxy clusters are sensitive to the volume element of the universe, given by
\[
\frac{dV}{dz \, d\Omega} = \frac{r^2(z)}{H(z)/c} \ ,
\]
where $dz$ and $d\Omega$ are the redshift and solid angle intervals, respectively. Similarly, some methods rely on measuring ages of galaxies, which requires knowledge of the age-redshift relation. The age of the universe for an arbitrary scale factor $a = 1/(1 + z)$ is given by
\[
t(a) = \int_0^a \frac{da'}{a' H(a')} \ .
\]

We will make one final point here. Notice from \eqref{eq:rz} that, when
calculating distance, the dark energy parameters $\Omega_\text{de}$ and $w$
are hidden behind an integral (and behind \emph{two} integrals when a general
$w(z)$ is considered; see \eqref{eq:rhode}). The Hubble parameter $H(z)$ is in the integrand of the distance formula and therefore requires one fewer integral to calculate; it depends more directly on the dark energy parameters. Therefore, direct measurements of
  the Hubble parameter, or of quantities that depend directly on $H(z)$, are
  nominally more sensitive to dark energy than observables that fundamentally
  depend on distance. Unfortunately, measurements of $H(z)$ are more difficult
  to achieve and/or are inferred somewhat indirectly, such as from
  differential distance measurements.

\subsection{Density fluctuations}

Next we turn to the growth of matter density fluctuations,
$\delta\equiv\delta\rho_m/\rho_m$. Assuming that general relativity holds, and
assuming small matter density fluctuations $|\delta| \ll 1$ on length scales
much smaller than the Hubble radius, the temporal evolution of the
fluctuations is given by
\begin{equation}
\ddot{\delta_k} + 2H\dot{\delta_k} - 4\pi G \rho_\text{m} \delta_k = 0 \ ,
\label{eq:growth}
\end{equation}
where $\delta_k$ is the Fourier component\footnote{Given our assumptions, each
  wavenumber evolves independently, though this does not always hold for modified
  theories of gravity, even in linear theory.} corresponding to the mode with
wavenumber $k \simeq 2\pi/\lambda$. In \eqref{eq:growth}, dark energy
enters twofold: in the friction term, where it affects $H$; and in the source 
term, where it reduces $\rho_\text{m}$. For $H(z)$ normalized at high redshift, dark energy
increases the expansion rate at $z \lesssim 1$, stunting the growth of density
fluctuations.

\begin{figure*}[ht]
\centering
\includegraphics[width=0.9\textwidth]{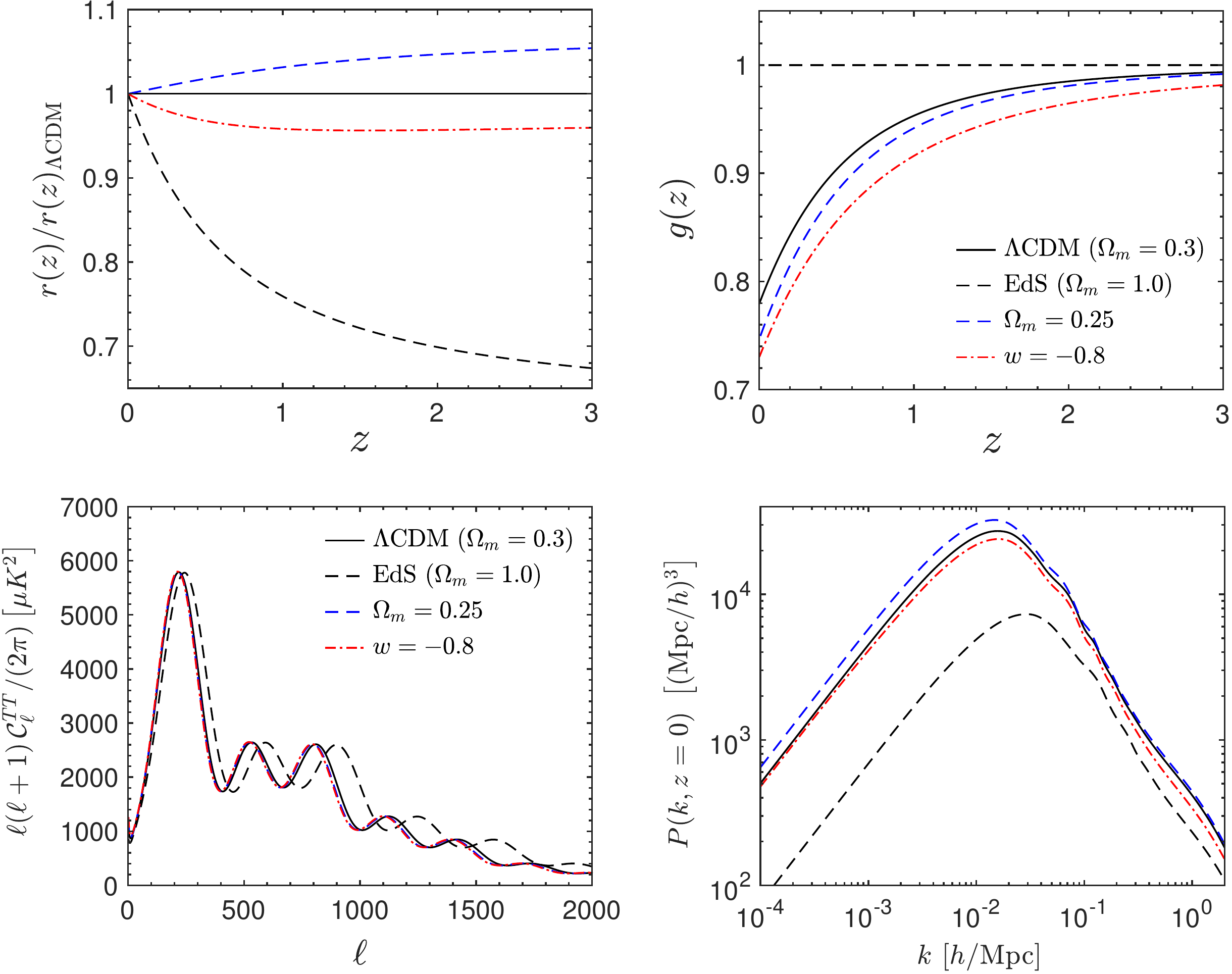}
\caption{Dependence of key cosmological observables on dark energy. The top
  left and right panels show, respectively, the comoving distance
  \eqref{eq:rz} and growth suppression (relative to the matter-only case) from
  \eqref{eq:Da_ga}. The bottom left and right panels show, respectively, the
  CMB angular power spectrum $\mathcal{C}_\ell$ as a function of multipole
  $\ell$ and the matter power spectrum $P(k)$ as a function of wavenumber
  $k$. For each observable, we indicate the prediction for a fiducial
  $\Lambda$CDM model ($\Omega_m = 0.3$, $w = -1$) and then illustrate the
  effect of varying the indicated parameter. In each case, we assume a flat
  universe and hold the combination $\Omega_m h^2$ fixed.}
\label{fig:obs}
\end{figure*}

The effect of dark energy on growth is illustrated in the top right panel of figure~\ref{fig:obs}, where we show the growth-suppression factor $g(z)$, which indicates the amount of growth relative to that in an Einstein-de Sitter universe, which contains no dark energy. It is implicitly defined with respect to the scaled linear growth of fluctuations,
\begin{equation}
D(a) \equiv \frac{\delta(a)}{\delta(1)} \equiv \frac{a \, g(a)}{g(1)}.
\label{eq:Da_ga}
\end{equation}
With only matter, $D(a) = a$ and $g(a) = 1$ at all times. In the presence of dark
energy, $g(a)$ falls below unity at late times. In the currently favored
$\Lambda$CDM model, $g(1) \simeq 0.78$. The value of the density fluctuation
$\delta$ at scale factor $a$, relative to the matter-only case, is suppressed
by a factor $g(a)$, while the two-point correlation function is suppressed by
$g^2$.

A useful alternative expression for the growth suppression is
\begin{equation}
g(a) = \exp \left[\int_0^a \frac{da'}{a'} \left(f(a') - 1 \right) \right] \ ,
\label{eq:ga}
\end{equation}
where
\begin{equation}
f(a) \equiv \frac{d\ln D}{d\ln a} \approx \Omega_m(a)^\gamma
\label{eq:fa}
\end{equation}
is the growth rate which, as we will see below, contains very important
sensitivity to both dark energy parameters and to modifications to gravity.
The latter, approximate equality in \eqref{eq:fa} is remarkably accurate
provided $\gamma\simeq 0.55$. While this functional form for $f(a)$ had been
noted long, in the context of the matter-only universe, before the discovery
of dark energy \cite{1980lssu.book.....P}, the formula remains percent-level
accurate even for a wide variety of dark energy models with varying equations
of state as long as we set $\gamma = 0.55 + 0.05 [1+w(z=1)]$
\cite{Linder_growth,LinCah07}.

The two point function is often phrased in terms of the Fourier
transform of the configuration-space two-point function --- the matter power
spectrum, defined via
\begin{equation}
\langle \delta_{\vec{k}} \, \delta_{\vec{k}'}^* \rangle = (2\pi)^3 \,
\delta^{(3)}(\vec{k}-\vec{k}')\,P(k)
\end{equation}
where we note that $P(\vec{k}) = P(k)$ due to homogeneity.  We can write the
general formula for the power spectrum of density fluctuations in the
dimensionless form $\Delta^2(k) \equiv k^3 P(k) / (2\pi^2)$ as
\begin{align}
\Delta^2(k, a) = A \, &\frac{4}{25} \frac{1}{\Omega_m^2} \left(\frac{k}{k_\text{piv}} \right)^{n - 1} \left(\frac{k}{H_0} \right)^4 \label{eq:Deltasq} \\[1ex]
&\times [a g(a)]^2 \, T^2(k) \, T_\text{nl}(k) \, , \nonumber
\end{align}
where $A$ is the normalization of the power spectrum (current constraints favor $A \approx 2.2 \times 10^{-9}$), $k_\text{piv}$ is the ``pivot'' wavenumber around which
we compute the spectral index $n$\footnote{The \textit{Planck} analysis uses $k_\text{piv} = 0.05~h \, \text{Mpc}^{-1}$, where $h$ is the dimensionless Hubble constant ($H_0 = 100 h~\text{km/s/Mpc}$), but beware that $k_\text{piv} = 0.002~\text{Mpc}^{-1}$ is occasionally used.}, and $[a g(a)]$ is the linear growth of perturbations. $T(k)$ is the transfer function, which is constant for modes that entered the horizon before the matter-radiation equality (comoving wavenumber $k \lesssim 0.01~h \, \text{Mpc}^{-1}$) and scales as $k^{-2}$ at smaller scales that entered the horizon during radiation domination. Finally, $T_\text{nl}$ indicates a prescription for the \textit{nonlinear} power spectrum, which is usually calibrated from N-body simulations. Recent analytic fitting formulae for this term were given in \cite{Takahashi:2012em,Mead:2016zqy}.

Finally, we outline the principal statistic that describes the distribution of hot and cold spots in the cosmic microwave background anisotropies. The angular power spectrum of the CMB
anisotropies is essentially a projection along the line of sight of the primordial matter power spectrum. Adopting the expansion of the temperature anisotropies on the sky in terms of the complex coefficients $a_{\ell m}$,
\begin{equation}
\frac{\delta T}{T}(\mathbf{\hat{n}}) = \sum_{\ell = 2}^\infty \sum_{m = -\ell}^\ell a_{\ell m} \, Y_{\ell m}(\mathbf{\hat{n}}) \, ,
\label{eq:harmonic}
\end{equation}
we can obtain the ensemble average of the two point correlation function of
the coefficients $C_\ell \equiv \langle |a_{\ell m}|^2 \rangle$ as
\begin{equation*}
C_\ell = 4\pi \int\Delta^2(k) j^2_\ell(kr_*) d\ln k \, ,
\end{equation*}
where $j_\ell$ is the spherical bessel function and $r_*$ is the radius of the
sphere onto which we are projecting (the comoving distance to
recombination); in the standard model, $r_* \simeq 14.4~\text{Gpc}$. Physical structures
that appear at angular separations $\theta$ roughly correspond to power at multipole $\ell \simeq \pi/\theta$.

Basic observables and their variation when a few basic parameters governing dark energy are varied are shown in figure~\ref{fig:obs}. To illustrate the effects
of variations in the dark energy model, we compare the following four models:
\begin{enumerate}
  \item Flat model with matter density $\Omega_m=0.3$, equation of state
    $w=-1$, and other parameters in agreement with the most recent
    cosmological constraints \cite{Ade:2015xua}.
 \item Same as (i), but with $\Omega_m = 1$. This is the Einstein-de Sitter
   model, flat and matter dominated with no dark energy. We hold all other
   parameters, including the combination $\Omega_m h^2$, fixed to their best-fit values in (i).
 \item Same as (i), except $\Omega_m = 0.25$.
 \item Same as (i), but with $ w = -0.8$.
\end{enumerate}

\section{Parametrizations of dark energy} \label{sec:parameters}

\subsection{Introduction}
Given the lack of a consensus model for cosmic acceleration, it is a challenge to provide a simple yet unbiased and sufficiently general description of dark energy.  The equation-of-state parameter $w$ has traditionally been identified as one useful phenomenological description; being the ratio of pressure to energy density, it is also
closely connected to the underlying physics.  Many more general
parametrizations exist, some of them with appealing statistical properties.  We
now review a variety of formalisms that have been used to describe and
constrain dark energy.

We first describe parametrizations used to describe the effects of dark energy
on observable quantities. We then discuss the reconstruction of the
dark-energy history; the principal-component description of it; the figures of
merit; and descriptions of more general possibilities beyond spatially smooth
dark energy, including modified gravity models. We end by outlining two
strategies to test the internal consistency of the currently favored
$\Lambda$CDM model.

\subsection{Parametrizations}
Assuming that dark energy is spatially smooth, its simplest parametrization is in terms of its
equation-of-state \cite{1988A&A...206..175L,Turner_White}
\begin{equation}
w \equiv \frac{p_\text{de}}{\rho_\text{de}} = \text{constant}.
\end{equation}
This form describes vacuum energy ($w = -1$) and topological defects ($w =
-N/3$, where $N$ is the integer dimension of the defect and takes the value 0, 1, or 2 for monopoles, cosmic strings, or textures, respectively). Together with $\Omega_\text{de}$, $w$ provides a two-parameter
description of the dark-energy sector. However, it does not describe models which have a time-varying $w$, such as scalar field dark energy or
modified gravity, although cosmological observables are often sufficiently
accurately described by a constant $w$ even for models with mildly varying
$w(z)$.

Promoting either the dark energy density or the equation of state to a general
function of redshift --- $\Omega_\text{de}(z)$ or $w(z)$ --- would be the most general way
to describe dark energy, still assuming its spatial homogeneity. In practice,
however, either of these functions formally corresponds to infinitely many parameters to
measure, and measuring even a few such parameters is a challenge. Perhaps not
surprisingly, therefore, the most popular parametrizations of $w$ have
involved two free parameters. One of the earliest and simplest such
parametrizations is linear evolution in redshift, $w(z) = w_0 + w' z$
\cite{Cooray_Huterer}.  Other low-dimensional parametrizations have been
proposed \cite{Gerke_Efstat}; for low redshift they are all essentially
equivalent, but for large $z$ they lead to different and often unphysical behavior.  The
parametrization \cite{Linder_wa,Chevallier_Polarski}
\begin{equation}
w(a) = w_0 + w_a (1 - a) = w_0 + w_a \, \frac{z}{1 + z} \ ,
\label{eq:w0wa}
\end{equation}
where $a = 1/(1 + z)$ is the scale factor, avoids this problem, and it fits many scalar
field and some modified gravity expansion histories. This therefore leads to the most commonly used
description of dark energy, namely the three-parameter set $\{\Omega_\text{de}, w_0, w_a \}$. The energy
density is then
\begin{equation}
\frac{\rho_\text{de}(a)}{\rho_{\text{crit}, 0}} = \Omega_\text{de} \, a^{-3 (1 + w_0 + w_a)} e^{-3 w_a (1 - a)}.
\end{equation}
Constraints on $w(z)$ derived from individual, marginalized constraints on $w_0$ and $w_a$ are shown in the left panel of figure~\ref{fig:wz}.

More general expressions have been proposed (e.g.\ \cite{Sahni:2002fz,Corasaniti_foundations}); however, introducing additional parameters makes the equation of state very difficult to measure, and such extra parameters are often \textit{ad hoc} and unmotivated from either a theoretical or empirical point of view.

\begin{figure*}[ht]
\centering
\includegraphics[width=0.49\textwidth]{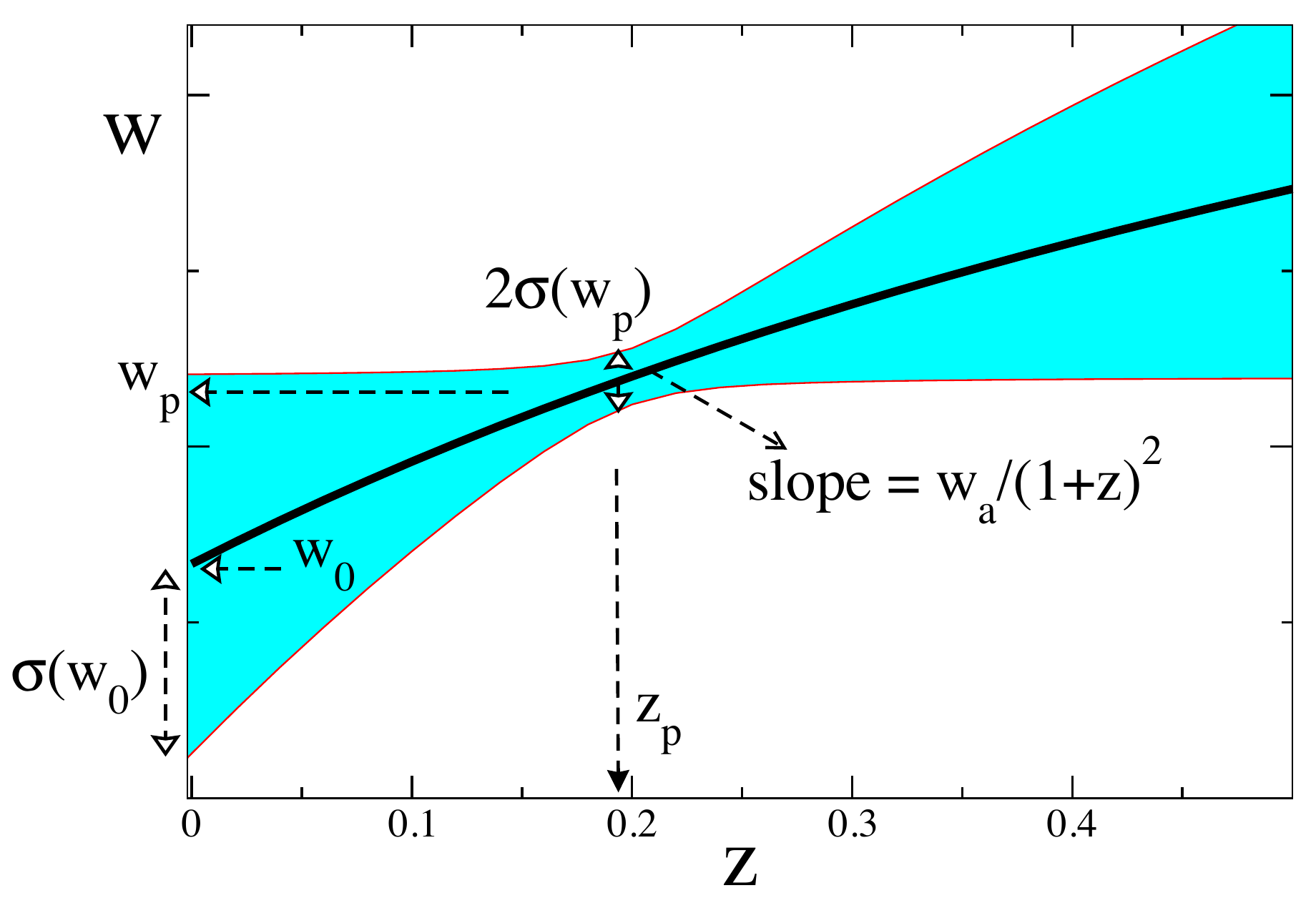}
\hspace{0.2cm}
\includegraphics[width=0.47\textwidth]{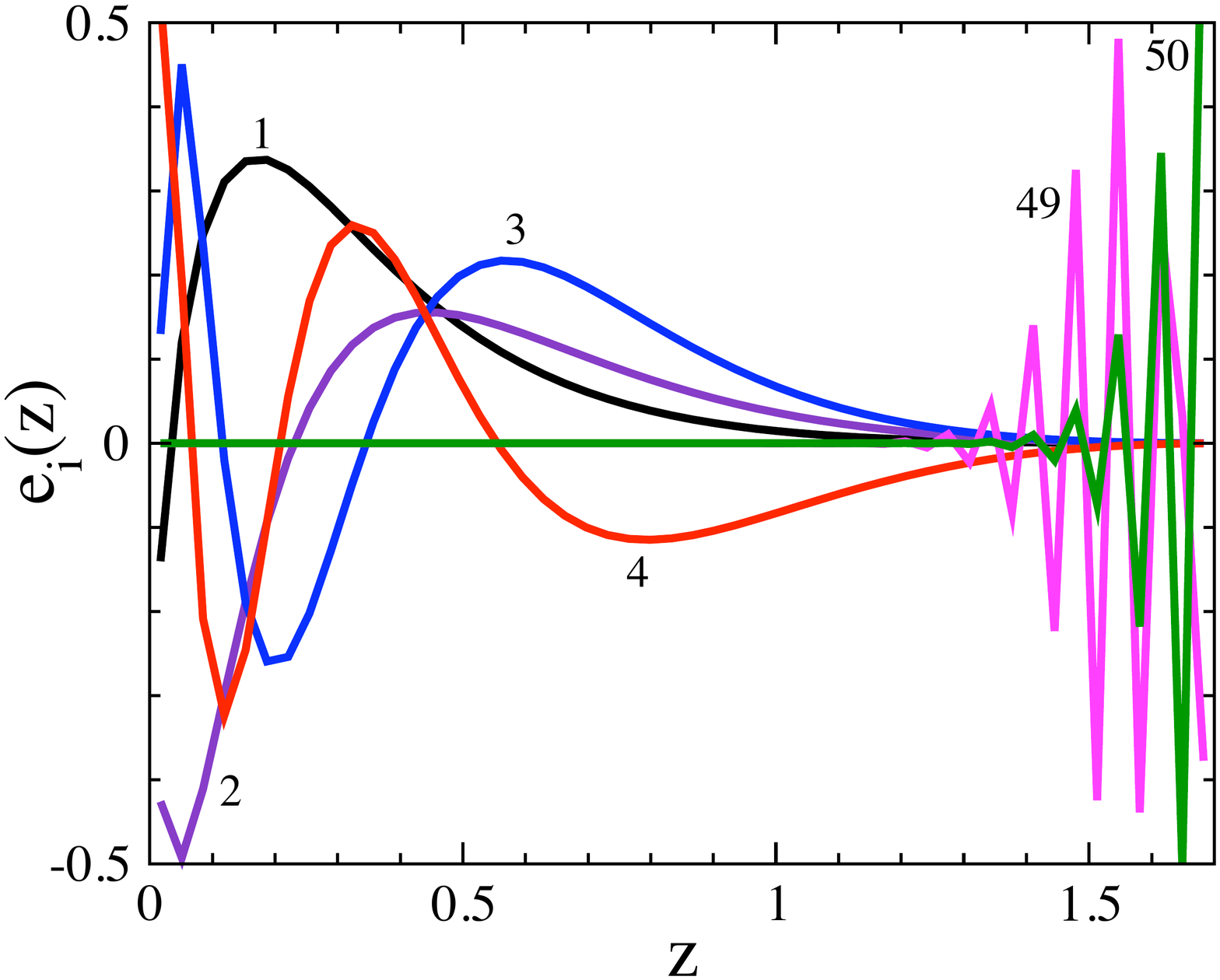}
\caption{\textit{Left panel}: Illustration of the main features of the popular parametrization of the equation of state \cite{Linder_wa,Chevallier_Polarski} given by $w(z) = w_0 + w_a \, z/(1 + z)$. We indicate the pivot redshift $z_p$, the corresponding value of the equation of state $w_p$, the intercept $w_0$, the slope (proportional to $w_a$), and a visual interpretation of the approximate uncertainties in $w_0$ and $w_p$. \textit{Right panel}: The four best-determined (labeled 1--4) and two worst-determined (labeled 49-50) principal components of $w(z)$ for a future SN Ia survey with several thousand SNe in the redshift range $0 < z < 1.7$; reproduced from \cite{Huterer_Starkman}.}
\label{fig:w0wp_PC}
\end{figure*}

\subsection{Pivot redshift}
Two-parameter descriptions of $w(z)$ that are linear in the parameters entail
the existence of a ``pivot'' redshift $z_p$ at which the measurements of the
two parameters (e.g.\ $w_0$ and $w_a$) are uncorrelated and the error in $w_p \equiv w(z_p)$ is
minimized. Essentially, $z_p$ indicates the redshift at which the error on
$w(z)$ is tightest, for fixed assumptions about the data. This is illustrated
in the left panel of figure~\ref{fig:w0wp_PC}. Writing the
equation of state in \eqref{eq:w0wa} in the form
\begin{equation}
w(a) = w_p + (a_p - a) w_a \ ,
\end{equation}
it is easy to translate constraints from the ($w_0, w_a$) to ($w_p, w_a$)
parametrization, as well as determine $a_p$ (or $z_p$), for any particular
data set. In particular, if $\mathbf{C}$ is the $2 \times 2$ covariance matrix
for $\{w_0, w_a \}$ (other parameters marginalzied over), then the pivot
redshift is given by \cite{Albrecht:2009ct}
\begin{equation}
z_p = -\frac{C_{w_0 w_a}}{C_{w_0 w_a} +  C_{w_a w_a}} \ ,
\end{equation}
while the variance at the pivot is given by
\begin{equation}
\sigma^2(w_p) = C_{w_0 w_0} - \frac{C_{w_0 w_a}^2}{C_{w_a w_a}} \, .
\end{equation}
Measurements of the equation of state at the
pivot point often  provides  the most useful information in ruling out models
(e.g.\ ruling out $w = -1$). Note that the pivot redshift (and all associated
quantities, such as $\sigma(w_p)$) depend on the choice of cosmological probes
and the specific data set used, therefore describing the quantities that are
best measured by that data.

\begin{figure*}[ht]
\centering
\includegraphics[width=0.9\textwidth]{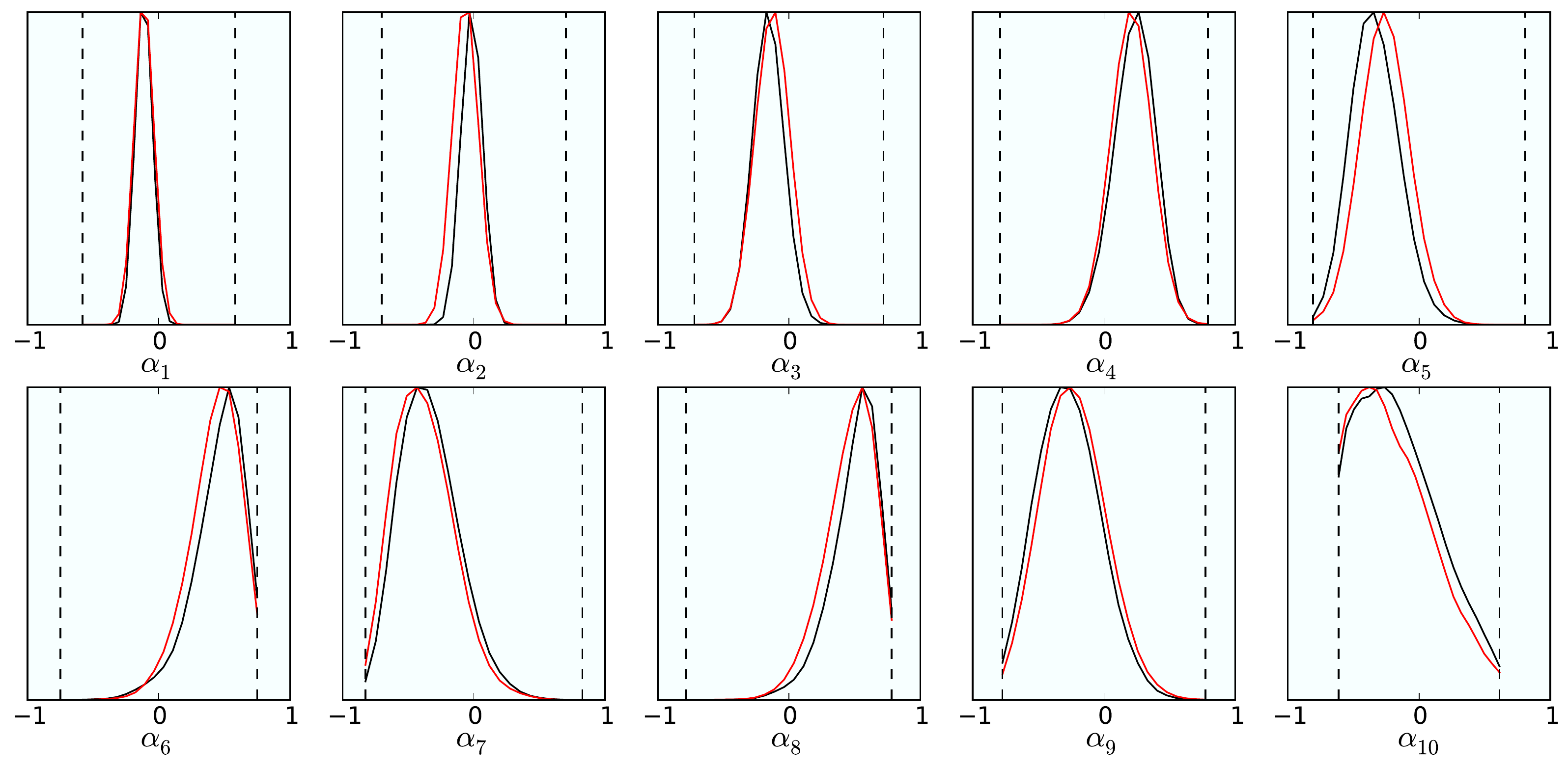}
\caption{Marginalized posterior likelihoods for the first 10 principal components of the dark energy equation of state, based on recent (2012) data (SN Ia + BAO + CMB) and reproduced from \cite{Ruiz:2012rc}. The dashed vertical lines represent the hard prior limits. Black curves indicate constraints when considering only the uncorrelated, statistical SN Ia uncertainties, while red curves correspond to an analysis using the full SN Ia covariance matrix, including all systematic uncertainties.}
\label{fig:pc_1d}
\end{figure*}

\subsection{Principal components}
The cosmological function that we would like to determine --- $w(z)$, $\rho_\text{de}(z)$, or $H(z)$ --- can be expanded in terms of principal components, a set of functions that are uncorrelated and orthogonal by construction \cite{Huterer_Starkman}. In this approach, the data determine
which parameters are measured best.

Suppose we parametrize $w(z)$ in terms of piecewise constant
values $w_i$ ($i = 1, \dots, N$), each defined over a narrow redshift range $z_i < z < z_i + \Delta z$). In the limit of small $\Delta z$ this recovers the
shape of an arbitrary dark energy history (in practice, $N \gtrsim 20$ is
sufficient \cite{Mortonson:2008qy}), but the estimates of the $w_i$ from a given dark energy probe will be very noisy. Principal component analysis (PCA) extracts from those noisy estimates the best-measured features of $w(z)$. One finds the eigenvectors $e_i(z)$ of the inverse covariance matrix for the parameters $w_i$ and the
corresponding eigenvalues $\lambda_i$. The equation-of-state parameter is then
expressed as
\begin{equation}
1 + w(z) = \sum_{i = 1}^N \alpha_i \, e_i(z) \ ,
\label{eq:w_expand}
\end{equation}
where the $e_i(z)$ are the principal components.  The coefficients $\alpha_i$,
which can be computed via the orthonormality condition $\alpha_i = \int (1 + w(z)) e_i(z) dz$, are each determined with an accuracy
$1/\sqrt{\lambda_i}$. Several of these components are shown for a future SN survey in the right panel of figure~\ref{fig:w0wp_PC}, while measurements of the first ten PCs of the equation of state from recent data are shown in figure~\ref{fig:pc_1d}.

There are multiple advantages to the PC approach for dark energy (when measuring either the equation of state $w(z)$ or $\rho_\text{de}(z)$ or $H(z)$). First, the method is
as close to model-independent as one can realistically get, as no information about the temporal dependence of these functions has been assumed \textit{a priori}\footnote{Of course, one still typically makes implicit assumptions about the speed of sound of dark energy, anisotropic stresses, etc., so the method is not truly model-independent.}. In essence, we are asking the data to tell us what we measure and how well we measure it; there are no arbitrary parametrizations imposed. Second, one can use this approach to optimize survey design --- for example, design a
survey that is most sensitive to the dark energy equation of state parameter in some specific redshift interval. Finally, PCs make it straightforward to quantify how many independent parameters can be measured by a given combination of cosmological probes (e.g.\, for how many PCs is $\sigma_{\alpha_i}$ or $\sigma_{\alpha_i}/\alpha_i$ less than some threshold value \cite{dePutter_Linder}).

\begin{figure*}[ht]
\centering
\includegraphics[width=0.49\textwidth]{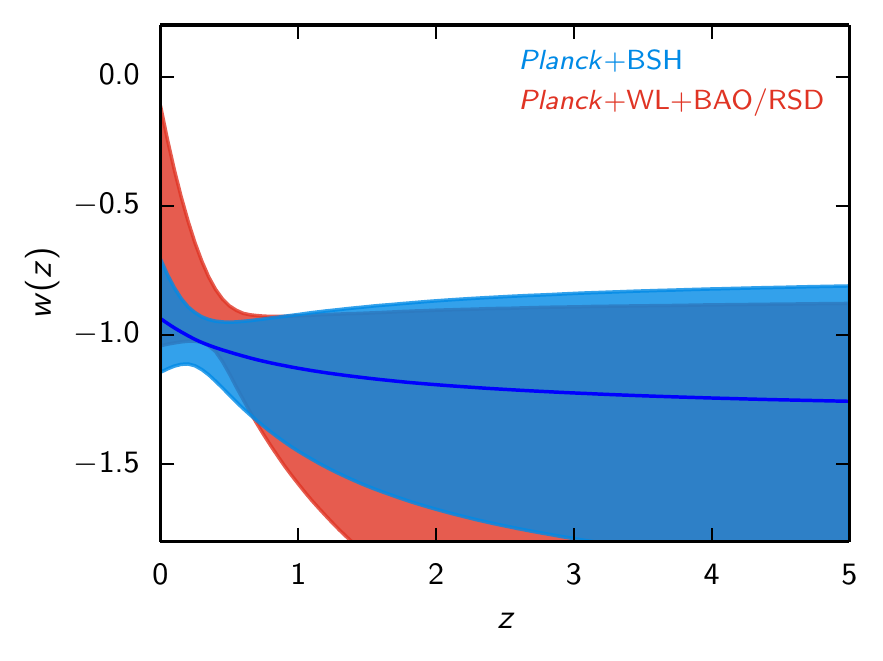}
\includegraphics[width=0.47\textwidth]{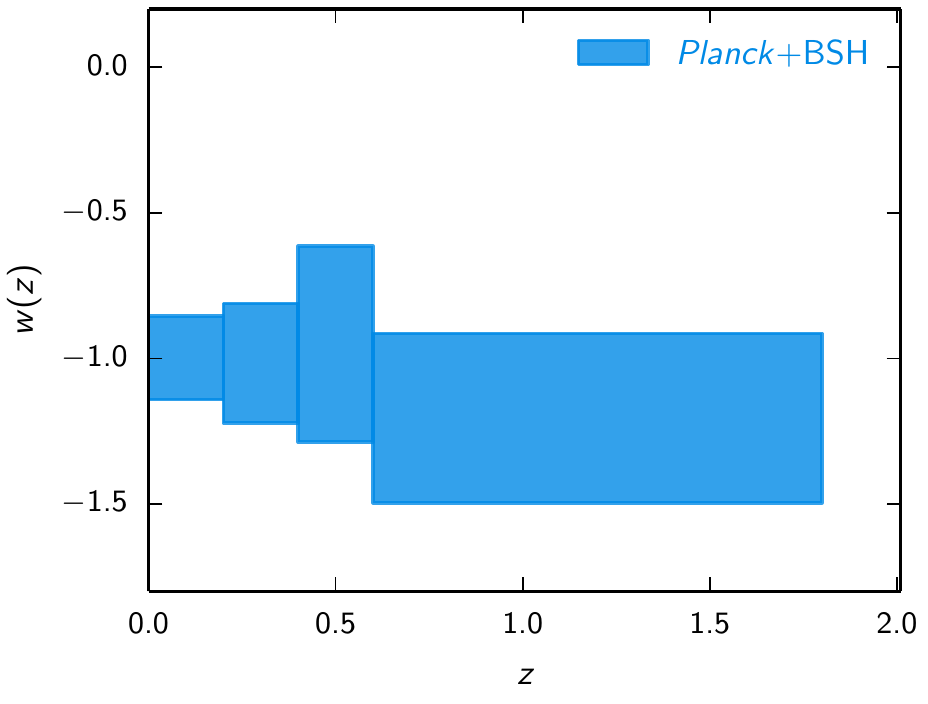}
\caption{Constraints on the redshift evolution of the dark energy equation of state, reproduced from the \textit{Planck} 2015 analysis \cite{Ade:2015rim}. \textit{Left panel}: Constraints on $w(z)$ assuming the parametrization $w(a) = w_0 + w_a (1 - a)$. The blue curve represents the best-fit model, and the shaded regions indicate models allowed at 95\% confidence level. \textit{Right panel}: Constraints on four uncorrelated bins of $w(z)$, using the formalism of \cite{Huterer_Cooray}. The shaded boxes indicate the mean and uncertainty for $w$ across each redshift range. In both cases, information from BAO, SN Ia, and Hubble constant (BSH) measurements is combined with the \textit{Planck} data. Also shown in the left panel is the result when information from measurements of weak lensing (WL) and redshift-space distortions (RSD) is used instead.}
\label{fig:wz}
\end{figure*}

There are a variety of extensions of the PCA method, including measurements of
the uncorrelated equation-of-state parameters \cite{Huterer_Cooray} or other
quantities such as the linear growth of density fluctuations \cite{Hu_PC} that also have the feature of being localized in redshift intervals, or generalizing principal components to functions in both redshift
$z$ and wavenumber $k$ \cite{Hojjati:2011xd,Zhao:2009fn}. The right
  panel of figure~\ref{fig:wz} shows constraints on four uncorrelated
  bins of $w(z)$ from an analysis that combines CMB, BAO, SN Ia, and 
  Hubble constant measurements \cite{Ade:2015rim}.

\subsection{Direct reconstruction}  It is tempting to consider the possibility that measurements of the comoving distance to a range of redshifts, such as those from SNe Ia, can be used to invert \eqref{eq:rz} and \eqref{eq:rhode} and obtain either $\rho_\text{de}(z)$ or $w(z)$ in full generality,
without using any parametrization. This program goes under the name of direct reconstruction \cite{Huterer:1998qv,Saini_reconstr, Nakamura_Chiba, Starobinsky}. The inversion is indeed analytic and the equation of state, for example, is given in terms of the first and second derivatives of the comoving distance as
\begin{equation}
1 + w(z) = \frac{1 + z}{3} \, \frac{3 H_0^2 \Omega_m (1 + z)^2 + 2 (d^2 r/d z^2)(dr/dz)^{-3}}{H_0^2 \, \Omega_m (1 + z)^3 - (dr/dz)^{-2}} \ .
\label{eq:wz_reconstr}
\end{equation}
Assuming that dark energy is due to a single rolling scalar field, the scalar
potential $V(\phi)$ can also be reconstructed.
\begin{align}
V \left[\phi(z) \right] &= \frac{1}{8\pi G} \left[\frac{3}{(dr/dz)^2} + (1 + z) \frac{d^2 r / dz^2}{(dr/dz)^3} \right] \\
&- \frac{3 \Omega_m H_0^2 (1 + z)^3}{16\pi G} \ . \nonumber
\label{eq:Vphi_reconstr}
\end{align}
One can also reconstruct the dark energy density \cite{Wang:2003gz, Wang_Teg_uncorr}, which depends only on the first derivative of distance with respect to redshift,
\begin{equation}
\rho_\text{de}(z) = \frac{3}{8\pi G} \left[\frac{1}{(dr/dz)^2} - \Omega_m H_0^2 (1 + z)^3 \right] \ .
\end{equation}

Direct reconstruction in its conceptual form is truly model-independent, in the sense that it does not require any assumptions about the functional form of time variation of dark energy density. However, to make
it possible in practice, one has to regularize the distance derivatives, since
these are derivatives of noisy data. One must fit the distance data with a smooth function (e.g., a polynomial, a Pad\'{e}
approximant, or a spline with tension), and the fitting process introduces
systematic biases. While a variety of methods have been pursued
\cite{Hut_Tur_00,Weller_Albrecht}, the consensus is that direct reconstruction
is simply not robust even with SN Ia data of excellent quality. Nevertheless,
sufficiently strong priors on the behavior of the equation of state, coupled
with advanced statistical treatments, can lead to successful (though somewhat
smoothing-model dependent) reconstructions of $w(z)$
\cite{Holsclaw:2010sk,Crittenden:2011aa,Shafieloo:2012ht,Zhao:2012aw,Shafieloo:2012ms}.

Although the expression for $\rho_\text{de}(z)$ involves only first derivatives of $r(z)$ and is therefore easier to reconstruct, it contains little information about the nature of dark energy. Dark energy reconstruction methods have been reviewed in \cite{Sahni_review}.

\subsection{Figures of Merit}
It is useful to quantify the power of some probe,
survey, or combination thereof, to measure dark energy properties. This is
typically achieved by defining some function of the error bars or covariances of
parameters describing dark energy and calling it the ``figure of merit'' (FoM). Such a quantity necessarily only paints a limited picture of the power of some
probe or experiment because it is a single number whose relative size depends on its very definition (for example, the weighting of dark
energy properties in redshift). Nevertheless, FoMs, if judiciously chosen, are useful
since they can dramatically simplify considerations
about various survey specifications or survey complementarities.

The most commonly adopted figure of merit is that proposed by the Dark Energy
Task Force (\cite{DETF}; see \cite{Hut_Tur_00} for the original
proposal). This DETF FoM is the inverse of the allowed area in the $w_0$--$w_a$
plane. For uncorrelated $w_0$ and $w_a$, this quantity would be $\propto
1/(\sigma_{w_0} \, \sigma_{w_a})$; because these two parameters are
typically correlated, the FoM can be defined as
\begin{equation}
\text{FoM} \equiv |\mathbf{C}|^{-1/2} \approx \frac{6.17 \pi}{A_{95}},
\label{eq:DETF_FoM}
\end{equation}
where $\mathbf{C}$ is the $2 \times 2$ covariance matrix for $(w_0, w_a)$ after marginalizing over all other parameters, and $A_{95}$ is the area of the 95.4\%~CL region in the $w_0$--$w_a$ plane. Note that the constant of proportionality is not important; when we compare FoMs for different surveys, we consider the FoM ratio in which the constant disappears.

While the DETF FoM defined in \eqref{eq:DETF_FoM} contains some
information about the dynamics of dark energy (that is, the time variation of $w(z)$),
several more general FoMs have been proposed. For example,
a more general FoM is inversely proportional to the volume of the
$n$-dimensional ellipsoid in the space of principal component parameters;
$\text{FoM}^\text{PC}_n \equiv \left(|\mathbf{C}_n| / |\mathbf{C}_n^\text{prior}| \right)^{-1/2}$ \cite{MHH_FoM}, where the prior covariance matrix is again unimportant since it cancels out when computing FoM ratios.

\subsection{Generalized dark energy phenomenology}
The simplest and by far the most studied class of models is dark energy that is spatially smooth
and its only degree of freedom is its energy density --- that is, it is fully
described by either $\rho_\text{de}(a)$ or $w(a) \sim -1$. More general possibilities
exist however, as the stress-energy tensor allows considerably more freedom
\cite{Hu_GDM,Hu:1998tj}.

One possibility is that dark energy has the speed of sound that allows
clustering at sub-horizon scales, that is, $c_s^2 \equiv \delta p_\text{de}/\delta \rho_\text{de} < 1$ (where $c_s$ is quoted in units of the speed of
light)
\cite{Erickson:2001bq,DeDeo:2003te,Weller:2003hw,Hannestad_sound}. Unfortunately,
the effects of the speed of sound are small, and become essentially negligible
in the limit when the equation of state of dark energy $w$ becomes close to
$-1$, and are difficult to discern with late-universe measurements even 
if $w$ deviates from the cosmological constant value at some epoch. It
will therefore be essentially impossible to measure the speed of sound even
with future surveys; see illustrations of the changes in the observables and
forecasts in \cite{dePutter:2010vy}.

Another possibility is the presence of ``early dark energy''
\cite{Wetterich:2004pv,Doran:2006kp,Linder:2006ud}, component that is
non-negligible at early times, typically around recombination or even earlier.
The early component is motivated by various theoretical models (e.g. scalar
fields \cite{Zlatev:1998tr}), and could imprint signatures via the early-time
Integrated Sachs-Wolfe effect. While the \textit{ acceleration} in the redshift
range $z\in [ 1, 10^5]$ is already ruled out \cite{Linder:2010wp}, of order a
percent contribution to the energy budget by early dark energy is still
allowed \cite{Reichardt:2011fv,Ade:2015rim}. In some models, this early
component this component acts like radiation in the early universe
\cite{Calabrese:2011hg}. Increasingly good constraints on models with early
dark energy are on the to-do list for upcoming cosmological probes.

Finally, there is a possibility that dark energy is coupled to dark matter, or
other components or particles (some of the early work is in
e.g.\ \cite{Gradwohl_Frieman,Amendola_99,Farrar_Peebles}). This is a much
richer --- though typically very model-dependent --- set of possibilities,
with many opportunities to test them using data; see
\cite{Wang:2016lxa} for a review.

As yet, there is no observational evidence for generalized dark energy beyond
the simplest model but, as with modified gravity, studying these extensions is
important to understand how dark energy phenomenology can be searched for by
cosmological probes.

\subsection{Descriptions of modified gravity}
Modifications of General Theory of relativity represent a fundamental alternative in describing the apparent
acceleration to the smooth fluid description with a negative equation of
state. In modified gravity (reviewed in this volume by P.\ Brax),
the modification of GR makes an order-unity change in the dynamics at
cosmological scales. At the solar-system scales, the modification of gravity
needs to have a very small or negligible effect --- usually satisfied by
invoking non-linear ``screening mechanisms'' which restore GR in high density
regions --- in order to respect the successful local tests of GR. There exists
a diverse set of proposed modified gravity theories, with very rich set of
potentially new cosmological signatures; for excellent reviews, see
\cite{Silvestri:2009hh,Joyce:2014kja,Joyce:2016vqv}.

Modified gravity affects the clustering of galaxies and changes how mass affects the propagation of light. One can write the metric perturbations via two
potentials $\Phi$ and $\Psi$ as 
\begin{equation}
\label{metric}
ds^2=\left(1+2\Psi\right ) dt^2
-\left(1-2\Phi\right ) a^2(t) d\textbf{x}^2.
\end{equation}

A fairly general way to parametrize modified gravity theories is to specify
the relation of the two metric potentials $\Phi$ and $\Psi$, which govern the
motion of matter and of light, respectively. One possible parametrization is
\cite{Daniel:2012kn}
\begin{align}
  \label{eq:Gmatter}
\nabla^2 \Psi &= 4\pi G_N a^2 \delta \rho \, G_\text{matter}  \\
\nabla^2(\Phi + \Psi) &= 8\pi G_N a^2 \delta \rho \, G_\text{light} 
\label{eq:Glight}
\end{align}
where deviations of dimensionless numbers $G_\text{matter}$ or
$G_\text{light}$ from unity indicate at the very least clustering of dark
energy, while $G_\text{matter}\neq G_\text{light}$ rather robustly alerts us
to possible modifications of General Relativity.  There is a surprisingly
large number of equivalent parametrization conventions in the literature; they
use different symbols, but all effectively describe the difference and ratio
of the two gravitational potentials
(e.g.\ \cite{Caldwell:2007cw,Bertschinger:2008zb,Daniel:2010ky}).  The scale-
and time-dependence of these parameters can be modeled with independent $(z,
k)$ bins \cite{Daniel:2012kn}, eigenmodes \cite{hojjati13062546}, or well
behaved functional forms
\cite{Bean:2010zq,Zhao:2011te,Dossett:2011tn,silvestri13021193}. Note that the
parametrization in equations \eqref{eq:Gmatter}-\eqref{eq:Glight} (and its
various equivalents) is valid on subhorizon scales and in the linear regime, and does
not capture the various screening mechanisms.

There exist various ways of testing gravity which stop short of modeling the
two gravitational potentials, and are therefore potentially simpler to
implement. The simplest such parametrization uses the growth index $\gamma$,
defined in \eqref{eq:fa}; any evidence for $\gamma \neq 0.55$ would point
to departures from the standard cosmological $\Lambda$CDM model
\cite{Hut_Lind_MMG}. Other examples are statistics constructed to be closely
related to the observables measured; for example, the $E_G$ statistic
\cite{Zhang:2007nk,Reyes:2010tr} is a suitably defined ratio of the
galaxy-galaxy lensing clustering amplitude to that of galaxy clustering, 
and it allows a relatively direct link to the modified gravity parameters.

\subsection{Consistency tests of the standard model}

Finally, there are powerful but more phenomenological methods of testing the
consistency of the current cosmological model that do not refer to explicit
parametrizations of modified gravity theory. The general idea behind such
tests is to begin with some widely adopted parametrization of the cosmological
model (say, the $\sim$5-parameter $\Lambda$CDM), then investigate whether
there exist observations that are inconsistent with the theoretical
predictions of the model.  Bayesian statistical tools
\cite{Marshall:2004zd,Trotta:2008qt,Grandis:2015qaa,
Raveri:2015maa,Charnock:2017vcd,Lin:2017ikq,Lin:2017bhs}
are particularly useful to quantify these consistency tests.
  
The simplest approach is to calculate predictions on cosmological functions
that can be measured that are consistent with current parameter constraints
\cite{Mortonson:2008qy,Mortonson:2009hk,Vanderveld:2012ec}. The predictions
depend on the class of models that one is trying to test; for example,
predictions for weak lensing shear power spectrum that assume an underlying
$\Lambda$CDM model are tighter than the weak-lensing predictions that assume
an evolving scalar field model where the equation of state of dark energy is a
free function of time.  Such model-dependent predictions for the observed
quantities are now routinely employed in cosmological data analysis, as they
provide a useful check of whether the newly obtained data fall within those
predictions (e.g.\ \cite{Alam:2016hwk}).

A complementary approach is to explicitly split the cosmological parameters
into those constrained by geometry (e.g,\ distances, as in SNIa and BAO), and
those constrained by the growth of structure (e.g.\ the evolution of
clustering amplitude in redshift)
\cite{Ishak_Upadhye,Wang:2007fsa,Zhang:2003ii}. In this approach, the equation
of state of dark energy $w$, for example, can be split into two separate
parameters, $w_\text{geom}$ and $w_\text{grow}$. These two parameters are then
employed in those terms in theory equations that are based on geometry and
growth, respectively. In this scheme, the principal hypothesis being tested is
whether $w_\text{geom} = w_\text{grow}$.  This so-called growth-geometry split
allows explicit insights into what the data is telling us in case there is
tension with $\Lambda$CDM, as this currently favored model makes very precise
predictions about the relation between the growth and geometry quantities. For
example, the currently discussed discrepancy between the measurement of the
amplitude of mass fluctuations between the CMB and weak lensing
(e.g.\ \cite{MacCrann:2014wfa}) can be understood more clearly as the fact
that the growth of structure --- from current data, and not (yet) at an
overwhelming statistical significance --- is even more suppressed than
predicted in the standard cosmological model, as the geometry-growth analysis
indicates \cite{Ruiz:2014hma,Bernal:2015zom}.  Future cosmological constraints
that incorporate an impressive range of probes with complementary physics
sensitive to dark energy will be a particularly good test bed for the
geometry-growth split analyses.

\section{Principal probes of dark energy} \label{sec:principal_probes}

In this section, we review the classic, principal cosmological probes of dark
energy. What criterion makes a probe 'primary' is admittedly somewhat
arbitrary; here we single out and describe the most mature probes of dark
energy: type Ia supernovae (SNe Ia), the baryon acoustic oscillations (BAO),
the cosmic microwave background (CMB), weak lensing, and galaxy clusters. We
 briefly review the history of these probes and discuss their current
status and future potential. In the following section
(section~\ref{sec:other_probes}), we will discuss other probes of dark energy.
Finally, in Table \ref{tab:probe_sys} we summarize the primary and secondary probes of
dark energy, along with their principal strengths and weaknesses. 

\subsection{Type Ia supernovae}

Type Ia supernovae (SNe Ia) are very bright standard candles (sometimes called \textit{standardizable} candles) useful for measuring cosmological distances. Below we discuss why standard candles are useful and then go on to review cosmology with SNe Ia, including a brief discussion of systematic errors and recent progress.

\subsubsection{Standard candles. }
Distances in astronomy are often notoriously difficult to measure. It is relatively straightforward to measure the angular location of an object in the sky, and we can often obtain a precise measurement of an object's redshift $z$ from its spectrum by observing the shift of known spectral lines due to the expansion of the universe ($1 + z \equiv \lambda_\text{obs}/\lambda_\text{emit}$). For a specified cosmological model, the distance-redshift relation (i.e.\ \eqref{eq:rz}) would then indicate the distance; however, since our goal is typically to \emph{infer} the cosmological model, we need an \emph{independent} distance measurement. Methods of independently measuring distance in astronomy typically involve uncertain empirical relationships. To measure the (absolute) distance to an object, such as a galaxy, astronomers have to construct a potentially unwieldy ``distance ladder.'' For instance, they may employ relatively direct parallax measurements (apparent shifts
due to Earth's motion around the Sun) to measure distances to nearby objects in our galaxy (e.g.\ Cepheid variable stars), then use those objects to measure distances to other nearby galaxies (for Cepheids, the empirical relation between pulsation period and intrinsic luminosity is the key). If systematic errors add up at each rung, the distance ladder will become flimsy.

Standard candles are idealized objects that have a fixed intrinsic luminosity or absolute magnitude \cite{Kowal}. Having standard candles would be very useful; they would allow us to infer distances to those objects using only the inverse square law for flux (recall that $f = L/(4\pi d_L^2)$, where $d_L$ is the luminosity distance). In fact, we do not even need to know the luminosity of the standard candle when determining \emph{relative} distances for a set of objects is sufficient. Observationally, flux is typically quantified logarithmically (the apparent magnitude), while luminosity is related to the absolute magnitude of the object. We therefore have the relation
\begin{equation}
m - M = 5\log_{10} \left(\frac{d_L}{10~\text{pc}} \right),
\label{eq:DM}
\end{equation}
where the quantity $m - M$ is known as the distance modulus. For an object that is 10~pc away, the distance modulus is zero. For a true standard candle, the absolute magnitude $M$ is the same for each object. Therefore, for each object, a measurement of the apparent magnitude provides direct information about the luminosity distance and therefore some information about the cosmological model.

\subsubsection{Cosmology with SNe Ia. }
Supernovae are energetic stellar explosions, often visible from distant
corners of the universe. Unlike other types of supernovae, which result from
the core collapse of a massive, dying star, a type Ia supernova is thought to
occur when a slowly rotating carbon-oxygen white dwarf accretes matter from a
companion star, eventually exceeds the Chandrasekhar mass limit
($\sim$1.4$M_\odot$), and subsequently collapses and explodes\footnote{While a
  white dwarf is always involved, other details of the progenitor system, or
  of the nuclear ignition and burning mechanism, are far from certain. It
  seems to be the case that many SNe Ia result from a merger between two white
  dwarfs (double degenerate progenitors), and there may be more diversity in
  the type of companion star than once thought (e.g.\ \cite{Maoz:2011iv,Wang:2012za}).}. The empirical SN classification scheme is based on spectral features, and type Ia SNe are characterized by a lack of hydrogen lines and the presence of a singly ionized silicon (Si II) line at 6150~\AA. The flux of light from SNe Ia increases and then fades over a period of about a month; at its peak flux, a SN can be as luminous as the entire galaxy in which it resides.

SNe Ia had been studied extensively by Fritz Zwicky (e.g.\ \cite{Baade_Zwicky}), who gave them their name and noted that SNe Ia have roughly uniform luminosities. The fact that SNe Ia can potentially be used as standard candles had been realized long ago, at least since the 1970s \cite{Wagoner,Colgate}. However, developing an observing strategy to detect SNe Ia before they reached peak flux was a major challenge. If we were to point a telescope at a single galaxy and wait for a SN to occur, we would have to wait $\sim$100~years, on average. A program in the 1980s to find SNe \cite{Norgaard_Nielsen} discovered only one, and even then, only after the peak of the light curve. Today, after many dedicated observational programs, thousands of SNe Ia have been observed, and nearly one thousand have been analyzed simultaneously for cosmological inference.

Of course, SNe Ia are not perfect standard candles; their peak magnitudes exhibit a scatter of $\sim 0.3~\text{mag}$, limiting their usefulness as distance indicators. We now understand that much of this scatter can be explained by empirical correlations between the SN peak magnitude and both the \textit{stretch} (broadness, decline time) of the light-curve and the SN color (e.g.\ the difference between magnitudes in two bands). Simply put, broader is brighter, and bluer is brighter. While the astrophysical mechanisms responsible for these relationships are somewhat uncertain, much of the color relation can be explained by dust extinction. After correcting the SN peak magnitudes for these relations, the intrinsic scatter decreases to $\lesssim 0.15~\text{mag}$, allowing distance measurements with $\sim \text{7--10}\%$ precision.

We can rewrite \eqref{eq:DM} and include the stretch and color corrections to the apparent magnitude:
\[
5\log_{10}\left[\frac{H_0}{c} \, d_L(z_i, \mathbf{p}) \right] = m_i + \alpha \, s_i - \beta \, C_i - \mathcal{M} \ ,
\]
where $m_i$, $s_i$, and $C_i$ are the observed peak magnitude, stretch, and
color, respectively, for the $i^\text{th}$ SN. The exact definitions of these
measures are specific to the light-curve fitting method employed (e.g.\ SALT2
\cite{Guy:2007dv}). Meanwhile, $\alpha$, $\beta$, and $\mathcal{M}$ are
``nuisance'' parameters that can be constrained simultaneously with the
cosmological parameters $\mathbf{p}$. The $\mathcal{M}$ parameter,
\[
\mathcal{M} \equiv M + 5\log_{10}\left[\frac{c}{H_0 \times 1~\text{Mpc}} \right] + 25 \ ,
\]
is the Hubble diagram offset, representing a combination of \emph{two} quantities which are unknown \textit{a priori}, the SN Ia absolute magnitude $M$ and the Hubble constant $H_0$. Their combination $\mathcal{M}$ can be constrained, often precisely, by SN Ia data alone, and one can marginalize over $\mathcal{M}$ to obtain constraints on the cosmological parameters $\mathbf{p}$. Note that $H_0$ and $M$ \emph{cannot} be individually constrained using SN data only, though external information about one of them allows a determination of the other.

Figure~\ref{fig:deceleration} is referred to as a Hubble diagram, and it illustrates the remarkable ability of SNe Ia to distinguish between various cosmological models that affect the expansion rate of the universe.

The original discovery of dark energy discussed in section~\ref{sec:history} involved the crucial addition of a higher-redshift SN sample to a separate low-redshift sample. Results since then have improved gradually as more and more SNe have been observed and analyzed simultaneously (e.g.\ \cite{hicken,Union2}). Meanwhile, other cosmological probes (e.g.\ CMB, BAO; see figure~\ref{fig:BAO_CMB_SN}) have matured and have independently confirmed the SN Ia results indicating the presence of a $\Lambda$-like dark energy fluid.

\subsubsection{Systematic errors and recent progress. }

Recent SN Ia analyses (e.g.\ \cite{Conley:2011ku}) have focused on carefully accounting for a number of systematic uncertainties. These uncertainties can typically be included as additional (off-diagonal) contributions to the covariance matrix of SN distance moduli. As the number of observed SNe grows and statistical errors shrink, reducing the systematic uncertainties is key for continued progress and precision dark energy measurements.

Photometric calibration errors are typically the largest contribution to current systematic uncertainty budgets. In order to compare peak magnitudes of different SNe and interpret the difference as a relative distance, it is crucial to precisely understand any variation in the fraction of photons, originating from the SNe, that ultimately reach the detector. This category includes both photometric bandpass uncertainties and zero-point uncertainties. Part of the challenge is that current SN compilations consist of multiple subsamples, each observed with different instruments and calibrated using a different photometric system. This is a limitation which future large, homogeneous SN surveys will likely overcome, though it is also possible to reduce this uncertainty through consistent, precise recalibration of the existing samples \cite{Supercal}.

Other contributions to the systematic error budget include uncertainties in
the correction of bias resulting from selection effects (e.g.\ Malmquist
bias), uncertainties in the correction for Milky Way dust extinction,
uncertainty accounting for possible intrinsic evolution of SNe Ia or of the
stretch and color relations, uncertainty due to contamination of the sample by
non-Ia SNe (important for photometrically-classified SNe),
uncertainty in K-corrections, gravitational lensing dispersion (primarily
affecting high-redshift SNe \cite{Holz_Linder}), peculiar velocities
(important for low-redshift SNe), and uncertainty in host galaxy
relations. Although there are numerous sources of systematic error, most are
currently a sub-dominant contribution to the error budget, and the systematic
effects themselves have by now been well studied. While systematic uncertainties are \emph{not} trivially reduced by obtaining a larger SN sample, future surveys featuring better observations will likely reduce these errors further.

Other recent efforts have focused on improving the analysis of SN Ia data in preparation for the large samples expected in the future (e.g.\ LSST, WFIRST). This work has included the development of Bayesian methods for properly estimating cosmological parameters from SN Ia data \cite{March:2011xa,Rubin:2015rza}, including methods applicable to large photometrically-classified samples that will be contaminated by non-Ia SNe, which would otherwise bias cosmological measurements \cite{Kunz:2006ik,Hlozek:2011wq,Jones:2016cnm}. There are also techniques employing rigorous simulations to correct for selection and other biases and more accurately model SN Ia uncertainties \cite{Scolnic:2016ukm,Kessler:2016uwi}. Meanwhile, new, detailed observations of individual SNe can help us identify subclasses of SNe Ia and determine the extent to which they may bias dark energy measurements \cite{Foley:2013bba,Graham:2014bva,Milne:2014rfa}. Finally, it may be possible to reduce the effective intrinsic scatter by identifying specific SNe which are more alike than others \cite{Fakhouri:2015mhg} or by understanding how other observables, such as host galaxy properties, affect inferred SN luminosities.

For present SN Ia analyses, known systematic uncertainties have been quantified and are comparable to, or less than, the statistical errors. The fact that other, independent probes (BAO and CMB; see below) agree quantitatively with SN Ia results is certainly reassuring. Indeed, even if one completely ignores the SN data, the combination of the CMB distance with BAO data firmly points to a nearly flat universe with a subcritical matter density, thereby indicating the presence of a dark energy component.

\subsection{Baryon Acoustic Oscillations} \label{sec:bao}

Baryon acoustic oscillations (BAO) refer to the wiggles in the matter power
spectrum due to the coherent oscillations in the baryon-photon fluid in the
epoch prior to recombination. The effect, first predicted nearly 50 years ago
\cite{Sunyaev:1970eu,Peebles:1970ag}, results in excess probability of a
galaxy having a neighbor separated by the sound-horizon distance. This
therefore implies a single acoustic peak in the configuration space clustering
of galaxies at separation $r_s \simeq 100~h^{-1}\text{Mpc}$ or, equivalently, several
$\sim$10\% oscillations in the Fourier transform of the correlation function, that is, the matter power spectrum.

The power of BAO to probe dark energy comes from their exquisite power to
measure the angular diameter distance to high redshift, as well as the Hubble
parameter $H(z)$, using the sound horizon as a ``standard ruler.'' The sound horizon
is the radiation-era distance covered by the speed of sound, which is $c/\sqrt{3}$ with a correction for the non-negligible presence of baryons:
\begin{align}
r_s = \int_0^{t_*} \frac{c_s}{a(t)} \, dt &= \frac{c}{\sqrt{3}} \int_0^{a_*} \frac{da}{a^2 H(a) \sqrt{1+ \frac{3\Omega_b}{4\Omega_\gamma} \, a}} \nonumber \\[1ex]
&= (144.6 \pm 0.5)~\text{Mpc} \, ,
\label{eq:rs}
\end{align}
where $a_* \sim 10^{-3}$ is the scale factor at recombination. The error quoted in \eqref{eq:rs} comes from \textit{Planck} \cite{Ade:2015xua}; it is known independently to such a high precision due to measurements of the physical
matter and baryon densities from the morphology of the peaks in the CMB angular power spectrum.

A pioneering detection of the BAO feature was made from analysis of the
Sloan Digital Sky Survey (SDSS) galaxy data \cite{Eisenstein:2005su}. Much improvement has been made in subsequent measurements \cite{Cole:2005sx,Padmanabhan:2006cia,Percival:2007yw,Blake:2011en,Beutler:2011hx,Padmanabhan:2012hf,Anderson:2013zyy,Ross:2014qpa}.

Measurement of the angular extent of the BAO feature, together with precise,
independent knowledge of the sound horizon, enables determination of both the
angular diameter distance to the redshift of the sample of galaxies and the Hubble parameter evaluated at that epoch
\cite{Blake03,Hu_rings,Seo_Eisenstein}.  More specifically, clustering of the
galaxies in the transverse direction can be used to measure the angular diameter
distance to the characteristic redshift of the galaxy sample,
\begin{equation}
\Delta \theta_{s} = \frac{r_s}{d_A(z)} \quad \text{(transverse modes)}.
\end{equation}
Meanwhile, clustering in the radial direction constrains the Hubble parameter at
the same redshift since the redshift extent of the BAO feature $\Delta z_s$ is
effectively observed; it is related to the Hubble parameter via
\begin{equation}
\Delta z_{s} = \frac{H(z) \, r_s}{c} \quad \text{(radial  modes)}.
\end{equation}
Radial modes are particularly helpful, as they provide localized information
about dark energy via the Hubble parameter at redshift of the galaxy sample.  However,
radial modes are also more difficult to measure than the transverse modes,
essentially because the transverse modes span a two-dimensional space while
radial modes live in only one dimension.  Up until recently, the BAO
measurements had sufficiently large statistical error that it was a good
approximation to constrain the generalized distance that combines the
transverse and radial information \cite{Eisenstein:2005su}
\begin{equation}
  D_V(z)_\equiv \left [(1+z)^2 d_A(z)\frac{cz}{H(z)}\right ]^{1/3}.
\end{equation}
With current or future data, separating into transverse and radial modes is
feasible, and enables extracting more information about dark energy.

The main strength of the BAO comes from its excellent theoretical foundation:
the physics of the acoustic oscillations is exceptionally well
understood. While the systematic errors do affect the amplitudes and, to a
lesser extent, positions of the acoustic peaks in the galaxy power spectrum,
these shifts are largely correctable. In particular, nonlinear clustering,
strongly subdominant at scales $\simeq 100\,h^{-1}\text{Mpc}$, shifts the peak
positions by only a fraction of one percent
\cite{Padmanabhan:2009yr,Mehta:2011xf}, and even that small shift can be
accurately predicted --- and therefore modeled --- using a combination of
theory and simulations \cite{Seo:2009fp}. Nevertheless, a mild concern remains
the possibility that galaxy density is modulated by non-gravitational effects
on scales of $\simeq 100\,h^{-1}\text{Mpc}$, which in principle shifts the peaks
by small but non-negligible amount. Such large-scale modulation could be
caused, for example, by the fact that galaxies are biased tracers of the
large-scale structure (for a review, see \cite{Desjacques:2016bnm}).

The most powerful BAO experiments necessarily need to be spectroscopic
surveys, as the required redshift accuracy in order not to smear the BAO
feature corresponds to a few percent of the BAO standard ruler $r_s$, meaning
a few megaparsecs or  $\delta z\lesssim 0.001$. [Low-resolution BAO measurements are
  possible with photometric surveys with sufficiently accurate photometric
  redshifts.] Another requirement is large volume, so that sufficiently many
samples of the sound-horizon feature can be mapped, and sample variance 
suppressed. Prime Focus Spectrograph (PFS; \cite{Ellis:2012rn}) and Dark
Energy Spectroscopic Instrument (DESI;
\cite{Aghamousa:2016zmz,Aghamousa:2016sne}) represent important future surveys
whose principal goal is to maximize the BAO science and obtain excellent
constraints on dark energy.

Tracers other than galaxies or quasars can be used to detect and utilize the
BAO feature.  For example, Lyman alpha forest is useful in mapping structure
in the universe; these are the ubiquitous absorption lines seen in
high-resolution spectra of distant quasars or galaxies due to hydrogen gas
clouds and filaments along the line of sight which show up as ``trees'' in the
forest.  Lyman-alpha systems are challenging to model since a variety of
physical processes, including hydrogen recombination, radiative heating, and
photo-ionization need to be known, often using simulations. However the BAO
feature, being at $\sim 100\,$Mpc scale, is considered more robust, and has
actually been detected in the Lyman-alpha forest and used to constrain the angular diameter distance and
Hubble parameter at $z\sim 2$, deep in the matter-dominated era
\cite{Busca:2012bu,Slosar:2013fi,Font-Ribera:2013wce,Delubac:2014aqe,Bautista:2017zgn}.

\begin{figure*}[!t]
\centering
\includegraphics[width=0.45\textwidth]{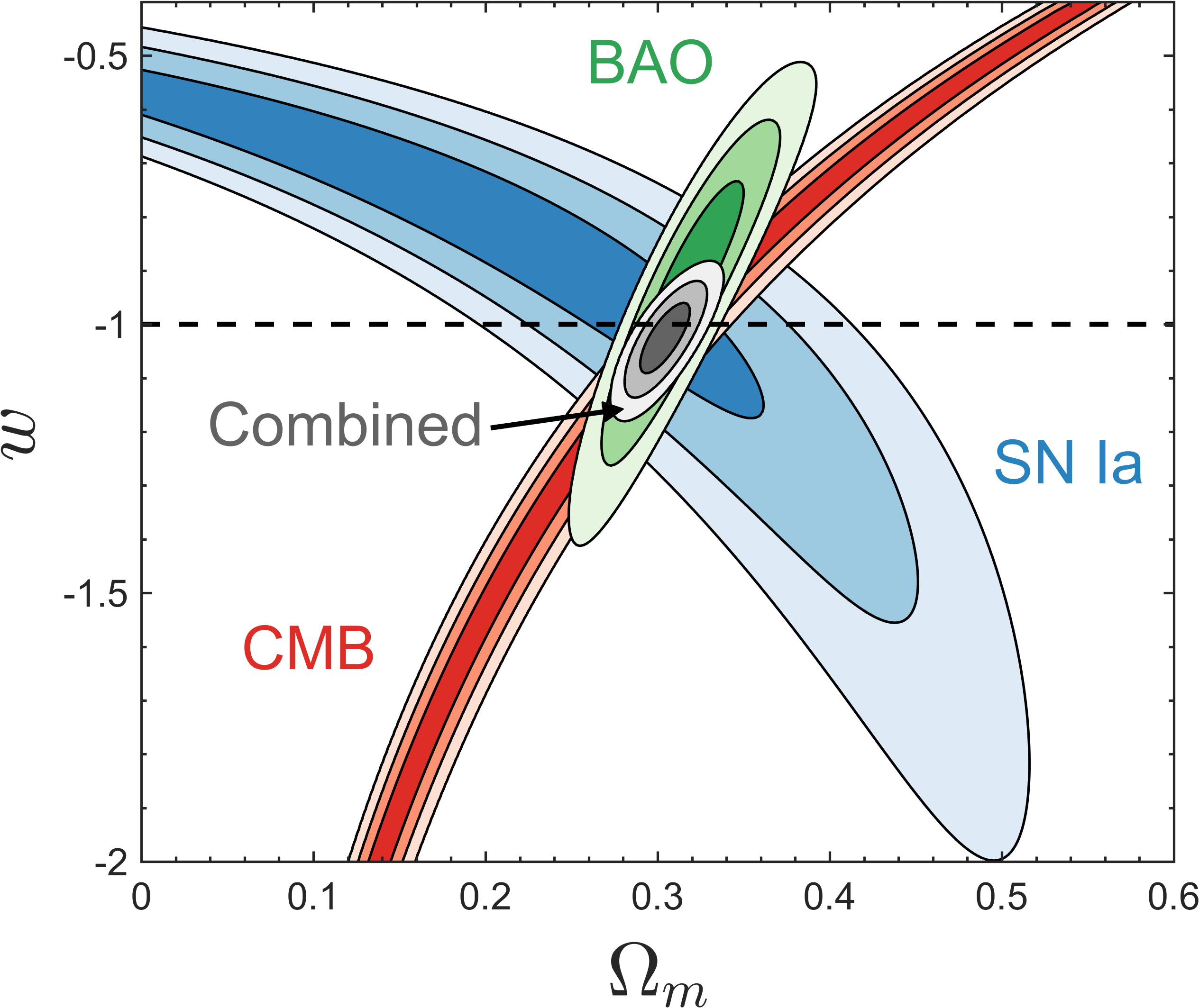}
\hspace{3mm}
\includegraphics[width=0.45\textwidth]{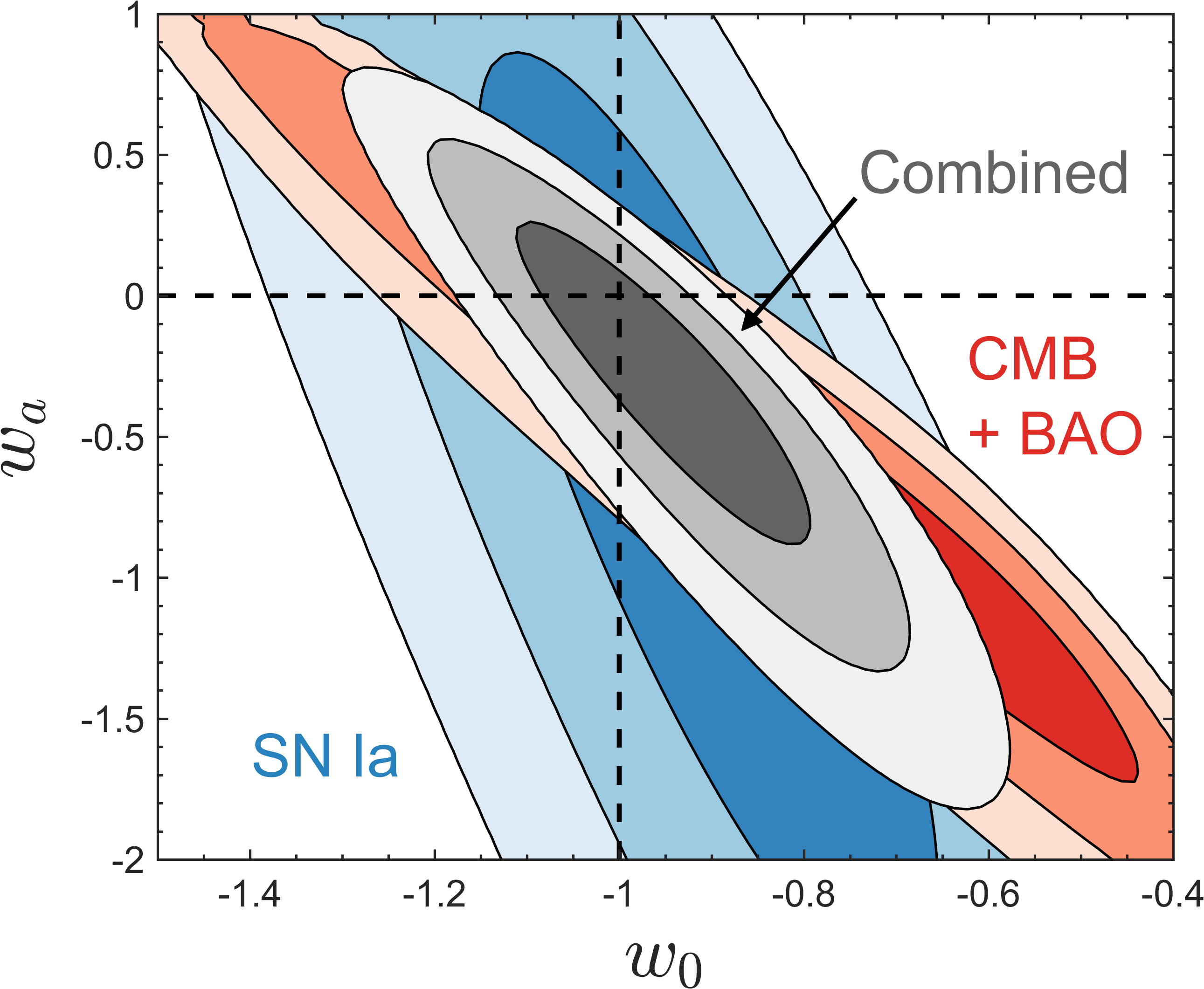}
\caption{Constraints on cosmological parameters from our analysis of current data from three principal probes: SN Ia (JLA \cite{Betoule:2014frx}; blue), BAO (BOSS DR12 \cite{Alam:2016hwk}; green), and CMB (\textit{Planck} 2015 \cite{Ade:2015xua}; red). We show constraints on $\Omega_m$ and constant $w$ (left panel) and on $w_0$ and $w_a$ in the parametrization from \eqref{eq:w0wa}, marginalized over $\Omega_m$ (right panel). The contours contain 68.3\%, 95.4\%, and 99.7\% of the likelihood, and we assume a flat universe in both cases.}
\label{fig:BAO_CMB_SN}
\end{figure*}

Finally, note that the BAO measurements provide an \textit{absolute} distance
measurement, in the limit when the sound horizon $r_s$ is perfectly known from
e.g.\ the morphology of the CMB peaks (recall that SNIa provide relative
distances since the vertical offset in the SN Hubble diagram, or equivalently
the Hubble constant, is marginalized over). This makes BAO not only
complementary to SNIa, but also powerful in connecting the low-z and high-z
measurements of the expansion history. Current BAO constraints on key dark
energy parameters are shown in figure~\ref{fig:BAO_CMB_SN}.

\subsection{Cosmic microwave background radiation}

While otherwise known as a Rosetta Stone of cosmology for its ability to
constrain cosmological parameters to spectacular precision \cite{1997PhT....50k..32B,Hu_Dodelson}, the cosmic microwave background at
first appears disappointingly insensitive to dark energy. This na\"ive
expectation is borne out because the physics of the CMB takes place in the
early universe, well before dark energy becomes important. There, baryons and photons are coupled due to the Coulomb coupling between protons and electrons and the Thomson scattering between
electrons and photons. This coupling leads to coherent oscillations, which in turn manifest themselves as wiggles in the observed power in the distribution of the hot and cold spots on the
microwave sky. The angular power spectrum that describes the statistical
distribution of the temperature anisotropies (see the lower left panel in figure~\ref{fig:obs}) therefore has rich structure that can be fully predicted
as a function of cosmological parameters to sub-percent-level accuracy. The
angular power spectrum is a superb source of information about, not only the inflationary
parameters, but also dark matter and even, as we discuss here, dark energy.

Dark energy affects the distance to the epoch of recombination, and therefore
the angular scale at which the CMB fluctuations are observed. This sensitivity
is precisely the reason why the CMB is in fact a very important complementary
probe of dark energy.  Given that the physics of the CMB takes place at the redshift of recombination when dark energy is presumably completely
negligible, the physical structure of CMB fluctuations is unaffected by dark
energy, as long as we do not consider the early dark energy models with significant
early contribution to the cosmic energy budget. The sound horizon $r_s$, defined in \eqref{eq:rs}, is projected to angle
\begin{equation}
\theta_* = \frac{r_s(z_*)}{r(z_*)} \, ,
\label{eq:lpeak}
\end{equation}
where $z_*$ is the recombination redshift and $r$ is the comoving distance \eqref{eq:rz}. The latter quantity is affected by dark energy at $z \lesssim 1$ (see figure~\ref{fig:obs}). Therefore, dark energy affects the angle at which
the features are observed --- that is, the horizontal location of the CMB
angular power spectrum peaks.  More dark energy (higher $\Omega_\text{de}$)
increases $d_A$ and therefore shifts the CMB pattern to smaller scales, and
vice versa.

To the extent that the CMB provides a single but \textit{very} precise measurement of the peak location, it provides a very important complementary constraint on
the dark energy parameters. In a flat universe, the CMB thus constrains a degenerate combination
of $\Omega_m$ and $w$ (and, optionally, $w_a$ or other parameters describing
the dark energy sector). While the CMB appears to constrain just another distance measurement --- much like SNe Ia or BAO, albeit at a very high redshift ($z_* \simeq 1000$) --- its key advantage is that the $d_A$
measurement comes with $\Omega_m h^2$ essentially fixed by features in the CMB power spectrum. In other words, the CMB essentially constrains the comoving distance to recombination with the physical matter
density $\Omega_m H_0^2$ fixed \cite{Bond:1997wr},
\begin{equation}
R \equiv \sqrt{\Omega_m H_0^2} \ r(z_*) \ ,
\end{equation}
which is sometimes referred to as the ``CMB shift parameter''
\cite{Melchiorri:2002ux,Wang:2003gz}. Because of the fact that $\Omega_m h^2$
is effectively factored out, the CMB probes a different combination of
dark energy parameters than SNe or BAO at any redshift. In particular, the combination
of $\Omega_m$ and $w$ constrained by the CMB is approximately
\cite{frieman_03} $D \equiv \Omega_m - 0.94 \, \overline{\Omega}_m \, (w - \overline{w})$ where $(\overline{\Omega}_m, \overline{w}) \simeq (0.3, -1)$. This combination is measured with few-percent-level precision by \textit{Planck}; see figure~\ref{fig:BAO_CMB_SN}. It drastically reduces the parameter errors when combined with other probes \cite{frieman_03} despite the fact that the CMB peak positions cannot constrain the dark energy parameters on their own (the lensing pattern in the CMB, however, is independently sensitive
to dark energy \cite{Sherwin:2011gv}). Important complementary constraints on
dark energy have been provided by several generations of CMB experiments,
including tBoomerang, Maxima and DASI \cite{Jaffe:2000tx,Sievers:2002tq}, \textit{WMAP} \cite{WMAP_1,WMAP_3,Komatsu:2008hk,Komatsu:2010fb,Hinshaw:2012aka}, \textit{Planck} \cite{Ade:2013zuv,Ade:2015xua}, and also CMB experiments that probe smaller angular scales such as ACT and SPT \cite{Hou:2012xq,Sievers:2013ica}.

Another, much weaker, effect of dark energy on the CMB power spectrum is
through the late-time Integrated Sachs Wolfe (ISW) effect
\cite{Sachs:1967er,Hu:1993xh}. The ISW is due to the change in the depth of
the potential wells when the universe is not matter dominated. One such epoch
--- the \textit{early-time} ISW effect --- occurs around recombination when radiation is not
yet completely negligible. The \emph{late-time} ISW effect occurs when dark energy becomes
important at $z\lesssim 1$. The late-time ISW produces additional power in the
CMB power spectrum at very large angular scales --- multipoles $\lesssim 20$,
corresponding to scales larger than about 10 degrees on the sky. There is an
additional dependence on the speed of sound of the dark energy fluid; however
this effect becomes negligible as $w \rightarrow -1$, leaving only the overall
effect of smooth dark energy \cite{Hu:2001fb,Bean:2003fb,Weller:2003hw}. Unfortunately the cosmic
  variance error is large at these scales, leading to very limited extent to
  which the late-time ISW can be measured. Nevertheless, it is important to
  account for the ISW when producing theory predictions of various dark energy
  models; for example, modified gravity explanations for the accelerating
  universe often predict specific ISW signatures \cite{Song:2006jk}.

\subsection{Weak Gravitational Lensing} \label{sec:wl}

Gravitational lensing --- bending of light by mass along the line of sight to
the observed source --- is theoretically well understood and also readily
observed, and therefore represents a powerful probe of both geometry and
structure in the universe. The principal  advantage of lensing
(relative to e.g.\ observations of galaxy clustering) is that lensing is
fundamentally independent of the prescription of how the observed halos or
galaxies trace the underlying dark matter --- the so-called 'bias'. While most of
the manifestations of gravitational lensing are sensitive to dark energy, we here describe
weak lensing as the principal probe. In section~\ref{sec:other_probes} we also
discuss the so-called strong lensing, galaxy-galaxy lensing, and counting of
the peaks in the shear field as additional, complementary lensing probes of the
accelerating universe.

\begin{figure*}[!t]
\centering
\includegraphics[width=0.4\textwidth]{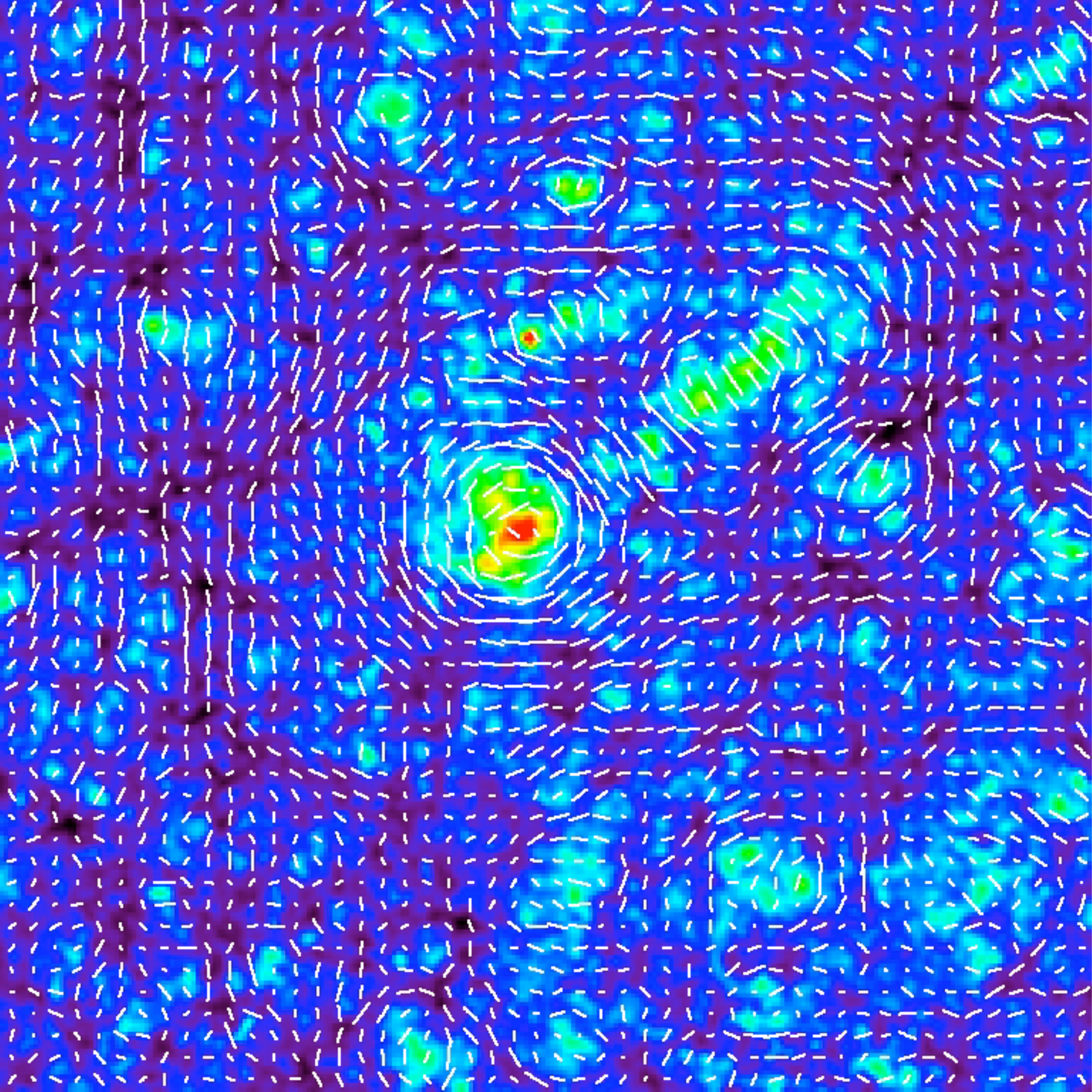}
\hspace{0.2cm}
\includegraphics[width=0.55\textwidth]{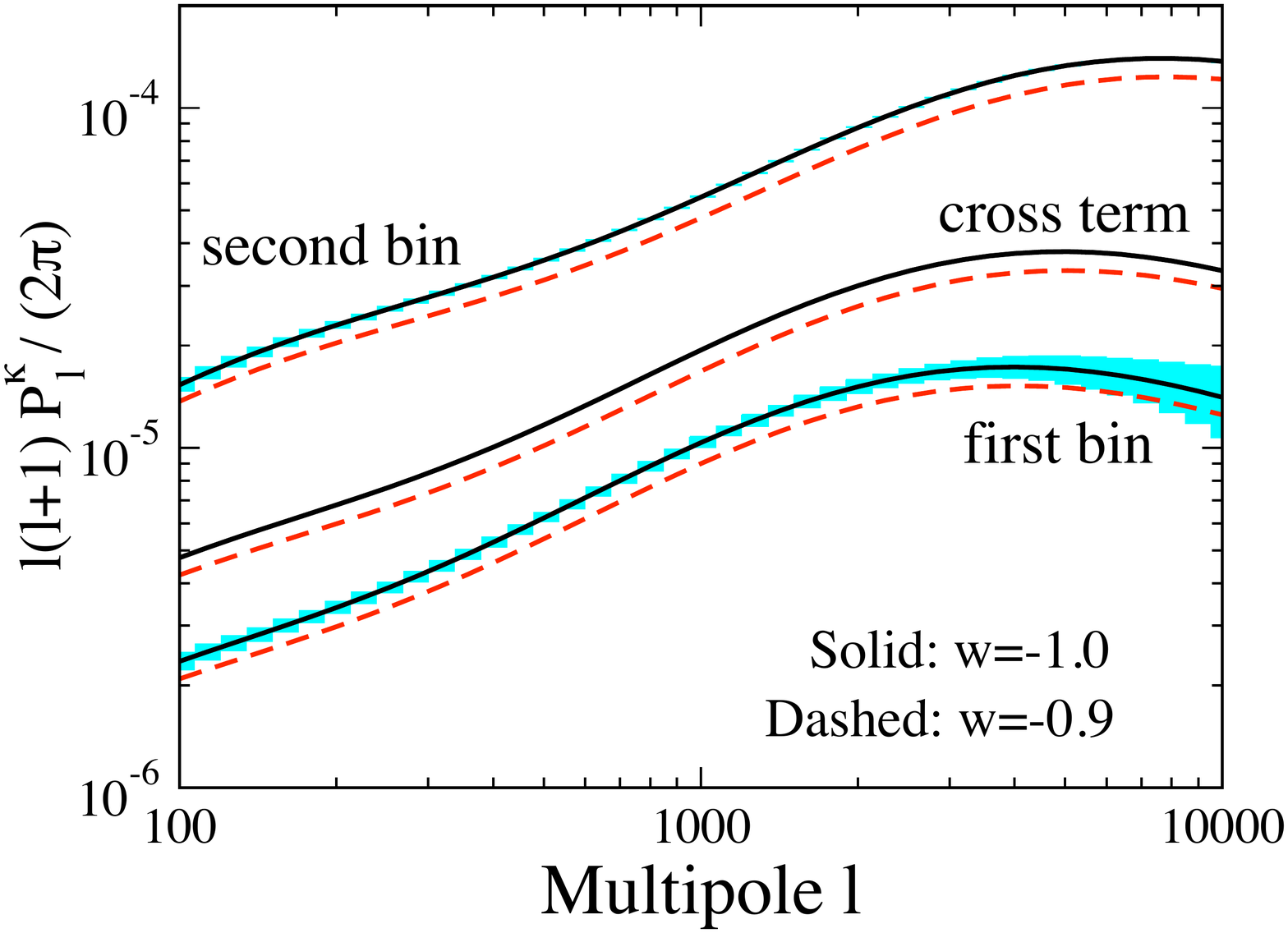}
\caption{\textit{Left panel}: Cosmic shear field (white whiskers) superimposed on the projected mass distribution from a
  cosmological $N$-body simulation, where overdense regions are bright and
  underdense regions are dark. Note how the shear field is correlated with the
  foreground mass distribution; the shears are azimuthal around
    overdensities and radial around underdensities. Figure courtesy of
  T.\ Hamana. \textit{Right panel}: Angular power spectrum of cosmic
    shear along with statistical errors expected for LSST for two values of the dark energy equation-of-state parameter. For illustration, results are shown for source galaxies in two broad redshift
  bins, $0 < z_s < 1$ (first bin) and $1 < z_s < 3$ (second bin); the
  cross-power spectrum between the two bins (cross term) is shown without the statistical errors.}
\label{fig:WL}
\end{figure*}

Weak gravitational lensing  is bending of light by
structures in the Universe; it leads to distorted or sheared images of distant galaxies,
see the left panel of figure~\ref{fig:WL}.  This distortion allows the
distribution of dark matter and its evolution with time to be measured,
thereby probing the influence of dark energy on the growth of structure (for a
detailed review, see e.g.\ \cite{Bartelmann_Schneider}; for 
brief reviews, see \cite{Hoekstra_Jain} and \cite{Huterer_GRG}).

Gravitational lensing produces distortions of images of background
galaxies. These distortions can be described as mapping between the
source plane ($S$) and image plane ($I$),
\begin{equation}
\delta x_i^S = A_{ij} \delta x_j^I \, ,
\end{equation}
where $\delta \mathbf{x}$ are the displacement vectors in the
two planes and $A$ is the distortion matrix,
\begin{equation}
A = \left( 
\begin{array}{cc}
1-\kappa-\gamma_1	&	-\gamma_2 \\
-\gamma_2		      & 1-\kappa+\gamma_1
\end{array}
\right).
\end{equation}
The deformation is described by the convergence $\kappa$ and complex shear
$(\gamma_1, \gamma_2)$; the total shear is defined as
$|\gamma|=\sqrt{\gamma_1^2+\gamma_2^2}$. We are interested in the limit of weak lensing, where $\kappa$, $|\gamma| \ll 1$. The magnification of the source, also given in terms of $\kappa$ and $\gamma_{1,2}$, is
\begin{equation}
\mu = \frac{1}{|1-\kappa|^2 - |\gamma|^2} \approx 1 + 2\kappa + O(\kappa^2, \gamma^2) \, ,
\end{equation}
where the second, approximate relation holds in the weak lensing limit.

Given a sample of sources with known redshift distribution and cosmological parameter values, the convergence and shear can be predicted from theory. The convergence $\kappa$ in any particular direction on the sky $\mathbf{\hat{n}}$ is given by the integral along the line of sight $\kappa(\mathbf{\hat{n}}, \chi) = \int_0^\chi W(\chi') \, \delta(\chi') \, d\chi'$, where $\delta$ is the relative perturbation in matter energy density and $W(\chi)$ is the geometric weight function describing the lensing efficiency of foreground galaxies. The most efficient lenses lie about halfway between us and the source galaxies whose shapes we measure.

The statistical signal due to gravitational lensing by large-scale structure is termed ``cosmic shear.'' To estimate the cosmic shear field at a given point in the sky, we locally average the shapes of large numbers of distant
galaxies.  The principal statistical measure of cosmic shear is the shear angular power spectrum, which chiefly depends on the source galaxy redshift $z_s$, and additional information can be obtained by measuring the correlations between shears at different redshifts or with foreground lensing galaxies, as well as the \textit{three}-point correlation function of cosmic shear \cite{Takada_Jain}.

The convergence can be transformed into multipole space $\kappa_{lm} = \int d\mathbf{\hat{n}} \, \kappa(\mathbf{\hat{n}}, \chi) \, Y_{lm}^*(\mathbf{\hat{n}})$, and the power spectrum is defined as the two-point correlation function (of convergence, in this case) $\langle \kappa_{\ell m} \kappa_{\ell'm'} \rangle = \delta_{\ell \ell'} \, \delta_{m m'} \, P_\ell^{\kappa}$. The convergence\footnote{At lowest order, the convergence power spectrum is equal to the shear power spectrum.} angular power spectrum is
\begin{equation}
P^\kappa_\ell(z_s) = \int_0^{z_s} \frac{dz}{H(z) d_A^2(z)} W(z)^2 P\left(k = \frac{\ell}{d_A(z)}; z \right),
\label{eqn:Limber}
\end{equation}
where $\ell$ denotes the angular multipole, $d_A(z) = (1 + z)^{-2} d_L(z)$ is the angular diameter distance, the weight function $W(z)$ is the efficiency for lensing a population of source galaxies and is determined by the distance
distributions of the source and lens galaxies, and $P(k,z)$ is the usual matter power spectrum. One important feature of \eqref{eqn:Limber} is the integral along the line of sight, which encodes the fact that weak lensing radially projects the density fluctuations between us and the sheared source galaxy. Additional information is obtained by
measuring the shear correlations between objects in different redshift bins;
this is referred to as weak lensing tomography \cite{Hu:1999ek} and
contains further useful information about the evolution of the growth of structure.

The dark energy sensitivity of the shear angular power
spectrum comes from two factors: 
\begin{itemize}
\item \textit{geometry} --- the Hubble parameter, the angular diameter distance,
  and the weight function $W(z)$; and
\item \textit{growth of structure} --- via the redshift evolution of the
  matter power spectrum $P(k)$ (more specifically, via the growth factor $D(z)$ in \eqref{eq:Da_ga}).
\end{itemize}
Due to this two-fold sensitivity to dark energy and, in recent
  years, the advent of better-quality observations and larger surveys, weak lensing now
  places increasingly competitive constraints on dark energy \cite{Jarvis:2005ck,Massey:2007gh,Schrabback:2009ba,Lin:2011bc,Heymans:2013fya,Huff:2011aa,Jee:2015jta,Hildebrandt:2016iqg,Troxel:2017xyo}.
	
The statistical uncertainty in measuring the shear power spectrum on large scales is 
\begin{equation}
\Delta P^\kappa_\ell = \sqrt{\frac{2}{(2\ell+1)f_\text{sky}}}
\left[ P^\kappa_\ell +\frac{\sigma^2(\gamma_i)}{n_\text{eff}} \right] \, ,
\label{eqn:power_error}
\end{equation}
where $f_\text{sky}$ is the fraction of sky area covered by the survey, $\sigma_{\gamma_i}$ is the
standard deviation in a single component of the (two-component) shear ($\sim 0.2$ for typical measurements), and $n_\text{eff}$ is the
effective number density per steradian of galaxies with well-measured shapes. The first term in the brackets represents sample
variance (also called \textit{cosmic variance}), which arises due to the fact that only a
finite number of independent samples of cosmic structures are
available in our survey. This term dominates on larger scales. The second term, which dominates on small scales, represents shot noise from both
the variance in galaxy ellipticities (``shape noise'') and the finite number of galaxies (hence the inverse proportionality to $n_\text{eff}$).

Systematic errors in weak lensing measurements principally come from the
limitations in accurately measuring galaxy shapes. These shear measurements are
complicated by a variety of thorny effects such as the atmospheric blurring of
the images, telescope distortions, charge transfer in CCDs, to name just a
few. More generally, a given measurement of the galaxy shear will be subject
to additive and multiplicative errors that affect the true shear
\cite{Huterer:2005ez,Heymans}. The weak lensing community has embarked on a
series of challenges to develop algorithms and techniques to ameliorate these
observational systematics (e.g.\ \cite{Mandelbaum:2013esa}).  There are also
systematic uncertainties due to limited knowledge of the redshifts of source
galaxies: because taking spectroscopic redshifts of most source galaxies will
be impossible (for upcoming surveys, that number will be of order a billion), the community has developed approximate photometric redshift techniques, where one obtains a noisy redshift estimate from multi-wavelength (i.e.\ multi-color) observations of each galaxy. In order for the
photometric redshift biases not to degrade future dark energy constraints,
their mean calibration at the 0.1\% level will be required \cite{Ma:2005rc,Hearin:2010jr}.

The interpretation of weak lensing measurements also faces theoretical
challenges, such as the need to have accurate predictions for clustering in
the non-linear regime from N-body simulations
\cite{Huterer_Takada,Rudd_Zentner_Kravtsov,Hearin:2009hz} and to account for
non-Gaussian errors on small angular scales
\cite{Takada:2008fn,Taylor:2012kz,Dodelson:2013uaa}. Finally, intrinsic
alignments of galaxy shapes, due to tidal gravitational fields, are a serious
contaminant which requires careful modeling as well as external astrophysics
input, such as observationally-inferred galaxy separation, type, and luminosity information; for a review, see \cite{Joachimi:2015mma}.

The right panel of figure~\ref{fig:WL} shows the weak lensing shear power
spectrum for two values of $w$, and the corresponding statistical errors
expected for a survey such as LSST, assuming a survey area of 15,000~deg$^2$ and effective source galaxy density of $n_\text{eff} = 30$ galaxies
per square arcminute, and divided into two radial slices. Current surveys cover
more modest hundreds of deg$^2$, although KIDS (450, and soon to be up to
1,500~deg$^2$ \cite{Hildebrandt:2016iqg}), Dark Energy Survey (about
1,500 and soon to be 5,000~deg$^2$ \cite{Abbott:2017wau}) and Hyper
Suprime-Cam (HSC; expected to be 1,500~deg$^2$ \cite{Aihara:2017tri})
are aiming to bring weak lensing to the forefront of dark energy
constraints. Note that the proportionality of errors to $f_\text{sky}^{-1/2}$
means that, as long as the systematic errors can be controlled, large sky
coverage is at a premium. Further improvement in dark energy constraints can be achieved by judiciously combining a photometric and a spectroscopic survey \cite{Cai:2011wj,Joachimi:2010xb,Font-Ribera:2013rwa,Eriksen:2014zua,vanUitert:2017ieu,Joudaki:2017zdt}.

The weak lensing signal can also be used to \textit{detect} and count massive
halos, particularly galaxy clusters. This method, pioneered recently
\cite{Wittman01,Wittman02}, can be used to obtain cluster samples whose masses
are reliably determined, avoiding the arguably more difficult signal-to-mass
conversions required with the X-ray or optical observations
\cite{Wittman06,Schirmer07,Dietrich07,Miyazaki07}.  Much important information
about the dark matter and gas content of galaxy clusters can be inferred with
the combined lensing, X-ray, and optical observations. This has recently been
demonstrated with observations of the ``Bullet Cluster'' \cite{Clowe_Bullet},
where the dark matter distribution inferred from weak lensing is clearly
offset from the hot gas inferred from the X-ray observations, indicating the
presence and distinctive fingerprints of dark matter.

\subsection{Galaxy clusters} \label{sec:dm}

Galaxy clusters --- the largest collapsed objects in the universe with mass
$\gtrsim 10^{14} M_\odot$ and size a few Mpc -- are just simple enough
that their spatial abundance and internal structure can be used to probe dark energy. Clusters are versatile
probes of cosmology and astrophysics and have had an important role in the
development of modern cosmology (for a review, see \cite{Allen:2011zs}). In
the context of dark energy, one can use the spatial abundance of clusters and compare
it to the theoretical expectation that includes the effects of dark energy. This
classic test is in principle very simple, since the number density of clusters
can be inferred from purely theoretical considerations or, more robustly, from
suites of numerical simulations. In practice, however, there are important challenges to overcome.

The number of halos in the mass range $[M, M+dM]$ in a patch of the sky
with solid angle $d\Omega$ and in redshift interval $[z, z + dz]$ is given by
\begin{equation}
\frac{d^2 N}{d\Omega dz} = \frac{r^2(z)}{H(z)} \, \frac{dn(M, z)}{dM} \, dM \, ,
\end{equation}
where $r^2/H = dV/(d\Omega dz)$ is the comoving volume element and
$n(M, z)$ is the number density (the ``mass function'') that is calibrated with
numerical simulations (e.g.\ \cite{Heitmann:2015xma}).

Assuming Gaussian initial conditions, the comoving number density of objects
in an interval $dM$ around mass $M$ is
\begin{equation}
\frac{dn}{d\ln M} = \frac{\rho_{M, 0}}{M} \left|\frac{dF(M)}{d\ln M} \right| \, ,
\end{equation}
where $\rho_{M, 0}$ is evaluated at the present time and $F(M)$ is the fraction of
collapsed objects. The original analysis of Press and Schechter \cite{Press_Schechter} assumed a Gaussian initial distribution of overdensities, leading to $F(M) = (1/2) \, \text{erfc}(\nu/\sqrt{2})$, where $\nu(M) \equiv \delta_c/\sigma(M)$ is the peak height and $\delta_c = 1.686$ is the critical threshold for collapse in the spherical top-hat model \cite{Gunn:1972sv}. The Press-Schechter formula also involves multiplying the mass function by the
notorious overall factor of two to account for underdensities as well as
overdensities. Subsequent work has put the theoretical estimates on considerably firmer footing (for a review, see \cite{Zentner:2006vw}), but the
most accurate results are based on fits to numerical simulations, which calibrate the mass function for the standard $\Lambda$CDM class of models to a precision of about 5\% \cite{Tinker}. Smooth dark energy models described by
the modified linear growth history via the equation of state $w(a)$ are still reasonably well fit with the standard $\Lambda$CDM formulae \cite{Linder:2003dr}, while modified gravity models sometimes predict
scale-dependent growth $D(a, k)$ even in the linear regime and must be calibrated by simulations specifically constructed for the given class of
modified gravity models (for a review, see \cite{Baldi:2012ky}).

\begin{figure*}[!t]
\centering
\includegraphics[width=0.51\textwidth]{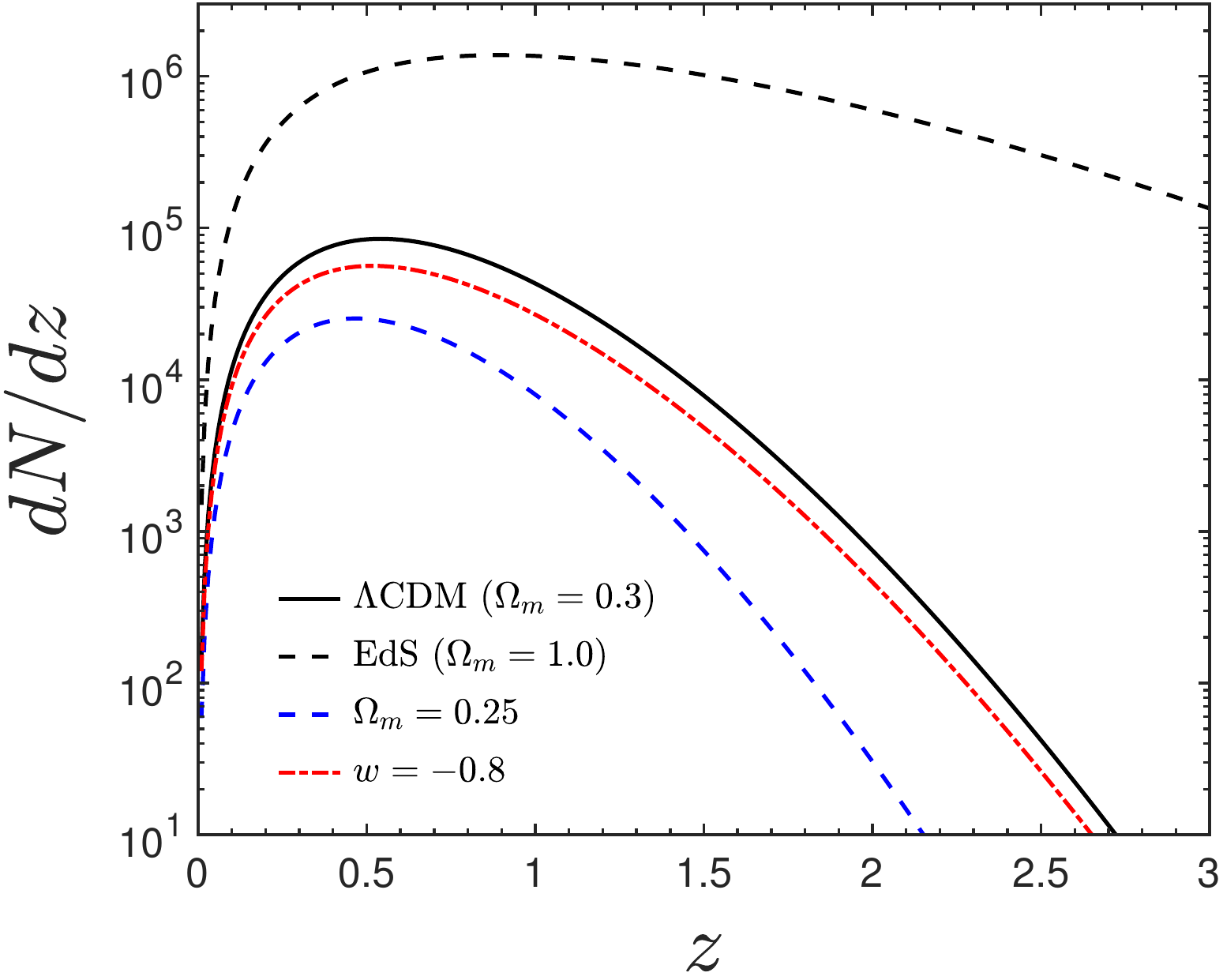}
\hspace{0.2cm}
\includegraphics[width=0.45\textwidth]{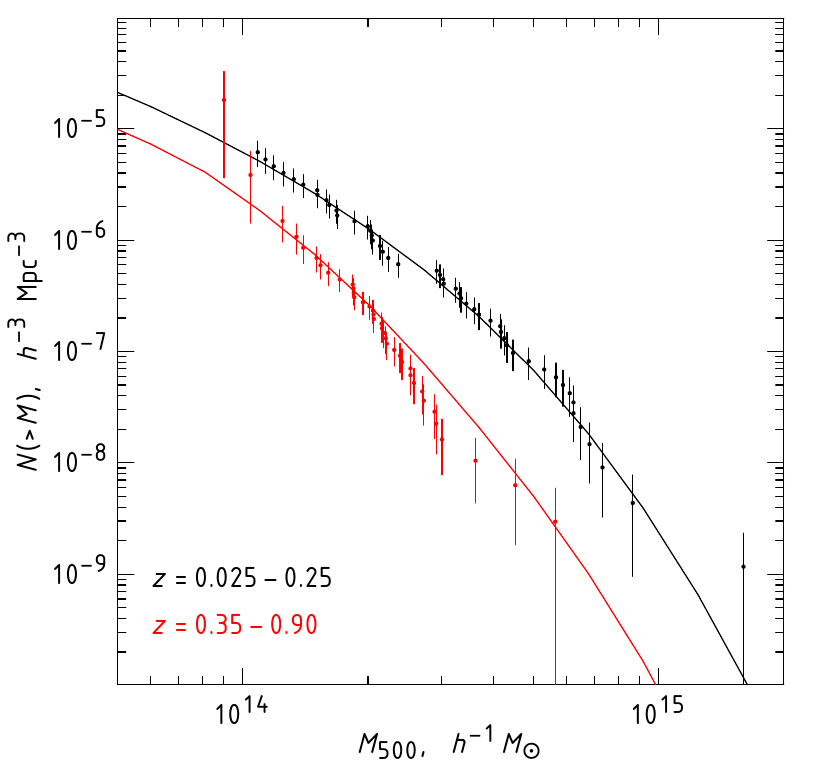}
\caption{\textit{Left panel}: Predicted cluster counts for a survey covering 5,000~deg$^2$ that is sensitive to halos more massive than $10^{14} \, h^{-1} M_\odot$, shown for a fiducial $\Lambda$CDM model as well as three variations as in figure~\ref{fig:obs}. \textit{Right panel}: Measured mass function --- $n(z, M_\text{min}(z))$, in our notation --- from the 400~deg$^2$ survey of ROSAT clusters followed up with the \textit{Chandra} space telescope; reproduced from \cite{Vikhlinin:2008ym}.}
\label{fig:massfun}
\end{figure*}

The absolute number of clusters in a survey of solid angle $\Omega_\text{survey}$
centered at redshift $z$ and in the shell of thickness $\Delta z$ is given by
\begin{equation}
N(z, \Delta z) = \Omega_\text{survey} \int_{z-\Delta z/2}^{z+\Delta z/2} n(z, M_\text{min}(z)) \, \frac{dV(z)}{d\Omega\,dz} \, dz \, ,
\label{eq:clustercount}
\end{equation}
where $M_\text{min}$ is the minimal mass of clusters in the survey. Note that
knowledge of the minimal mass is extremely important, since the mass function
$n(z, M_\text{min}(z))$ decreases exponentially with $M$ such that most of the contribution comes from a small range of masses just above $M_\text{min}$. Recent cluster observations typically do not have enough
signal-to-noise to determine the cluster masses directly; instead,
forward-modeling can be applied to the mass function to recast the theory in the
space of observable quantities \cite{Evrard:2014cea}. One commonly used proxy
for the cluster mass is the optical ``richness'' --- the number of
galaxies per cluster --- which is straightforward to measure from observations \cite{Rykoff:2011xi,Rykoff:2013ovv}.

The sensitivity of cluster counts to dark energy arises from the same two factors as in the case of
weak lensing:
\begin{itemize}
\item \textit{geometry} --- the term $dV(z)/(d\Omega \, dz)$ in \eqref{eq:clustercount}, which is the comoving volume element; and
\item \textit{growth of structure} --- $n(z, M_\text{min}(z))$ depends on the evolution of density perturbations.
\end{itemize}

The mass function's near-exponential dependence on the power spectrum in the high-mass limit is at the root of the power of clusters to probe the growth of density fluctuations. Specifically, the mass function is very sensitive to the amplitude of mass fluctuations smoothed on some scale $R$ \textit{calculated assuming linear theory}. That is,
\begin{equation}
\sigma^2(R, z) = \int_0^\infty \Delta^2(k, z) \left(\frac{3 j_1(kR)}{kR} \right)^2 d\ln k \, ,
\end{equation}
where $\Delta^2$ is the linear version of the power spectrum from \eqref{eq:Deltasq} and $R$ is traditionally taken to be $8 \, h^{-1}\text{Mpc}$ at $z = 0$, roughly corresponding to the characteristic size of galaxy clusters. The term in parentheses in the integrand is the Fourier
transform of the top-hat window which averages out the perturbations over
regions of radius $R$. The left panel of figure~\ref{fig:massfun} shows the
sensitivity of the cluster counts to the dark energy equation-of-state
parameter, while the right panel shows measurements of the mass function based on X-ray observations \cite{Vikhlinin:2008ym}.

There are other ways in which clusters can be used to probe dark energy. For
example, their two-point correlation function probes the matter power spectrum
as well as the growth and geometry factors sensitive to dark energy. Clusters
can also be correlated with background galaxies to probe the growth
(\cite{Oguri:2010vi}; this is
essentially a version of galaxy-galaxy lensing discussed in section
\ref{sec:other_probes}). While these two tests can also be carried out using the much
more numerous galaxies, clusters have the advantage of having more accurate
individual photometric redshifts.

Clusters can be detected using light in the X-ray, optical, or millimeter
waveband, or else using weak gravitational lensing of background galaxies
behind the cluster. Some of these methods suffer from contamination due to the
projected mass, as large-scale structures between us and the cluster
contribute to the signal and can, in extreme cases, conspire to create
appearance of a cluster from radially aligned, but dispersed, collections of
numerous low-mass halos. This particularly affects detection of clusters via
lensing (e.g.\ \cite{Hu_Keeton}).  It is therefore necessary to use N-body
simulations to calibrate purity (contribution of false detections) and
completeness (fraction of detections relative to the truth) in these lensing
observations \cite{Marian_Smith_Bernstein,Dietrich_Hartlap}. However, the
possibility of cluster finding and mass inference in multiple wavebands is
also a great strength of this probe, as it allows cross-checks and
cross-calibrations, in particular of the cluster masses. Over the past decade,
the wealth of ways to detect and characterize clusters, combined with
improving ways to characterize the relation between their observable
properties and mass, has led to increasingly interesting constraints on dark
energy parameters
\cite{Vikhlinin:2008ym,Rozo:2009jj,Mantz:2009fw,Allen:2011zs,Tinker:2011pv,Mantz:2014xba,deHaan:2016qvy}. Comparisons
between dynamical and lensing cluster mass estimates are also sensitive to
modifications of gravity \cite{Schmidt:2010jr}.

Regardless of how the clusters are detected, the principal systematic concern is how to relate the observable quantity (X-ray flux, Sunyaev-Zeldovich signal, lensing
signature) to the mass of the cluster. In the past, the principal mass proxies
have used X-ray observations and assumed hydrostatic equilibrium. Recurring
concerns about the latter assumption imply significant statistical and
systematic errors, and only tens of percent mass accuracy per cluster are achieved with
these traditional approaches. Arguably the most secure method for determining
the mass is weak gravitational lensing of source galaxies behind the cluster
which, when possible, is combined with their strong lensing signatures. Such
lensing efforts already enable a better than 10\% mass accuracy per cluster
\cite{Becker:2010xj,Rasia:2012jz,vonderLinden:2012kh}. The requirement on the individual
mass precision, in order to be sufficient for future surveys, is approximately
2--5\% \cite{Weinberg:2012es}.

A secondary source of systematics is the photometric redshifts of galaxy cluster members, which are combined to determine the redshift of the cluster.
Averaging of individual redshifts fortunately leads to fairly accurate
photometric redshift estimates, $\sigma_z/(1+z) \simeq 0.01$
(e.g.\ \cite{Rykoff:2016trm}), such that only moderate improvement is required
for future dark energy constraints \cite{Huterer:2004rf,Lima:2007kx}.

Like other probes, clusters are amenable to determining the parameters describing the systematic
errors internally from the data, the process known as self-calibration
\cite{Levine:2002uq,Majumdar:2003mw,Lima:2005tt}. While any nuisance
parameters can be self-calibrated, the most important uncertainty is typically
tied to parameters that describe the scaling relations between mass and
observable properties of the cluster (e.g.\ flux, temperature). Key to progress on the control of cluster systematics, as well as the program of
self-calibration, is a multi-wavelength view of the clusters. Analyses that use a combination of weak and strong lensing signatures, as well as detections and observations in the optical, X-ray, and microwave (via the SZ
effect), open many avenues for the robust use of clusters to probe geometry and growth evolution (e.g.\ \cite{Rozo:2012xa}).

\Table{\label{tab:probe_sys}Comparison of dark energy probes. The five primary
probes are the most mature, but a variety of other probes
offer complementary information and have potential to provide important constraints
on dark energy.}
\br
\textbf{Probe/Method} & \textbf{Strengths} & \textbf{Weaknesses} \\
\mr
\centre{3}{\textbf{Primary probes of dark energy}} \\
\mr
SN Ia & pure geometry, & calibration, \\
& model-independent, & evolution, \\
& mature & dust extinction \\[1ex]
BAO & pure geometry, & requires millions \\
& low systematics & of spectra \\[1ex]
CMB & breaks degeneracy, & single distance \\
& precise, & only \\
& low systematics & \\[1ex]
Weak lensing & growth \& geometry, & measuring shapes, \\
& no bias &  baryons, photo-$z$ \\[1ex]
Cluster counts & growth \& geometry, & mass-observable, \\
& X-ray, SZ, \& optical & selection function \\[1ex]
\mr
\centre{3}{\textbf{Other probes of dark energy}} \\
\mr
Gal-gal lensing    &  high S/N  & bias,  \\
&                  &  baryons \\[1ex]
Strong lensing   & unique combination & lens modeling, \\
                 & of distances       &  structure along los\ \\[1ex]
RSD           & lots of modes,  & theoretical modeling \\
              & probes growth   & \\[1ex]
Peculiar velocities & probes growth,   & selection effects, \\
                    & modified gravity & need distances \\[1ex]
Hubble constant & breaks degeneracy, & distance ladder \\
                & model-independent & systematics \\[1ex]
Cosmic voids & nearly linear, & galaxy tracer fidelity, \\
             & easy to find   & consistent definition \\
             &                & and selection \\[1ex]
Shear peaks & probes beyond   & theoretical modeling \\
            & 2-pt           & vs.\ projection \\[1ex]
Galaxy ages & Sensitive to $H(z)$ & galaxy evolution, \\
            &                     & larger systematics \\[1ex]
Standard sirens &  high z,          & optical counterpart \\
                & absolute distance & needed for redshift, \\
                &                   &  lensing \\[1ex]
Redshift drift & clean interpretation & tiny signal, \\
               &                      &  huge telescope, \\
               &                      &  stability \\[1ex]
GRB \& quasars & very high $z$ & standardizable? \\[1ex]
\br
\endTable

\section{Other probes of dark energy} \label{sec:other_probes}

There are a number of powerful secondary probes of dark
energy. While they do not quite have the power to individually impose strong
constraints on dark energy without major concerns about the systematic errors,
they provide complementary information, often hold a lot of promise, and sometimes
come ``for free'' with astrophysical or cosmological observations in surveys.
Here we review some of the most promising of these methods.

\subsection{Galaxy-galaxy lensing}
Another effective application of weak lensing is to measure the correlation of the shear of background galaxies with the mass of the foreground galaxies. This
method, which is referred to as ``galaxy-galaxy lensing''
\cite{Brainerd95,Fischer,Hoek_Yee_Glad,
  Sheldon:2003xj,Mandelbaum:2005nx,Johnston:2007uc,Choi:2012kf,Velander:2013jga,Leauthaud:2016jdb,Prat:2017goa}, essentially probes the galaxy-shear correlation function across the sky. Galaxy-galaxy lensing measures the surface mass density contrast $\Delta\Sigma(R)$,
\begin{equation}
\Delta \Sigma(R)\equiv \overline{\Sigma}(<R) - \overline{\Sigma}(R) = \Sigma_\text{crit} \times \gamma_t(R) \, ,
\label{dsigma}
\end{equation}
where $\overline{\Sigma}(< R)$ is the mean surface density within proper radius $R$,
$\overline{\Sigma}(R)$ is the azimuthally averaged surface density at radius
$R$ (e.g.\ \cite{Miralda-Escude91,Wilson:2001}), $\gamma_t$ is the
tangentially-projected shear, and $\Sigma_\text{crit}$ is the critical surface density, a known function of the distances to the source and the lens.

Current measurements constrain the density profiles and bias of dark matter halos
\cite{Sheldon:2003xj,Kleinheinrich04,Mandelbaum_profiles,Johnston07} as well as the
relation between their masses and luminosities \cite{Leauthaud_ML,Sheldon09_ML}. In the future, galaxy-shear correlations
have the potential to constrain dark energy models \cite{Hu_Jain} and modified gravity models for the accelerating universe \cite{Schmidt_MGWL}.

\subsection{Strong gravitational lensing}
Distant galaxies and quasars
 occasionally get multiply imaged due to intervening structure along the
 line of sight. While relatively rare --- about one in a thousand  objects
 is multiply imaged --- strong lensing has the nice feature that, like weak
 lensing, it is sensitive to \textit{all} matter in the universe and not
 just the visible part. There is  a  long history of trying to use counts of
 strongly lensed system to constrain the cosmological parameters
 \cite{Kochanek:1995ap,Huterer:2003uf,Chae:2002mb}; however, its strong
 dependence on the independent knowledge of the density profile of lenses
 makes robustness of this approach extremely challenging to
 achieve. Instead, strong lensing time delays between images of the same
 source object offer a more promising way to constrain the Hubble constant
 \cite{Refsdal:1964nw} but also dark energy
 (e.g.\ \cite{Linder:2011dr,Suyu:2012aa}). Time delays are sensitive to a
 unique combination of distances, sometimes called the time-delay distance \cite{Suyu:2009by}
\[
D_{\Delta t} \equiv (1 + z_l)\,\frac{d_A(z_l) \, d_A(z_s)}{d_A(z_l, z_s)} = \frac{\Delta t}{\Delta\phi} \ ,
\]
where $z_l$ and $z_s$ are the lens and source redshift respectively, $\Delta t$ is the time delay, and $\Delta \phi$ is the so-called Fermat potential
difference evaluated between different image locations. Because the Fermat
potential can be constrained by lens modeling, the time-delay measurements
measure $D_{\Delta t}$, which in turn offers dark energy parameter sensitivity that is complementary
to that of other cosmological probes \cite{Linder:2004hx}. Additional
information can be obtained by measurements of the velocity dispersion of
lens galaxies, which effectively determine its mass; this, plus measurements
of the gravitational potential, determine the size of the lens, which can
then be used as a standard ruler and provide information about $d_A(z_l)$
\cite{Paraficz:2009xj,Jee:2014uxa} and thus dark energy \cite{Jee:2015yra}.
Strong lensing time delays are reviewed in \cite{Treu:2016ljm}.

\subsection{Redshift-space distortions (RSD)}
\begin{figure*}[t]
\centering
\includegraphics[width=0.7\textwidth]{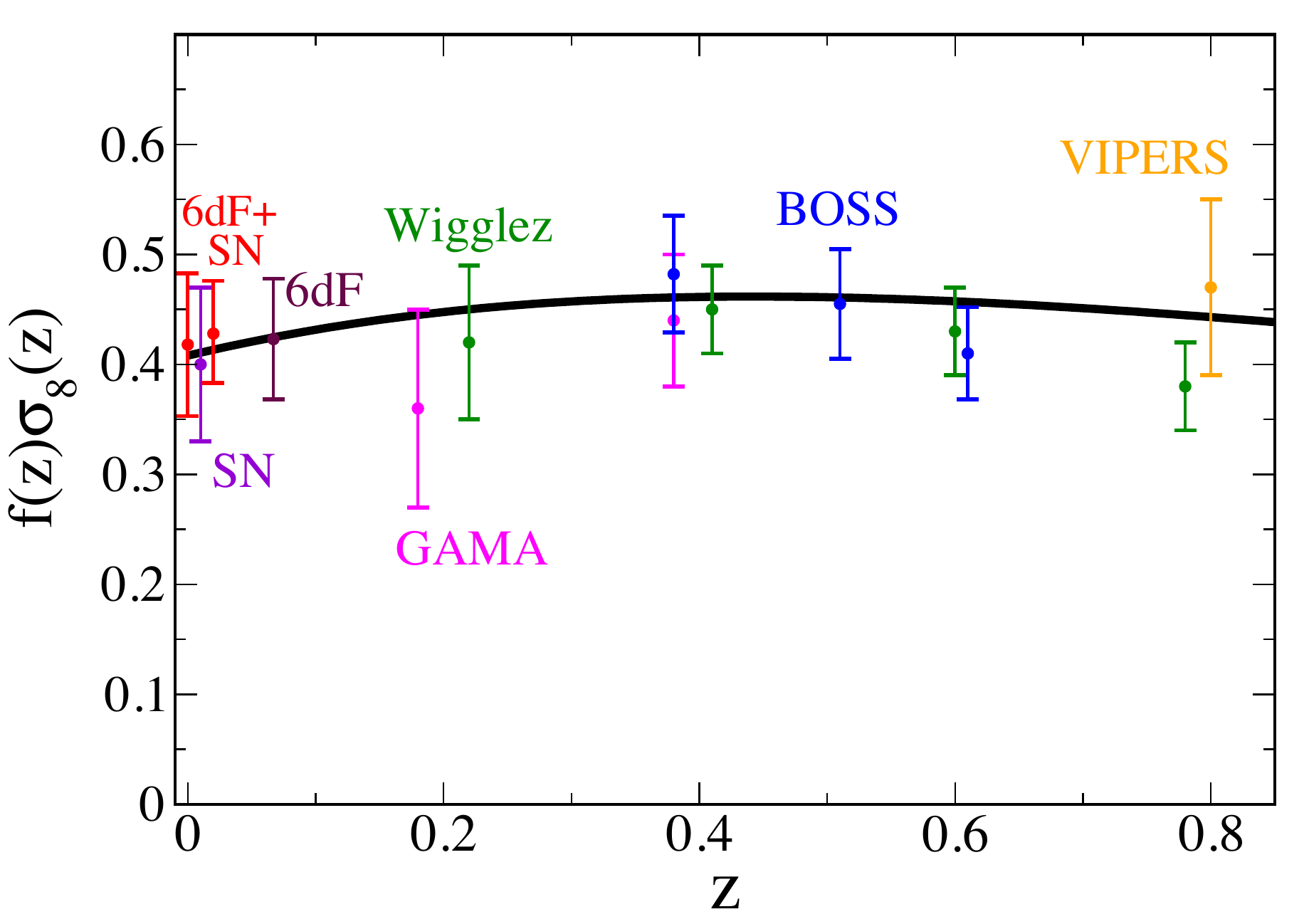}
\caption{Constraints on the quantity $f\sigma_8$ at different redshifts from RSD and peculiar velocity surveys. At the lowest redshifts $z \approx 0$, peculiar velocities from galaxies and SNe Ia (leftmost \cite{Johnson:2014kaa} and rightmost \cite{Huterer:2016uyq} red points) and SNe Ia alone (purple data point; \cite{Turnbull:2011ty}) constrain the velocity power spectrum and effectively the quantity $f\sigma_8$. At higher redshifts, constraints on $f\sigma_8$ come from the RSD analyses from 6dFGS (maroon at $z = 0.067$; \cite{Beutler:2012px}), GAMA (pink points; \cite{Blake:2013nif}), WiggleZ (dark green; \cite{Blake:2011rj}), BOSS (dark blue; \cite{Beutler:2016arn}), and VIPERS (orange; \cite{delaTorre:2013rpa}). The solid line shows the prediction corresponding to the currently favored flat $\Lambda$CDM cosmology.}
\label{fig:fsig8}
\end{figure*}

On large scales, peculiar velocities of galaxies are affected by gravitational potential of the
    large-scale structures in a coherent, quantifiable way. In linear theory,
    the gradient of the velocity is proportional to the overdensity,
    $\nabla\cdot {\bf v}({\bf r}) = -(aH)f\delta({\bf r})$, where $f\equiv d\ln
    D/d\ln a$ is the growth rate introduced in \eqref{eq:ga}; the
    line-of-sight component of the peculiar velocity of a given galaxy
    directly affect the measured redshift (hence redshift-space distortions,
    or RSD). Still assuming linear theory, the two-point correlation function
    of galaxies measured \textit{ in redshift space}, $P^s$, is related to the
    usual configuration-space power spectrum $P(k)$ via the
    Kaiser formula \cite{Kaiser:1987qv}
    \begin{equation}
      P^s(k, \mu) = P(k) \left [b+f\mu^2\right ]^2
      \label{eq:Kaiser_RSD}
    \end{equation}
    where $b$ is the bias of galaxies and $\mu$ is cosine of the angle made by
    wavevector ${\bf k}$ and the line-of-sight
    direction. \eqref{eq:Kaiser_RSD} predicts the general shape of the
    correlation of function measured as a function of the angle between galaxy
    pairs and the line-of-sight. Sensitivity to dark energy mainly comes from
    the factor $f(a)$ --- more precisely, the combination $f(a)\sigma_8(a)$
    \cite{Song:2008qt,Percival:2008sh} --- which is, as mentioned in
    section~\ref{sec:basic}, very sensitive not only to dark energy parameters
    but also to modifications of gravity \cite{Linder:2007nu}. The RSD signal
    has been measured and used to constrain the parameter $f\sigma_8$ out to
    redshift $\simeq 1$, and finds good agreement with the currently favored
    $\Lambda$CDM model \cite{delaTorre:2013rpa,Beutler:2013yhm}; see
    figure~\ref{fig:fsig8}. A challenge is the theoretical modeling of the RSD; the
    Kaiser formula in \eqref{eq:Kaiser_RSD} breaks down at small scales
    where the velocities become non-linear, requiring complex modeling
    combined with careful validation with numerical simulations.

\subsection{Peculiar velocities}
Galaxies respond to the gravitational pull
  of large-scale structure, leading to the so-called peculiar
  velocities. These motions lead to the Doppler effect: $(1 + z_\text{obs}) =
  (1 + z)(1 + v_\parallel /c)$, where $z$ and $z_\text{obs}$ are the true and
  observed redshift and $v_\parallel$ is the peculiar velocity projected along
  the line of sight. Roughly speaking, objects physically close to each other
  are being pulled by the same large-scale structures, and are therefore more
  likely to have similar velocities. The statistical properties of the
  velocity field are straightforwardly related to the matter power spectrum
  \cite{Kaiser:1989kb,Gorski_etal}. As with the RSD, the fact that the
  velocity is related to density via $\nabla\cdot \mathbf{v}(\mathbf{r}) =
  -(aH)f\delta(\mathbf{r})$ implies that the peculiar velocities are sensitive to the
  quantity $f\sigma_8$, where $f\equiv d\ln D/d\ln a$ is the growth rate.
  Peculiar velocities typically determine $f\sigma_8$ at very low redshift,
  $z<0.1$, and thus provide an important complementary test of both dark
  energy and modified gravity. There has been a lot of activity in using
  velocities to to test for consistency with expectations from the
  $\Lambda$CDM model
  \cite{Haugboelle:2006uc,Gordon:2007zw,Ma:2010ps,Dai:2011xm,Nusser:2011tu,Weyant:2011hs,Ma:2012tt,
    Rathaus:2013ut,Feindt:2013pma,Ma:2013oja,Johnson:2014kaa,Huterer:2016uyq}
  and to measure cosmological parameters
  \cite{Turnbull:2011ty,Carrick:2015xza}; see figure~\ref{fig:fsig8}.  Chief
  concerns include the reliability of distance indicators which are required
  in order to infer the peculiar velocity.

\subsection{Hubble constant}
Direct measurements of the Hubble constant
  offer useful complementary information that helps break degeneracy between
  dark energy and other cosmological parameters. This is because precise CMB
  measurements effectively fix high-redshift parameters including the physical matter
  density $\Omega_m h^2$; independent measurements of $H_0$ (i.e.\ $h$) therefore help determine
  $\Omega_m$ which is degenerate with the dark energy equation of state. Current $\gtrsim 3\sigma$ tension between the most precise direct measurements of $H_0$ from the Cepheid distance ladder \cite{Riess:2016jrr,Bernal:2016gxb} and the indirect $\Lambda$CDM determination from the CMB \cite{Ade:2015xua} is partially, but not fully, relieved by allowing phantom dark energy ($w < -1$) or extra relativistic degrees of freedom \cite{Riess:2016jrr}. Future measurements of the Hubble constant, expected to be at the 1\% level, will not only serve as a powerful test of the $\Lambda$CDM model, but will also provide extra leverage for dark energy measurements \cite{Hu:2004kn}.
  
\subsection{Cosmic voids}
It has been suggested that counting the cosmic voids ---
  large underdense regions of size up to $\sim 100\,$Mpc --- is an effective way
  to probe dark energy \cite{Lee:2007kq,Lavaux:2011yh}. Counting voids is
  similar in spirit as counting clusters of galaxies, but voids offer some
  advantages --- they are ``more linear'' than the clusters (largely thanks to
  the mathematical requirement that $\delta\rho/\rho\geq -1$), and therefore
  arguably more robustly modeled in numerical simulations. On the flip side,
  one typically uses galaxy surveys to find voids which is a challenge, given
  that the latter are defined as regions that are mostly devoid of galaxies.
  Recent work includes void catalogs extracted from the Sloan Digital Sky
  Survey \cite{Sutter:2012wh,Nadathur:2013bba,Leclercq:2014pga} and even
  constraints on the basic $\Lambda$CDM parameters \cite{Hamaus:2016wka} but
  concerns remain about the robustness of the void definition in
  simulations, as well as their correspondence to void counts in the data
  \cite{Nadathur:2013bba,Sutter:2013yda}.

\subsection{Shear peaks}
Another method that is conceptually similar to counting clusters of galaxies
is to count the peaks in the matter density field. Because the weak lensing
shear is directly proportional to the matter (baryonic and dark) projected
along the line of sight, counting the peaks in weak lensing maps enables this
method in practice
\cite{Jain:1999nu,Hamana:2003ts,Hennawi:2004ai,Marian:2006zp,Dietrich:2009jq,Kratochvil:2009wh}. While
primarily sensitive to the amount and distribution of matter, the method
generally constrains the cosmological model, including the dark energy
parameters. This probe has developed rather rapidly over the past decade, in
parallel to increased quality and area of available weak lensing shear
maps. Current constraints are broadly consistent with theoretical expectation
the standard $\Lambda$CDM model
\cite{Liu:2014fzc,Hamana:2015bwa,Kacprzak:2016vir}. The advantage of the
method is that it is sensitive to non-Gaussian aspects of the lensing field,
and thus provides additional information than the angular power spectrum.
Principal systematics include accurately calibrating the effects of shear
projection from multiple halos along the line of sight, which dominates for
all except the highest peaks \cite{Dietrich:2009jq,Liu:2016xjb} and needs to
be carefully calibrated using numerical simulations
\cite{Jain:1999nu,Hamana:2003ts,Hennawi:2004ai,Marian:2009wi,Lin:2014dua} and
measured using optimized estimators \cite{Schmidt:2010ex}.

\subsection{Relative ages of galaxies}
If the relative ages of galaxies at different redshifts can be determined reliably, then they provide a measurement of $dt/dz$. Since
\begin{equation}
t(z) = \int_0^{t(z)} dt' = \int_z^\infty \frac{dz'}{(1 + z') H(z')} \, ,
\label{eq:cosmic_time}
\end{equation}
one can then measure the expansion history directly \cite{Jimenez_Loeb}. Age has already been employed in cosmological constraints across a wide redshift range \cite{Stern:2009ep,Moresco:2012jh,Moresco:2016mzx}. However, the presence of systematic errors due to galaxy evolution and star formation remains a serious concern.

\subsection{Standard sirens}
The recently detected gravitational radiation from inspiraling binary
  neutron stars or black holes can,  in the future, enable these sources to serve as
  ``standard sirens'' \cite{Schutz:1986gp,Holz_Hughes_05}. From the observed
  waveform of each inspiral event, one can solve for the orbit's angular
  velocity, its rate of change, and the orbital velocity, in order to determine the
  luminosity of the object and hence its (absolute) luminosity distance.  If
  the electromagnetic counterpart to the observed gravitational wave signature
  can be unambiguously identified, then the redshift of the host galaxy can be
  determined, and the inspiral can be used to probe dark energy through the
  Hubble diagram \cite{Dalal06}. This potentially very complementary probe is
  still in the early stages of development, but holds promise to provide
  strong constraints on dark energy \cite{Cutler:2009qv}, out to potentially
  very high redshifts. Key to its success, beyond finding inspiral events at
  cosmological distances, is ability to localize the sources in three
  dimensions in order to get their redshifts \cite{Chen:2016tys}.
  
\subsection{Redshift drift}
The redshift drift \cite{Sandage62,1997fpc..book.....L,Loeb} refers to the redshift change of
  an object due to expansion, observed over a human timescale. The expected
  change of a quasar or galaxy at cosmological redshift, observed over a
  period of $dt_0 \sim$ 10--20 years, is tiny,
\begin{equation}
dz = \left[H_0 (1 + z) - H(z) \right] dt_0 \sim 10^{-9},
\end{equation}
but can potentially be measured using very high-resolution spectroscopy
\cite{Liske:2008ph}. The redshift drift method is fairly unique in its direct sensitivity to $H(z)$ across a wide redshift range and may someday contribute significantly to constraining the expansion history
\cite{Cor_Hut_Mel,Quercellini:2010zr,Kim:2014uha}. While a measurement of
  the redshift drift requires a very high
  telescope stability over a period of about a decade, there are proposals to
  use the intensity mapping of the 21~cm emission signal --- which involves a
  different set of systematics --- to detect the redshift-drift signal
  \cite{Yu:2013bia,Klockner:2015rqa}.

\subsection{Other standard candles/rulers}
A wide variety of astronomical objects have been proposed as standardizable
candles or rulers, useful for inferring cosmological distances via
semi-empirical relations. Notable examples include radio galaxies as
standardizable rulers \cite{1994ApJ...426...38D,Daly:2002kn} and quasars as
standardizable candles, most recently via the non-linear relation between UV
and X-ray luminosities \cite{Risaliti:2015zla}. Long-duration gamma-ray bursts
(GRBs) are an attractive possibility \cite{schaefer03} because their ability
to be detected at very high redshifts ($z \sim 6$ or higher) means they would
probe a redshift range beyond that of SNe Ia. Several different relationships
for GRBs have been proposed, most famously the Amati relation
\cite{Amati:2002ny,Amati:2006ky} between the peak energy of the integrated
spectrum and the isotropic-equivalent total energy output of the GRB. Some
analyses have employed several of these relations simultaneously
\cite{Schaefer:2006pa}. Due to the relatively small number of (useful) GRBs
and the substantial scatter about the relations, as well as concerns about the
presence of serious systematic errors, GRBs have not yielded competitive
cosmological constraints. It remains to be seen whether the aforementioned
relations hold over such large spans of cosmological time and can be
calibrated and understood to a sufficient accuracy.
  
\subsection{Observation of unexpected features}
When interpreted in the context of a cosmological model (e.g.\ LCDM),
observation of unexpected features in cosmological observations or existence
of objects at high statistical significance can be used to rule out the model
in question. High-redshift, high-mass clusters of galaxies have been
particularly discussed in this context: observation of clusters had been used
to disfavor the matter-only universe \cite{Bahcall:1998ur} while, more
recently, there has been a discussion of whether the existence of the observed
high-mass, high-redshift ``pink elephant'' clusters are in conflict with the
currently dominant LCDM paradigm (e.g.\ \cite{Holz:2010ck}). However such
analyses requires a careful accounting of all sources of \textit{statistical}
error closely related to the precise way in which the observations have been
carried out \cite{Mortonson:2010mj,Hotchkiss:2011ms}. Thus, while the
observation of unexpected features can be used to rule out aspects of the dark
energy paradigm, its \textit{a posteriori} nature implies that independent
confirmation that uses other cosmological probes will be required.

\section{The accelerating universe: Summary} \label{sec:summary}

In this article, we have briefly reviewed the developments leading to the discovery of dark energy and the accelerating universe. We have discussed the current status of dark energy, described parametrizations of the equation of state and physical aspects that can be measured, and reviewed both primary and secondary cosmological probes that allow us to study this mysterious component. In summary, there are a few important things to know about dark energy:
\begin{itemize}
\item Dark energy has negative pressure. It can be described by its present-day energy density relative to critical $\Omega_\text{de}$ and equation of state $w \equiv p_\text{de}/\rho_\text{de}$. For a cosmological constant, corresponding to vacuum energy, $w = -1$ precisely and at all times. More general explanations for dark energy typically lead to a time-dependent equation of state.
\item Current observational data constrain the equation of state to be $w \approx -1$ to within about 5\%. Measuring $w$ and any time dependence --- as well as searching for hints of any other, as yet unseen, properties of dark energy --- will help us understand the physical nature of this mysterious component, a key goal of modern cosmology.
\item Dark energy is spatially smooth. It quenches the gravitational collapse of large-scale structures and suppresses the growth of density perturbations; whenever dark energy dominates, structures do not grow.
\item Only relatively recently ($z \lesssim 0.5$) has dark energy come to dominate the energy budget of the universe. At earlier epochs, the dark energy density is small relative to that of matter and radiation, although a $\sim$1\% contribution by dark energy at early times is still allowed by the data.
\item Dark energy affects both the geometry (distances in the universe) and the growth of structure (clustering and abundance of galaxies and galaxy clusters). Separately measuring geometry and growth is an excellent way, not only to measure dark energy parameters, but also to differentiate between separate classes of dark energy models.
\item Dark energy can be studied using a variety of cosmological probes that
  span a wide range of spatial and temporal scales and involve a wide variety
  of observable quantities. Control of systematic errors in these individual cosmological
  probes is key to their ability to discriminate testable predictions of theoretical models. The worldwide effort in theoretically modeling and observationally measuring dark energy reflects a vibrant field with many fruitful avenues that still remain to be explored.
\end{itemize}

\ack
We thank Chris Blake, Gus Evrard, Eric Linder, and Fabian Schmidt for detailed comments on earlier versions of the manuscript. DH is supported by NSF grant AST-1210974, DOE grant DEFG02-95ER40899, and NASA grant NNX16AI41G.

\newcommand{\newblock}{} 
\bibliographystyle{JHEP} 
\bibliography{ropp,wl}

\providecommand{\href}[2]{#2}\begingroup\raggedright\begin{thebibliography}{100}

\bibitem{riess98}
{\bf Supernova Search Team} Collaboration, A.~G. Riess et~al., {\it
  {Observational evidence from supernovae for an accelerating universe and a
  cosmological constant}},  {\em Astron. J.} {\bf 116} (1998) 1009--1038,
  [\href{http://arxiv.org/abs/astro-ph/9805201}{{\tt astro-ph/9805201}}].

\bibitem{perlmutter99}
{\bf Supernova Cosmology Project} Collaboration, S.~Perlmutter et~al., {\it
  {Measurements of Omega and Lambda from 42 high redshift supernovae}},  {\em
  Astrophys. J.} {\bf 517} (1999) 565--586,
  [\href{http://arxiv.org/abs/astro-ph/9812133}{{\tt astro-ph/9812133}}].

\bibitem{Phillips_93}
M.~M. {Phillips}, {\it {The absolute magnitudes of Type Ia supernovae}},  {\em
  \apjl} {\bf 413} (Aug., 1993) L105--L108.

\bibitem{Kim:1997rm}
A.~G. Kim, {\em {The Discovery of high redshift supernovae and their
  cosmological implications}}.
\newblock PhD thesis, UC, Berkeley, 1997.

\bibitem{Copeland_review}
E.~J. Copeland, M.~Sami, and S.~Tsujikawa, {\it {Dynamics of dark energy}},
  {\em Int. J. Mod. Phys.} {\bf D15} (2006) 1753--1936,
  [\href{http://arxiv.org/abs/hep-th/0603057}{{\tt hep-th/0603057}}].

\bibitem{Padmanabhan_review}
T.~Padmanabhan, {\it Cosmological constant: The weight of the vacuum},  {\em
  Phys. Rept.} {\bf 380} (2003) 235--320,
  [\href{http://arxiv.org/abs/hep-th/0212290}{{\tt hep-th/0212290}}].

\bibitem{Li:2011sd}
M.~Li, X.-D. Li, S.~Wang, and Y.~Wang, {\it {Dark Energy}},  {\em Commun.
  Theor. Phys.} {\bf 56} (2011) 525--604,
  [\href{http://arxiv.org/abs/1103.5870}{{\tt arXiv:1103.5870}}].

\bibitem{Peebles_Ratra_03}
P.~J.~E. Peebles and B.~Ratra, {\it The cosmological constant and dark energy},
   {\em Rev. Mod. Phys.} {\bf 75} (2003) 559--606,
  [\href{http://arxiv.org/abs/astro-ph/0207347}{{\tt astro-ph/0207347}}].

\bibitem{Uzan_06}
J.-P. Uzan, {\it The acceleration of the universe and the physics behind it},
  {\em Gen. Rel. Grav.} {\bf 39} (2007) 307--342,
  [\href{http://arxiv.org/abs/astro-ph/0605313}{{\tt astro-ph/0605313}}].

\bibitem{Hut_Tur_00}
D.~Huterer and M.~S. Turner, {\it Probing the dark energy: Methods and
  strategies},  {\em Phys. Rev.} {\bf D64} (2001) 123527,
  [\href{http://arxiv.org/abs/astro-ph/0012510}{{\tt astro-ph/0012510}}].

\bibitem{Weinberg:2012es}
D.~H. Weinberg, M.~J. Mortonson, D.~J. Eisenstein, C.~Hirata, A.~G. Riess, and
  E.~Rozo, {\it {Observational Probes of Cosmic Acceleration}},  {\em Phys.
  Rept.} {\bf 530} (2013) 87--255, [\href{http://arxiv.org/abs/1201.2434}{{\tt
  arXiv:1201.2434}}].

\bibitem{Sahni_review}
V.~{Sahni} and A.~{Starobinsky}, {\it {Reconstructing Dark Energy}},  {\em
  astro-ph/0610026} (Oct., 2006)
  [\href{http://arxiv.org/abs/astro-ph/0610026}{{\tt astro-ph/0610026}}].

\bibitem{Linder_review}
E.~V. Linder, {\it {The Dynamics of Quintessence, The Quintessence of
  Dynamics}},  {\em Gen. Rel. Grav.} {\bf 40} (2008) 329--356,
  [\href{http://arxiv.org/abs/0704.2064}{{\tt arXiv:0704.2064}}].

\bibitem{Carroll_LivRevRel}
S.~M. Carroll, {\it The cosmological constant},  {\em Living Rev. Rel.} {\bf 4}
  (2001) 1, [\href{http://arxiv.org/abs/astro-ph/0004075}{{\tt
  astro-ph/0004075}}].

\bibitem{WeinbergRMP}
S.~Weinberg, {\it The cosmological constant problem},  {\em Rev. Mod. Phys.}
  {\bf 61} (1989) 1--23.

\bibitem{Frieman:2008sn}
J.~Frieman, M.~Turner, and D.~Huterer, {\it {Dark Energy and the Accelerating
  Universe}},  {\em Ann. Rev. Astron. Astrophys.} {\bf 46} (2008) 385--432,
  [\href{http://arxiv.org/abs/0803.0982}{{\tt arXiv:0803.0982}}].

\bibitem{Guth:1980zm}
A.~H. Guth, {\it {The Inflationary Universe: A Possible Solution to the Horizon
  and Flatness Problems}},  {\em Phys. Rev.} {\bf D23} (1981) 347--356.

\bibitem{Linde:1981mu}
A.~D. Linde, {\it {A New Inflationary Universe Scenario: A Possible Solution of
  the Horizon, Flatness, Homogeneity, Isotropy and Primordial Monopole
  Problems}},  {\em Phys. Lett.} {\bf B108} (1982) 389--393.

\bibitem{Albrecht:1982wi}
A.~Albrecht and P.~J. Steinhardt, {\it {Cosmology for Grand Unified Theories
  with Radiatively Induced Symmetry Breaking}},  {\em Phys. Rev. Lett.} {\bf
  48} (1982) 1220--1223.

\bibitem{1991MNRAS.253P..29F}
A.~C. {Fabian}, {\it {On the baryon content of the Shapley Supercluster}},
  {\em \mnras} {\bf 253} (Dec., 1991) 29P.

\bibitem{1991ApJ...379...52W}
S.~D.~M. {White} and C.~S. {Frenk}, {\it {Galaxy formation through hierarchical
  clustering}},  {\em \apj} {\bf 379} (Sept., 1991) 52--79.

\bibitem{White:1993wm}
S.~D.~M. White, J.~F. Navarro, A.~E. Evrard, and C.~S. Frenk, {\it {The Baryon
  content of galaxy clusters: A Challenge to cosmological orthodoxy}},  {\em
  Nature} {\bf 366} (1993) 429--433.

\bibitem{Maddox:1990hb}
S.~J. Maddox, G.~Efstathiou, W.~J. Sutherland, and J.~Loveday, {\it {Galaxy
  correlations on large scales}},  {\em Mon. Not. Roy. Astron. Soc.} {\bf 242}
  (1990) 43--49.

\bibitem{Efstathiou:1990xe}
G.~Efstathiou, W.~J. Sutherland, and S.~J. Maddox, {\it {The cosmological
  constant and cold dark matter}},  {\em Nature} {\bf 348} (1990) 705--707.

\bibitem{Freedman:1994fc}
{\bf Hubble Cepheid} Collaboration, W.~L. Freedman et~al., {\it {Distance to
  the Virgo cluster galaxy M100 from Hubble space telescope observations of
  Cepheids}},  {\em Nature} {\bf 371} (1994) 757--762.

\bibitem{Krauss_Chaboyer}
L.~M. Krauss and B.~Chaboyer, {\it Age estimates of globular clusters in the
  milky way: Constraints on cosmology},  {\em Science} {\bf 299} (2003) 65--70.

\bibitem{Donahue:1997sp}
M.~Donahue, G.~M. Voit, I.~M. Gioia, G.~Luppino, J.~P. Hughes, and J.~T.
  Stocke, {\it {A very hot, high redshift cluster of galaxies: more trouble for
  $\Omega_0 = 1$}},  {\em Astrophys. J.} {\bf 502} (1998) 550,
  [\href{http://arxiv.org/abs/astro-ph/9707010}{{\tt astro-ph/9707010}}].

\bibitem{Bahcall:1998ur}
N.~A. Bahcall and X.-h. Fan, {\it {The Most massive distant clusters:
  Determining $\Omega$ and $\sigma_8$}},  {\em Astrophys. J.} {\bf 504} (1998)
  1, [\href{http://arxiv.org/abs/astro-ph/9803277}{{\tt astro-ph/9803277}}].

\bibitem{Supercal}
D.~Scolnic et~al., {\it {SUPERCAL: Cross=Calibration of Multiple Photometric
  Systems to Improve Cosmological Measurements with type Ia Supernovae}},  {\em
  Astrophys. J.} {\bf 815} (2015), no.~2 117.

\bibitem{Alam:2016hwk}
{\bf BOSS} Collaboration, S.~Alam et~al., {\it {The clustering of galaxies in
  the completed SDSS-III Baryon Oscillation Spectroscopic Survey: cosmological
  analysis of the DR12 galaxy sample}},  {\em Submitted to: Mon. Not. Roy.
  Astron. Soc.} (2016) [\href{http://arxiv.org/abs/1607.03155}{{\tt
  arXiv:1607.03155}}].

\bibitem{Loh:1986wg}
E.~D. Loh and E.~J. Spillar, {\it {A measurement of the mass density of the
  universe}},  {\em Astrophys. J.} {\bf 307} (1986) L1.

\bibitem{1993ApJ...405..437N}
A.~{Nusser} and A.~{Dekel}, {\it {Omega and the initial fluctuations from
  velocity and density fields}},  {\em Astrophys. J.} {\bf 405} (Mar., 1993)
  437--448.

\bibitem{Bernardeau:1994vz}
F.~Bernardeau, R.~Juszkiewicz, A.~Dekel, and F.~R. Bouchet, {\it {Omega from
  the skewness of the cosmic velocity divergence}},  {\em Mon. Not. Roy.
  Astron. Soc.} {\bf 274} (1995) 20--26,
  [\href{http://arxiv.org/abs/astro-ph/9404052}{{\tt astro-ph/9404052}}].

\bibitem{Dekel:1993hq}
A.~Dekel and M.~J. Rees, {\it {Omega from velocities in voids}},  {\em
  Astrophys. J.} {\bf 422} (1994) L1,
  [\href{http://arxiv.org/abs/astro-ph/9308029}{{\tt astro-ph/9308029}}].

\bibitem{Bucher:1994gb}
M.~Bucher, A.~S. Goldhaber, and N.~Turok, {\it {An open universe from
  inflation}},  {\em Phys. Rev.} {\bf D52} (1995) 3314--3337,
  [\href{http://arxiv.org/abs/hep-ph/9411206}{{\tt hep-ph/9411206}}].

\bibitem{Ratra:1994vw}
B.~Ratra and P.~J.~E. Peebles, {\it {Inflation in an open universe}},  {\em
  Phys. Rev.} {\bf D52} (1995) 1837--1894.

\bibitem{Bartlett:1994je}
J.~G. Bartlett, A.~Blanchard, J.~Silk, and M.~S. Turner, {\it {The Case for a
  Hubble constant of 30kms$^{-1}$Mpc$^{-1}$}},  {\em Science} {\bf 267} (1995)
  980--983, [\href{http://arxiv.org/abs/astro-ph/9407061}{{\tt
  astro-ph/9407061}}].

\bibitem{Perlmutter_97}
{\bf Supernova Cosmology Project} Collaboration, S.~Perlmutter et~al., {\it
  {Measurements of the cosmological parameters Omega and Lambda from the first
  7 supernovae at $z>=0.35$}},  {\em Astrophys. J.} {\bf 483} (1997) 565,
  [\href{http://arxiv.org/abs/astro-ph/9608192}{{\tt astro-ph/9608192}}].

\bibitem{Einstein:1917ce}
A.~Einstein, {\it {Cosmological Considerations in the General Theory of
  Relativity}},  {\em Sitzungsber. Preuss. Akad. Wiss. Berlin (Math. Phys.)}
  {\bf 1917} (1917) 142--152.

\bibitem{peebles84}
P.~J.~E. {Peebles}, {\it {Tests of cosmological models constrained by
  inflation}},  {\em \apj} {\bf 284} (Sept., 1984) 439--444.

\bibitem{turner84}
M.~S. {Turner}, G.~{Steigman}, and L.~M. {Krauss}, {\it {Flatness of the
  universe - Reconciling theoretical prejudices with observational data}},
  {\em \prl} {\bf 52} (June, 1984) 2090--2093.

\bibitem{kofman93}
L.~A. {Kofman}, N.~Y. {Gnedin}, and N.~A. {Bahcall}, {\it {Cosmological
  constant, COBE cosmic microwave background anisotropy, and large-scale
  clustering}},  {\em \apj} {\bf 413} (Aug., 1993) 1--9.

\bibitem{Krauss:1995yb}
L.~M. Krauss and M.~S. Turner, {\it {The Cosmological constant is back}},  {\em
  Gen. Rel. Grav.} {\bf 27} (1995) 1137--1144,
  [\href{http://arxiv.org/abs/astro-ph/9504003}{{\tt astro-ph/9504003}}].

\bibitem{Ostriker:1995su}
J.~P. Ostriker and P.~J. Steinhardt, {\it {The Observational case for a low
  density universe with a nonzero cosmological constant}},  {\em Nature} {\bf
  377} (1995) 600--602.

\bibitem{Frieman_PNGB}
J.~A. Frieman, C.~T. Hill, A.~Stebbins, and I.~Waga, {\it Cosmology with
  ultralight pseudo nambu-goldstone bosons},  {\em Phys. Rev. Lett.} {\bf 75}
  (1995) 2077--2080, [\href{http://arxiv.org/abs/astro-ph/9505060}{{\tt
  astro-ph/9505060}}].

\bibitem{Stompor:1995jd}
R.~Stompor, K.~M. Gorski, and A.~J. Banday, {\it {COBE - DMR normalization for
  inflationary flat dark matter models}},  {\em Mon. Not. Roy. Astron. Soc.}
  {\bf 277} (1995) 1225, [\href{http://arxiv.org/abs/astro-ph/9506088}{{\tt
  astro-ph/9506088}}].

\bibitem{Coble}
K.~Coble, S.~Dodelson, and J.~A. Frieman, {\it Dynamical lambda models of
  structure formation},  {\em Phys. Rev.} {\bf D55} (1997) 1851--1859,
  [\href{http://arxiv.org/abs/astro-ph/9608122}{{\tt astro-ph/9608122}}].

\bibitem{Liddle:1995pd}
A.~R. Liddle, D.~H. Lyth, P.~T.~P. Viana, and M.~J. White, {\it {Cold dark
  matter models with a cosmological constant}},  {\em Mon. Not. Roy. Astron.
  Soc.} {\bf 282} (1996) 281,
  [\href{http://arxiv.org/abs/astro-ph/9512102}{{\tt astro-ph/9512102}}].

\bibitem{Perlmutter_Schmidt}
S.~{Perlmutter} and B.~P. {Schmidt}, {\it {Measuring Cosmology with
  Supernovae}},  in {\em Supernovae and Gamma-Ray Bursters} (K.~{Weiler}, ed.),
  vol.~598 of {\em Lecture Notes in Physics, Berlin Springer Verlag},
  pp.~195--217, 2003.

\bibitem{Riess:1996pa}
A.~G. Riess, W.~H. Press, and R.~P. Kirshner, {\it {A Precise distance
  indicator: Type Ia supernova multicolor light curve shapes}},  {\em
  Astrophys. J.} {\bf 473} (1996) 88,
  [\href{http://arxiv.org/abs/astro-ph/9604143}{{\tt astro-ph/9604143}}].

\bibitem{BTC}
D.~M. {Wittman}, J.~A. {Tyson}, G.~M. {Bernstein}, R.~W. {Lee}, I.~P.
  {dell'Antonio}, P.~{Fischer}, D.~R. {Smith}, and M.~M. {Blouke}, {\it {Big
  Throughput Camera: the first year}},  in {\em Optical Astronomical
  Instrumentation} (S.~{D'Odorico}, ed.), vol.~3355 of {\em \procspie},
  pp.~626--634, July, 1998.

\bibitem{Rockosi2002}
C.~M. {Rockosi}, J.~E. {Gunn}, M.~A. {Carr}, M.~{Sekiguchi}, Z.~{Ivezic}, and
  J.~A. {Munn}, {\it {Sloan Digital Sky Survey imaging camera: design and
  performance}},  in {\em Survey and Other Telescope Technologies and
  Discoveries} (J.~A. {Tyson} and S.~{Wolff}, eds.), vol.~4836 of {\em
  \procspie}, pp.~180--188, Dec., 2002.

\bibitem{Miyazaki:2002wa}
S.~Miyazaki et~al., {\it {Subaru prime focus camera: Suprime-Cam}},  {\em Publ.
  Astron. Soc. Jap.} {\bf 54} (2002) 833--853,
  [\href{http://arxiv.org/abs/astro-ph/0211006}{{\tt astro-ph/0211006}}].

\bibitem{LSST_camera}
S.~M. {Kahn}, N.~{Kurita}, K.~{Gilmore}, M.~{Nordby}, P.~{O'Connor},
  R.~{Schindler}, J.~{Oliver}, R.~{Van Berg}, S.~{Olivier}, V.~{Riot},
  P.~{Antilogus}, T.~{Schalk}, M.~{Huffer}, G.~{Bowden}, J.~{Singal}, and
  M.~{Foss}, {\it {Design and development of the 3.2 gigapixel camera for the
  Large Synoptic Survey Telescope}},  in {\em Ground-based and Airborne
  Instrumentation for Astronomy III}, vol.~7735 of {\em \procspie}, p.~77350J,
  July, 2010.

\bibitem{panstarrs_camera}
P.~{Onaka}, C.~{Rae}, S.~{Isani}, J.~L. {Tonry}, A.~{Lee}, R.~{Uyeshiro},
  L.~{Robertson}, and G.~{Ching}, {\it {GPC1 and GPC2: the Pan-STARRS 1.4
  gigapixel mosaic focal plane CCD cameras with an on-sky on-CCD tip-tilt image
  compensation}},  in {\em High Energy, Optical, and Infrared Detectors for
  Astronomy V}, vol.~8453 of {\em \procspie}, p.~84530K, July, 2012.

\bibitem{MMT}
B.~{McLeod}, J.~{Geary}, M.~{Conroy}, D.~{Fabricant}, M.~{Ordway},
  A.~{Szentgyorgyi}, S.~{Amato}, M.~{Ashby}, N.~{Caldwell}, D.~{Curley},
  T.~{Gauron}, M.~{Holman}, T.~{Norton}, M.~{Pieri}, J.~{Roll}, D.~{Weaver},
  J.~{Zajac}, P.~{Palunas}, and D.~{Osip}, {\it {Megacam: A Wide-Field CCD
  Imager for the MMT and Magellan}},  {\em \pasp} {\bf 127} (Apr., 2015) 366.

\bibitem{Flaugher:2015pxc}
{\bf DES} Collaboration, B.~Flaugher et~al., {\it {The Dark Energy Camera}},
  {\em Astron. J.} {\bf 150} (2015) 150,
  [\href{http://arxiv.org/abs/1504.02900}{{\tt arXiv:1504.02900}}].

\bibitem{Knop:2003iy}
{\bf Supernova Cosmology Project} Collaboration, R.~A. Knop et~al., {\it {New
  constraints on Omega(M), Omega(lambda), and w from an independent set of
  eleven high-redshift supernovae observed with HST}},  {\em Astrophys. J.}
  {\bf 598} (2003) 102, [\href{http://arxiv.org/abs/astro-ph/0309368}{{\tt
  astro-ph/0309368}}].

\bibitem{Astier:2005qq}
{\bf SNLS} Collaboration, P.~Astier et~al., {\it {The Supernova legacy survey:
  Measurement of omega(m), omega(lambda) and W from the first year data set}},
  {\em Astron. Astrophys.} {\bf 447} (2006) 31--48,
  [\href{http://arxiv.org/abs/astro-ph/0510447}{{\tt astro-ph/0510447}}].

\bibitem{WoodVasey:2007jb}
{\bf ESSENCE} Collaboration, W.~M. Wood-Vasey et~al., {\it {Observational
  Constraints on the Nature of the Dark Energy: First Cosmological Results from
  the ESSENCE Supernova Survey}},  {\em Astrophys. J.} {\bf 666} (2007)
  694--715, [\href{http://arxiv.org/abs/astro-ph/0701041}{{\tt
  astro-ph/0701041}}].

\bibitem{Miknaitis:2007jd}
G.~Miknaitis et~al., {\it {The ESSENCE Supernova Survey: Survey Optimization,
  Observations, and Supernova Photometry}},  {\em Astrophys. J.} {\bf 666}
  (2007) 674--693, [\href{http://arxiv.org/abs/astro-ph/0701043}{{\tt
  astro-ph/0701043}}].

\bibitem{Kowalski:2008ez}
{\bf Supernova Cosmology Project} Collaboration, M.~Kowalski et~al., {\it
  {Improved Cosmological Constraints from New, Old and Combined Supernova
  Datasets}},  {\em Astrophys. J.} {\bf 686} (2008) 749--778,
  [\href{http://arxiv.org/abs/0804.4142}{{\tt arXiv:0804.4142}}].

\bibitem{Riess:2001gk}
{\bf Supernova Search Team} Collaboration, A.~G. Riess et~al., {\it {The
  farthest known supernova: support for an accelerating universe and a glimpse
  of the epoch of deceleration}},  {\em Astrophys. J.} {\bf 560} (2001) 49--71,
  [\href{http://arxiv.org/abs/astro-ph/0104455}{{\tt astro-ph/0104455}}].

\bibitem{Riess:2004nr}
{\bf Supernova Search Team} Collaboration, A.~G. Riess et~al., {\it {Type Ia
  supernova discoveries at $z>1$ from the Hubble Space Telescope: Evidence for
  past deceleration and constraints on dark energy evolution}},  {\em
  Astrophys. J.} {\bf 607} (2004) 665--687,
  [\href{http://arxiv.org/abs/astro-ph/0402512}{{\tt astro-ph/0402512}}].

\bibitem{Riess:2006fw}
A.~G. Riess et~al., {\it {New Hubble Space Telescope Discoveries of Type Ia
  Supernovae at $z\ge 1$: Narrowing Constraints on the Early Behavior of Dark
  Energy}},  {\em Astrophys. J.} {\bf 659} (2007) 98--121,
  [\href{http://arxiv.org/abs/astro-ph/0611572}{{\tt astro-ph/0611572}}].

\bibitem{Huterer:1998qv}
D.~Huterer and M.~S. Turner, {\it {Prospects for probing the dark energy via
  supernova distance measurements}},  {\em Phys. Rev.} {\bf D60} (1999) 081301,
  [\href{http://arxiv.org/abs/astro-ph/9808133}{{\tt astro-ph/9808133}}].

\bibitem{Scrimgeour:2012wt}
M.~Scrimgeour et~al., {\it {The WiggleZ Dark Energy Survey: the transition to
  large-scale cosmic homogeneity}},  {\em Mon. Not. Roy. Astron. Soc.} {\bf
  425} (2012) 116--134, [\href{http://arxiv.org/abs/1205.6812}{{\tt
  arXiv:1205.6812}}].

\bibitem{Laurent:2016eqo}
P.~Laurent et~al., {\it {A 14 $h^{-3}$ Gpc$^3$ study of cosmic homogeneity
  using BOSS DR12 quasar sample}},  {\em JCAP} {\bf 1611} (2016), no.~11 060,
  [\href{http://arxiv.org/abs/1602.09010}{{\tt arXiv:1602.09010}}].

\bibitem{1980lssu.book.....P}
P.~J.~E. {Peebles}, {\em {The large-scale structure of the universe}}.
\newblock 1980.

\bibitem{Linder_growth}
E.~V. Linder, {\it Cosmic growth history and expansion history},  {\em Phys.
  Rev.} {\bf D72} (2005) 043529,
  [\href{http://arxiv.org/abs/astro-ph/0507263}{{\tt astro-ph/0507263}}].

\bibitem{LinCah07}
E.~V. {Linder} and R.~N. {Cahn}, {\it {Parameterized beyond-Einstein growth}},
  {\em Astroparticle Physics} {\bf 28} (Dec., 2007) 481--488,
  [\href{http://arxiv.org/abs/astro-ph/0701317}{{\tt astro-ph/0701317}}].

\bibitem{Takahashi:2012em}
R.~Takahashi, M.~Sato, T.~Nishimichi, A.~Taruya, and M.~Oguri, {\it {Revising
  the Halofit Model for the Nonlinear Matter Power Spectrum}},  {\em Astrophys.
  J.} {\bf 761} (2012) 152, [\href{http://arxiv.org/abs/1208.2701}{{\tt
  arXiv:1208.2701}}].

\bibitem{Mead:2016zqy}
A.~Mead, C.~Heymans, L.~Lombriser, J.~Peacock, O.~Steele, and H.~Winther, {\it
  {Accurate halo-model matter power spectra with dark energy, massive neutrinos
  and modified gravitational forces}},  {\em Mon. Not. Roy. Astron. Soc.} {\bf
  459} (2016), no.~2 1468--1488, [\href{http://arxiv.org/abs/1602.02154}{{\tt
  arXiv:1602.02154}}].

\bibitem{Ade:2015xua}
{\bf Planck} Collaboration, P.~A.~R. Ade et~al., {\it {Planck 2015 results.
  XIII. Cosmological parameters}},  {\em Astron. Astrophys.} {\bf 594} (2016)
  A13, [\href{http://arxiv.org/abs/1502.01589}{{\tt arXiv:1502.01589}}].

\bibitem{1988A&A...206..175L}
E.~V. {Linder}, {\it {Cosmological tests of generalized Friedmann models}},
  {\em \aap} {\bf 206} (Nov., 1988) 175--189.

\bibitem{Turner_White}
M.~S. Turner and M.~J. White, {\it Cdm models with a smooth component},  {\em
  Phys. Rev.} {\bf D56} (1997) 4439--4443,
  [\href{http://arxiv.org/abs/astro-ph/9701138}{{\tt astro-ph/9701138}}].

\bibitem{Cooray_Huterer}
A.~R. Cooray and D.~Huterer, {\it Gravitational lensing as a probe of
  quintessence},  {\em \apj} {\bf 513} (1999) L95--L98,
  [\href{http://arxiv.org/abs/astro-ph/9901097}{{\tt astro-ph/9901097}}].

\bibitem{Gerke_Efstat}
B.~F. Gerke and G.~Efstathiou, {\it Probing quintessence: Reconstruction and
  parameter estimation from supernovae},  {\em Mon. Not. Roy. Astron. Soc.}
  {\bf 335} (2002) 33, [\href{http://arxiv.org/abs/astro-ph/0201336}{{\tt
  astro-ph/0201336}}].

\bibitem{Linder_wa}
E.~V. Linder, {\it Exploring the expansion history of the universe},  {\em
  Phys. Rev. Lett.} {\bf 90} (2003) 091301,
  [\href{http://arxiv.org/abs/astro-ph/0208512}{{\tt astro-ph/0208512}}].

\bibitem{Chevallier_Polarski}
M.~Chevallier and D.~Polarski, {\it Accelerating universes with scaling dark
  matter},  {\em Int. J. Mod. Phys.} {\bf D10} (2001) 213--224,
  [\href{http://arxiv.org/abs/gr-qc/0009008}{{\tt gr-qc/0009008}}].

\bibitem{Sahni:2002fz}
V.~Sahni, T.~D. Saini, A.~A. Starobinsky, and U.~Alam, {\it {Statefinder: A New
  geometrical diagnostic of dark energy}},  {\em JETP Lett.} {\bf 77} (2003)
  201--206, [\href{http://arxiv.org/abs/astro-ph/0201498}{{\tt
  astro-ph/0201498}}]. [Pisma Zh. Eksp. Teor. Fiz.77,249(2003)].

\bibitem{Corasaniti_foundations}
P.~S. Corasaniti, M.~Kunz, D.~Parkinson, E.~J. Copeland, and B.~A. Bassett,
  {\it The foundations of observing dark energy dynamics with the wilkinson
  microwave anisotropy probe},  {\em Phys. Rev.} {\bf D70} (2004) 083006,
  [\href{http://arxiv.org/abs/astro-ph/0406608}{{\tt astro-ph/0406608}}].

\bibitem{Huterer_Starkman}
D.~Huterer and G.~Starkman, {\it Parameterization of dark-energy properties: A
  principal- component approach},  {\em Phys. Rev. Lett.} {\bf 90} (2003)
  031301, [\href{http://arxiv.org/abs/astro-ph/0207517}{{\tt
  astro-ph/0207517}}].

\bibitem{Albrecht:2009ct}
A.~Albrecht et~al., {\it {Findings of the Joint Dark Energy Mission Figure of
  Merit Science Working Group}},  \href{http://arxiv.org/abs/0901.0721}{{\tt
  arXiv:0901.0721}}.

\bibitem{Ruiz:2012rc}
E.~J. Ruiz, D.~L. Shafer, D.~Huterer, and A.~Conley, {\it {Principal components
  of dark energy with SNLS supernovae: the effects of systematic errors}},
  {\em Phys. Rev.} {\bf D86} (2012) 103004,
  [\href{http://arxiv.org/abs/1207.4781}{{\tt arXiv:1207.4781}}].

\bibitem{Mortonson:2008qy}
M.~J. Mortonson, W.~Hu, and D.~Huterer, {\it {Falsifying Paradigms for Cosmic
  Acceleration}},  {\em Phys. Rev.} {\bf D79} (2009) 023004,
  [\href{http://arxiv.org/abs/0810.1744}{{\tt arXiv:0810.1744}}].

\bibitem{dePutter_Linder}
R.~de~Putter and E.~V. Linder, {\it {To Bin or Not To Bin: Decorrelating the
  Cosmic Equation of State}},  {\em Astropart. Phys.} {\bf 29} (2008) 424,
  [\href{http://arxiv.org/abs/0710.0373}{{\tt arXiv:0710.0373}}].

\bibitem{Huterer_Cooray}
D.~Huterer and A.~Cooray, {\it Uncorrelated estimates of dark energy
  evolution},  {\em Phys. Rev.} {\bf D71} (2005) 023506,
  [\href{http://arxiv.org/abs/astro-ph/0404062}{{\tt astro-ph/0404062}}].

\bibitem{Hu_PC}
W.~Hu, {\it Dark energy and matter evolution from lensing tomography},  {\em
  Phys. Rev.} {\bf D66} (2002) 083515,
  [\href{http://arxiv.org/abs/astro-ph/0208093}{{\tt astro-ph/0208093}}].

\bibitem{Hojjati:2011xd}
A.~Hojjati, G.-B. Zhao, L.~Pogosian, A.~Silvestri, R.~Crittenden, and
  K.~Koyama, {\it {Cosmological tests of General Relativity: a principal
  component analysis}},  {\em Phys. Rev.} {\bf D85} (2012) 043508,
  [\href{http://arxiv.org/abs/1111.3960}{{\tt arXiv:1111.3960}}].

\bibitem{Zhao:2009fn}
G.-B. Zhao, L.~Pogosian, A.~Silvestri, and J.~Zylberberg, {\it {Cosmological
  Tests of General Relativity with Future Tomographic Surveys}},  {\em Phys.
  Rev. Lett.} {\bf 103} (2009) 241301,
  [\href{http://arxiv.org/abs/0905.1326}{{\tt arXiv:0905.1326}}].

\bibitem{Ade:2015rim}
{\bf Planck} Collaboration, P.~A.~R. Ade et~al., {\it {Planck 2015 results.
  XIV. Dark energy and modified gravity}},  {\em Astron. Astrophys.} {\bf 594}
  (2016) A14, [\href{http://arxiv.org/abs/1502.01590}{{\tt arXiv:1502.01590}}].

\bibitem{Saini_reconstr}
T.~D. Saini, S.~Raychaudhury, V.~Sahni, and A.~A. Starobinsky, {\it
  Reconstructing the cosmic equation of state from supernova distances},  {\em
  Phys. Rev. Lett.} {\bf 85} (2000) 1162--1165,
  [\href{http://arxiv.org/abs/astro-ph/9910231}{{\tt astro-ph/9910231}}].

\bibitem{Nakamura_Chiba}
T.~Nakamura and T.~Chiba, {\it Determining the equation of state of the
  expanding universe: Inverse problem in cosmology},  {\em Mon. Not. Roy.
  Astron. Soc.} {\bf 306} (1999) 696--700,
  [\href{http://arxiv.org/abs/astro-ph/9810447}{{\tt astro-ph/9810447}}].

\bibitem{Starobinsky}
A.~A. Starobinsky, {\it How to determine an effective potential for a variable
  cosmological term},  {\em JETP Lett.} {\bf 68} (1998) 757--763,
  [\href{http://arxiv.org/abs/astro-ph/9810431}{{\tt astro-ph/9810431}}].

\bibitem{Wang:2003gz}
Y.~Wang and P.~Mukherjee, {\it {Model - independent constraints on dark energy
  density from flux - averaging analysis of type Ia supernova data}},  {\em
  Astrophys. J.} {\bf 606} (2004) 654--663,
  [\href{http://arxiv.org/abs/astro-ph/0312192}{{\tt astro-ph/0312192}}].

\bibitem{Wang_Teg_uncorr}
Y.~Wang and M.~Tegmark, {\it Uncorrelated measurements of the cosmic expansion
  history and dark energy from supernovae},  {\em Phys. Rev.} {\bf D71} (2005)
  103513, [\href{http://arxiv.org/abs/astro-ph/0501351}{{\tt
  astro-ph/0501351}}].

\bibitem{Weller_Albrecht}
J.~Weller and A.~Albrecht, {\it Future supernovae observations as a probe of
  dark energy},  {\em Phys. Rev.} {\bf D65} (2002) 103512,
  [\href{http://arxiv.org/abs/astro-ph/0106079}{{\tt astro-ph/0106079}}].

\bibitem{Holsclaw:2010sk}
T.~Holsclaw, U.~Alam, B.~Sanso, H.~Lee, K.~Heitmann, S.~Habib, and D.~Higdon,
  {\it {Nonparametric Dark Energy Reconstruction from Supernova Data}},  {\em
  Phys. Rev. Lett.} {\bf 105} (2010) 241302,
  [\href{http://arxiv.org/abs/1011.3079}{{\tt arXiv:1011.3079}}].

\bibitem{Crittenden:2011aa}
R.~G. Crittenden, G.-B. Zhao, L.~Pogosian, L.~Samushia, and X.~Zhang, {\it
  {Fables of reconstruction: controlling bias in the dark energy equation of
  state}},  {\em JCAP} {\bf 1202} (2012) 048,
  [\href{http://arxiv.org/abs/1112.1693}{{\tt arXiv:1112.1693}}].

\bibitem{Shafieloo:2012ht}
A.~Shafieloo, A.~G. Kim, and E.~V. Linder, {\it {Gaussian Process
  Cosmography}},  {\em Phys. Rev.} {\bf D85} (2012) 123530,
  [\href{http://arxiv.org/abs/1204.2272}{{\tt arXiv:1204.2272}}].

\bibitem{Zhao:2012aw}
G.-B. Zhao, R.~G. Crittenden, L.~Pogosian, and X.~Zhang, {\it {Examining the
  evidence for dynamical dark energy}},  {\em Phys. Rev. Lett.} {\bf 109}
  (2012) 171301, [\href{http://arxiv.org/abs/1207.3804}{{\tt
  arXiv:1207.3804}}].

\bibitem{Shafieloo:2012ms}
A.~Shafieloo, A.~G. Kim, and E.~V. Linder, {\it {Model independent tests of
  cosmic growth versus expansion}},  {\em Phys. Rev.} {\bf D87} (2013), no.~2
  023520, [\href{http://arxiv.org/abs/1211.6128}{{\tt arXiv:1211.6128}}].

\bibitem{DETF}
A.~{Albrecht}, G.~{Bernstein}, R.~{Cahn}, W.~L. {Freedman}, J.~{Hewitt},
  W.~{Hu}, J.~{Huth}, M.~{Kamionkowski}, E.~W. {Kolb}, L.~{Knox}, J.~C.
  {Mather}, S.~{Staggs}, and N.~B. {Suntzeff}, {\it {Report of the Dark Energy
  Task Force}},  {\em astro-ph/0609591} (Sept., 2006).

\bibitem{MHH_FoM}
M.~J. Mortonson, D.~Huterer, and W.~Hu, {\it {Figures of merit for present and
  future dark energy probes}},  {\em Phys. Rev.} {\bf D82} (2010) 063004,
  [\href{http://arxiv.org/abs/1004.0236}{{\tt arXiv:1004.0236}}].

\bibitem{Hu_GDM}
W.~Hu, {\it Structure formation with generalized dark matter},  {\em \apj} {\bf
  506} (1998) 485--494, [\href{http://arxiv.org/abs/astro-ph/9801234}{{\tt
  astro-ph/9801234}}].

\bibitem{Hu:1998tj}
W.~Hu and D.~J. Eisenstein, {\it {The Structure of structure formation
  theories}},  {\em Phys. Rev.} {\bf D59} (1999) 083509,
  [\href{http://arxiv.org/abs/astro-ph/9809368}{{\tt astro-ph/9809368}}].

\bibitem{Erickson:2001bq}
J.~K. Erickson, R.~R. Caldwell, P.~J. Steinhardt, C.~Armendariz-Picon, and
  V.~F. Mukhanov, {\it {Measuring the speed of sound of quintessence}},  {\em
  Phys. Rev. Lett.} {\bf 88} (2002) 121301,
  [\href{http://arxiv.org/abs/astro-ph/0112438}{{\tt astro-ph/0112438}}].

\bibitem{DeDeo:2003te}
S.~DeDeo, R.~R. Caldwell, and P.~J. Steinhardt, {\it {Effects of the sound
  speed of quintessence on the microwave background and large scale
  structure}},  {\em Phys. Rev.} {\bf D67} (2003) 103509,
  [\href{http://arxiv.org/abs/astro-ph/0301284}{{\tt astro-ph/0301284}}].
  [Erratum: Phys. Rev.D69,129902(2004)].

\bibitem{Weller:2003hw}
J.~Weller and A.~M. Lewis, {\it {Large scale cosmic microwave background
  anisotropies and dark energy}},  {\em Mon. Not. Roy. Astron. Soc.} {\bf 346}
  (2003) 987--993, [\href{http://arxiv.org/abs/astro-ph/0307104}{{\tt
  astro-ph/0307104}}].

\bibitem{Hannestad_sound}
S.~Hannestad, {\it Constraints on the sound speed of dark energy},  {\em Phys.
  Rev.} {\bf D71} (2005) 103519,
  [\href{http://arxiv.org/abs/astro-ph/0504017}{{\tt astro-ph/0504017}}].

\bibitem{dePutter:2010vy}
R.~de~Putter, D.~Huterer, and E.~V. Linder, {\it {Measuring the Speed of Dark:
  Detecting Dark Energy Perturbations}},  {\em Phys. Rev.} {\bf D81} (2010)
  103513, [\href{http://arxiv.org/abs/1002.1311}{{\tt arXiv:1002.1311}}].

\bibitem{Wetterich:2004pv}
C.~Wetterich, {\it {Phenomenological parameterization of quintessence}},  {\em
  Phys. Lett.} {\bf B594} (2004) 17--22,
  [\href{http://arxiv.org/abs/astro-ph/0403289}{{\tt astro-ph/0403289}}].

\bibitem{Doran:2006kp}
M.~Doran and G.~Robbers, {\it {Early dark energy cosmologies}},  {\em JCAP}
  {\bf 0606} (2006) 026, [\href{http://arxiv.org/abs/astro-ph/0601544}{{\tt
  astro-ph/0601544}}].

\bibitem{Linder:2006ud}
E.~V. Linder, {\it {Dark Energy in the Dark Ages}},  {\em Astropart. Phys.}
  {\bf 26} (2006) 16--21, [\href{http://arxiv.org/abs/astro-ph/0603584}{{\tt
  astro-ph/0603584}}].

\bibitem{Zlatev:1998tr}
I.~Zlatev, L.-M. Wang, and P.~J. Steinhardt, {\it {Quintessence, cosmic
  coincidence, and the cosmological constant}},  {\em Phys. Rev. Lett.} {\bf
  82} (1999) 896--899, [\href{http://arxiv.org/abs/astro-ph/9807002}{{\tt
  astro-ph/9807002}}].

\bibitem{Linder:2010wp}
E.~V. Linder and T.~L. Smith, {\it {Dark Before Light: Testing the Cosmic
  Expansion History through the Cosmic Microwave Background}},  {\em JCAP} {\bf
  1104} (2011) 001, [\href{http://arxiv.org/abs/1009.3500}{{\tt
  arXiv:1009.3500}}].

\bibitem{Reichardt:2011fv}
C.~L. Reichardt, R.~de~Putter, O.~Zahn, and Z.~Hou, {\it {New limits on Early
  Dark Energy from the South Pole Telescope}},  {\em Astrophys. J.} {\bf 749}
  (2012) L9, [\href{http://arxiv.org/abs/1110.5328}{{\tt arXiv:1110.5328}}].

\bibitem{Calabrese:2011hg}
E.~Calabrese, D.~Huterer, E.~V. Linder, A.~Melchiorri, and L.~Pagano, {\it
  {Limits on Dark Radiation, Early Dark Energy, and Relativistic Degrees of
  Freedom}},  {\em Phys. Rev.} {\bf D83} (2011) 123504,
  [\href{http://arxiv.org/abs/1103.4132}{{\tt arXiv:1103.4132}}].

\bibitem{Gradwohl_Frieman}
B.-A. Gradwohl and J.~A. Frieman, {\it Dark matter, long range forces, and
  large scale structure},  {\em \apj} {\bf 398} (1992) 407--424.

\bibitem{Amendola_99}
L.~Amendola, {\it Coupled quintessence},  {\em Phys. Rev.} {\bf D62} (2000)
  043511, [\href{http://arxiv.org/abs/astro-ph/9908023}{{\tt
  astro-ph/9908023}}].

\bibitem{Farrar_Peebles}
G.~R. Farrar and P.~J.~E. Peebles, {\it Interacting dark matter and dark
  energy},  {\em \apj} {\bf 604} (2004) 1--11,
  [\href{http://arxiv.org/abs/astro-ph/0307316}{{\tt astro-ph/0307316}}].

\bibitem{Wang:2016lxa}
B.~Wang, E.~Abdalla, F.~Atrio-Barandela, and D.~Pavon, {\it {Dark Matter and
  Dark Energy Interactions: Theoretical Challenges, Cosmological Implications
  and Observational Signatures}},  {\em Rept. Prog. Phys.} {\bf 79} (2016),
  no.~9 096901, [\href{http://arxiv.org/abs/1603.08299}{{\tt
  arXiv:1603.08299}}].

\bibitem{Silvestri:2009hh}
A.~Silvestri and M.~Trodden, {\it {Approaches to Understanding Cosmic
  Acceleration}},  {\em Rept. Prog. Phys.} {\bf 72} (2009) 096901,
  [\href{http://arxiv.org/abs/0904.0024}{{\tt arXiv:0904.0024}}].

\bibitem{Joyce:2014kja}
A.~Joyce, B.~Jain, J.~Khoury, and M.~Trodden, {\it {Beyond the Cosmological
  Standard Model}},  {\em Phys. Rept.} {\bf 568} (2015) 1--98,
  [\href{http://arxiv.org/abs/1407.0059}{{\tt arXiv:1407.0059}}].

\bibitem{Joyce:2016vqv}
A.~Joyce, L.~Lombriser, and F.~Schmidt, {\it {Dark Energy Versus Modified
  Gravity}},  {\em Ann. Rev. Nucl. Part. Sci.} {\bf 66} (2016) 95--122,
  [\href{http://arxiv.org/abs/1601.06133}{{\tt arXiv:1601.06133}}].

\bibitem{Daniel:2012kn}
S.~F. Daniel and E.~V. Linder, {\it {Constraining Cosmic Expansion and Gravity
  with Galaxy Redshift Surveys}},  {\em JCAP} {\bf 1302} (2013) 007,
  [\href{http://arxiv.org/abs/1212.0009}{{\tt arXiv:1212.0009}}].

\bibitem{Caldwell:2007cw}
R.~Caldwell, A.~Cooray, and A.~Melchiorri, {\it {Constraints on a New
  Post-General Relativity Cosmological Parameter}},  {\em Phys. Rev.} {\bf D76}
  (2007) 023507, [\href{http://arxiv.org/abs/astro-ph/0703375}{{\tt
  astro-ph/0703375}}].

\bibitem{Bertschinger:2008zb}
E.~Bertschinger and P.~Zukin, {\it {Distinguishing Modified Gravity from Dark
  Energy}},  {\em Phys. Rev.} {\bf D78} (2008) 024015,
  [\href{http://arxiv.org/abs/0801.2431}{{\tt arXiv:0801.2431}}].

\bibitem{Daniel:2010ky}
S.~F. Daniel, E.~V. Linder, T.~L. Smith, R.~R. Caldwell, A.~Cooray,
  A.~Leauthaud, and L.~Lombriser, {\it {Testing General Relativity with Current
  Cosmological Data}},  {\em Phys. Rev.} {\bf D81} (2010) 123508,
  [\href{http://arxiv.org/abs/1002.1962}{{\tt arXiv:1002.1962}}].

\bibitem{hojjati13062546}
S.~Asaba, C.~Hikage, K.~Koyama, G.-B. Zhao, A.~Hojjati, et~al., {\it {Principal
  Component Analysis of Modified Gravity using Weak Lensing and Peculiar
  Velocity Measurements}},  {\em JCAP} {\bf 1308} (2013) 029,
  [\href{http://arxiv.org/abs/1306.2546}{{\tt arXiv:1306.2546}}].

\bibitem{Bean:2010zq}
R.~Bean and M.~Tangmatitham, {\it {Current constraints on the cosmic growth
  history}},  {\em Phys. Rev.} {\bf D81} (2010) 083534,
  [\href{http://arxiv.org/abs/1002.4197}{{\tt arXiv:1002.4197}}].

\bibitem{Zhao:2011te}
G.-B. Zhao, H.~Li, E.~V. Linder, K.~Koyama, D.~J. Bacon, et~al., {\it {Testing
  Einstein Gravity with Cosmic Growth and Expansion}},  {\em Phys.Rev.} {\bf
  D85} (2012) 123546, [\href{http://arxiv.org/abs/1109.1846}{{\tt
  arXiv:1109.1846}}].

\bibitem{Dossett:2011tn}
J.~N. Dossett, M.~Ishak, and J.~Moldenhauer, {\it {Testing General Relativity
  at Cosmological Scales: Implementation and Parameter Correlations}},  {\em
  Phys.Rev.} {\bf D84} (2011) 123001,
  [\href{http://arxiv.org/abs/1109.4583}{{\tt arXiv:1109.4583}}].

\bibitem{silvestri13021193}
A.~{Silvestri}, L.~{Pogosian}, and R.~V. {Buniy}, {\it {Practical approach to
  cosmological perturbations in modified gravity}},  {\em Phys.Rev.D} {\bf 87}
  (May, 2013) 104015, [\href{http://arxiv.org/abs/1302.1193}{{\tt
  arXiv:1302.1193}}].

\bibitem{Hut_Lind_MMG}
D.~Huterer and E.~V. Linder, {\it Separating dark physics from physical
  darkness: Minimalist modified gravity vs. dark energy},  {\em Phys. Rev.}
  {\bf D75} (2007) 023519, [\href{http://arxiv.org/abs/astro-ph/0608681}{{\tt
  astro-ph/0608681}}].

\bibitem{Zhang:2007nk}
P.~Zhang, M.~Liguori, R.~Bean, and S.~Dodelson, {\it {Probing Gravity at
  Cosmological Scales by Measurements which Test the Relationship between
  Gravitational Lensing and Matter Overdensity}},  {\em Phys. Rev. Lett.} {\bf
  99} (2007) 141302, [\href{http://arxiv.org/abs/0704.1932}{{\tt
  arXiv:0704.1932}}].

\bibitem{Reyes:2010tr}
R.~Reyes, R.~Mandelbaum, U.~Seljak, T.~Baldauf, J.~E. Gunn, L.~Lombriser, and
  R.~E. Smith, {\it {Confirmation of general relativity on large scales from
  weak lensing and galaxy velocities}},  {\em Nature} {\bf 464} (2010)
  256--258, [\href{http://arxiv.org/abs/1003.2185}{{\tt arXiv:1003.2185}}].

\bibitem{Marshall:2004zd}
P.~Marshall, N.~Rajguru, and A.~Slosar, {\it {Bayesian evidence as a tool for
  comparing datasets}},  {\em Phys. Rev.} {\bf D73} (2006) 067302,
  [\href{http://arxiv.org/abs/astro-ph/0412535}{{\tt astro-ph/0412535}}].

\bibitem{Trotta:2008qt}
R.~Trotta, {\it {Bayes in the sky: Bayesian inference and model selection in
  cosmology}},  {\em Contemp. Phys.} {\bf 49} (2008) 71--104,
  [\href{http://arxiv.org/abs/0803.4089}{{\tt arXiv:0803.4089}}].

\bibitem{Grandis:2015qaa}
S.~Grandis, S.~Seehars, A.~Refregier, A.~Amara, and A.~Nicola, {\it
  {Information Gains from Cosmological Probes}},  {\em JCAP} {\bf 1605} (2016),
  no.~05 034, [\href{http://arxiv.org/abs/1510.06422}{{\tt arXiv:1510.06422}}].

\bibitem{Raveri:2015maa}
M.~Raveri, {\it {Are cosmological data sets consistent with each other within
  the $\Lambda$ cold dark matter model?}},  {\em Phys. Rev.} {\bf D93} (2016),
  no.~4 043522, [\href{http://arxiv.org/abs/1510.00688}{{\tt
  arXiv:1510.00688}}].

\bibitem{Charnock:2017vcd}
T.~Charnock, R.~A. Battye, and A.~Moss, {\it {Planck data versus large scale
  structure}},  {\em Phys. Rev.} {\bf D95} (2017), no.~12 123535,
  [\href{http://arxiv.org/abs/1703.05959}{{\tt arXiv:1703.05959}}].

\bibitem{Lin:2017ikq}
W.~Lin and M.~Ishak, {\it {Cosmological discordances: A new measure,
  marginalization effects, and application to geometry versus growth current
  data sets}},  {\em Phys. Rev.} {\bf D96} (2017), no.~2 023532,
  [\href{http://arxiv.org/abs/1705.05303}{{\tt arXiv:1705.05303}}].

\bibitem{Lin:2017bhs}
W.~Lin and M.~Ishak, {\it {Cosmological discordances II: Hubble constant,
  Planck and large-scale-structure data sets}},  {\em Phys. Rev.} {\bf D96}
  (2017), no.~8 083532, [\href{http://arxiv.org/abs/1708.09813}{{\tt
  arXiv:1708.09813}}].

\bibitem{Mortonson:2009hk}
M.~J. Mortonson, W.~Hu, and D.~Huterer, {\it {Testable dark energy predictions
  from current data}},  {\em Phys. Rev.} {\bf D81} (2010) 063007,
  [\href{http://arxiv.org/abs/0912.3816}{{\tt arXiv:0912.3816}}].

\bibitem{Vanderveld:2012ec}
R.~A. Vanderveld, M.~J. Mortonson, W.~Hu, and T.~Eifler, {\it {Testing dark
  energy paradigms with weak gravitational lensing}},  {\em Phys. Rev.} {\bf
  D85} (2012) 103518, [\href{http://arxiv.org/abs/1203.3195}{{\tt
  arXiv:1203.3195}}].

\bibitem{Ishak_Upadhye}
M.~Ishak, A.~Upadhye, and D.~N. Spergel, {\it Probing cosmic acceleration
  beyond the equation of state: Distinguishing between dark energy and modified
  gravity models},  {\em Phys. Rev.} {\bf D74} (2006) 043513,
  [\href{http://arxiv.org/abs/astro-ph/0507184}{{\tt astro-ph/0507184}}].

\bibitem{Wang:2007fsa}
S.~Wang, L.~Hui, M.~May, and Z.~Haiman, {\it {Is Modified Gravity Required by
  Observations? An Empirical Consistency Test of Dark Energy Models}},  {\em
  Phys. Rev.} {\bf D76} (2007) 063503,
  [\href{http://arxiv.org/abs/0705.0165}{{\tt arXiv:0705.0165}}].

\bibitem{Zhang:2003ii}
J.~Zhang, L.~Hui, and A.~Stebbins, {\it {Isolating geometry in weak lensing
  measurements}},  {\em Astrophys. J.} {\bf 635} (2005) 806--820,
  [\href{http://arxiv.org/abs/astro-ph/0312348}{{\tt astro-ph/0312348}}].

\bibitem{MacCrann:2014wfa}
N.~MacCrann, J.~Zuntz, S.~Bridle, B.~Jain, and M.~R. Becker, {\it {Cosmic
  Discordance: Are Planck CMB and CFHTLenS weak lensing measurements out of
  tune?}},  {\em Mon. Not. Roy. Astron. Soc.} {\bf 451} (2015), no.~3
  2877--2888, [\href{http://arxiv.org/abs/1408.4742}{{\tt arXiv:1408.4742}}].

\bibitem{Ruiz:2014hma}
E.~J. Ruiz and D.~Huterer, {\it {Testing the dark energy consistency with
  geometry and growth}},  {\em Phys. Rev.} {\bf D91} (2015) 063009,
  [\href{http://arxiv.org/abs/1410.5832}{{\tt arXiv:1410.5832}}].

\bibitem{Bernal:2015zom}
J.~L. Bernal, L.~Verde, and A.~J. Cuesta, {\it {Parameter splitting in dark
  energy: is dark energy the same in the background and in the cosmic
  structures?}},  {\em JCAP} {\bf 1602} (2016), no.~02 059,
  [\href{http://arxiv.org/abs/1511.03049}{{\tt arXiv:1511.03049}}].

\bibitem{Kowal}
C.~T. {Kowal}, {\it {Absolute magnitudes of supernovae.}},  {\em \aj} {\bf 73}
  (Dec., 1968) 1021--1024.

\bibitem{Maoz:2011iv}
D.~Maoz and F.~Mannucci, {\it {Type-Ia supernova rates and the progenitor
  problem, a review}},  {\em Publ. Astron. Soc. Austral.} {\bf 29} (2012) 447,
  [\href{http://arxiv.org/abs/1111.4492}{{\tt arXiv:1111.4492}}].

\bibitem{Wang:2012za}
B.~Wang and Z.~Han, {\it {Progenitors of type Ia supernovae}},  {\em New
  Astron. Rev.} {\bf 56} (2012) 122--141,
  [\href{http://arxiv.org/abs/1204.1155}{{\tt arXiv:1204.1155}}].

\bibitem{Baade_Zwicky}
W.~{Baade} and F.~{Zwicky}, {\it {On Super-novae}},  {\em Proceedings of the
  National Academy of Science} {\bf 20} (May, 1934) 254--259.

\bibitem{Wagoner}
R.~V. {Wagoner}, {\it {Determining q0 from Supernovae}},  {\em \apjl} {\bf 214}
  (May, 1977) L5+.

\bibitem{Colgate}
S.~A. {Colgate}, {\it {Supernovae as a standard candle for cosmology}},  {\em
  \apj} {\bf 232} (Sept., 1979) 404--408.

\bibitem{Norgaard_Nielsen}
H.~U. {Norgaard-Nielsen}, L.~{Hansen}, H.~E. {Jorgensen}, A.~{Aragon
  Salamanca}, and R.~S. {Ellis}, {\it {The discovery of a type IA supernova at
  a redshift of 0.31}},  {\em Nature} {\bf 339} (June, 1989) 523--525.

\bibitem{Guy:2007dv}
{\bf SNLS} Collaboration, J.~Guy et~al., {\it {SALT2: Using distant supernovae
  to improve the use of Type Ia supernovae as distance indicators}},  {\em
  Astron. Astrophys.} {\bf 466} (2007) 11--21,
  [\href{http://arxiv.org/abs/astro-ph/0701828}{{\tt astro-ph/0701828}}].

\bibitem{hicken}
M.~{Hicken} et~al., {\it {Improved Dark Energy Constraints from \~{}100 New CfA
  Supernova Type Ia Light Curves}},  {\em \apj} {\bf 700} (Aug., 2009)
  1097--1140, [\href{http://arxiv.org/abs/0901.4804}{{\tt arXiv:0901.4804}}].

\bibitem{Union2}
R.~Amanullah et~al., {\it {Spectra and Light Curves of Six Type Ia Supernovae
  at $0.511 < z < 1.12$ and the Union2 Compilation}},  {\em Astrophys. J.} {\bf
  716} (2010) 712--738, [\href{http://arxiv.org/abs/1004.1711}{{\tt
  arXiv:1004.1711}}].

\bibitem{Conley:2011ku}
{\bf SNLS} Collaboration, A.~Conley et~al., {\it {Supernova Constraints and
  Systematic Uncertainties from the First 3 Years of the Supernova Legacy
  Survey}},  {\em Astrophys. J. Suppl.} {\bf 192} (2011) 1,
  [\href{http://arxiv.org/abs/1104.1443}{{\tt arXiv:1104.1443}}].

\bibitem{Holz_Linder}
D.~E. Holz and E.~V. Linder, {\it Safety in numbers: Gravitational lensing
  degradation of the luminosity distance-redshift relation},  {\em \apj} {\bf
  631} (2005) 678--688, [\href{http://arxiv.org/abs/astro-ph/0412173}{{\tt
  astro-ph/0412173}}].

\bibitem{March:2011xa}
M.~C. March, R.~Trotta, P.~Berkes, G.~D. Starkman, and P.~M. Vaudrevange, {\it
  {Improved constraints on cosmological parameters from SNIa data}},  {\em Mon.
  Not. Roy. Astron. Soc.} {\bf 418} (2011) 2308--2329,
  [\href{http://arxiv.org/abs/1102.3237}{{\tt arXiv:1102.3237}}]. [Mon. Not.
  Roy. Astron. Soc.418,4(2011)].

\bibitem{Rubin:2015rza}
{\bf Supernova Cosmology Project} Collaboration, D.~Rubin et~al., {\it {Unity:
  Confronting Supernova Cosmology’s Statistical and Systematic Uncertainties
  in a Unified Bayesian Framework}},  {\em Astrophys. J.} {\bf 813} (2015),
  no.~2 137, [\href{http://arxiv.org/abs/1507.01602}{{\tt arXiv:1507.01602}}].

\bibitem{Kunz:2006ik}
M.~Kunz, B.~A. Bassett, and R.~Hlozek, {\it {Bayesian Estimation Applied to
  Multiple Species: Towards cosmology with a million supernovae}},  {\em Phys.
  Rev.} {\bf D75} (2007) 103508,
  [\href{http://arxiv.org/abs/astro-ph/0611004}{{\tt astro-ph/0611004}}].

\bibitem{Hlozek:2011wq}
R.~Hlozek et~al., {\it {Photometric Supernova Cosmology with BEAMS and
  SDSS-II}},  {\em Astrophys. J.} {\bf 752} (2012) 79,
  [\href{http://arxiv.org/abs/1111.5328}{{\tt arXiv:1111.5328}}].

\bibitem{Jones:2016cnm}
D.~O. Jones et~al., {\it {Measuring the Properties of Dark Energy with
  Photometrically Classified Pan-STARRS Supernovae. I. Systematic Uncertainty
  from Core-Collapse Supernova Contamination}},  {\em Astrophys. J.} {\bf 843}
  (2017), no.~1 6, [\href{http://arxiv.org/abs/1611.07042}{{\tt
  arXiv:1611.07042}}].

\bibitem{Scolnic:2016ukm}
D.~Scolnic and R.~Kessler, {\it {Measuring Type Ia Supernova Populations of
  Stretch and Color and Predicting Distance Biases}},  {\em Astrophys. J.} {\bf
  822} (2016), no.~2 L35, [\href{http://arxiv.org/abs/1603.01559}{{\tt
  arXiv:1603.01559}}].

\bibitem{Kessler:2016uwi}
R.~Kessler and D.~Scolnic, {\it {Correcting Type Ia Supernova Distances for
  Selection Biases and Contamination in Photometrically Identified Samples}},
  {\em Astrophys. J.} {\bf 836} (2017), no.~1 56,
  [\href{http://arxiv.org/abs/1610.04677}{{\tt arXiv:1610.04677}}].

\bibitem{Foley:2013bba}
R.~J. Foley and R.~P. Kirshner, {\it {Metallicity Differences in Type Ia
  Supernova Progenitors Inferred from Ultraviolet Spectra}},  {\em Astrophys.
  J.} {\bf 769} (2013) L1, [\href{http://arxiv.org/abs/1302.4479}{{\tt
  arXiv:1302.4479}}].

\bibitem{Graham:2014bva}
M.~L. Graham, R.~J. Foley, W.~Zheng, P.~L. Kelly, I.~Shivvers, J.~M. Silverman,
  A.~V. Filippenko, K.~I. Clubb, and M.~Ganeshalingam, {\it {Twins for life? A
  comparative analysis of the Type Ia supernovae 2011fe and 2011by}},  {\em
  Mon. Not. Roy. Astron. Soc.} {\bf 446} (2015) 2073--2088,
  [\href{http://arxiv.org/abs/1408.2651}{{\tt arXiv:1408.2651}}].

\bibitem{Milne:2014rfa}
P.~A. Milne, R.~J. Foley, P.~J. Brown, and G.~Narayan, {\it {The Changing
  Fractions of Type ia Supernova Nuv–optical Subclasses With Redshift}},
  {\em Astrophys. J.} {\bf 803} (2015), no.~1 20,
  [\href{http://arxiv.org/abs/1408.1706}{{\tt arXiv:1408.1706}}].

\bibitem{Fakhouri:2015mhg}
{\bf Nearby Supernova Factory} Collaboration, H.~K. Fakhouri et~al., {\it
  {Improving Cosmological Distance Measurements Using Twin Type Ia
  Supernovae}},  {\em Astrophys. J.} {\bf 815} (2015), no.~1 58,
  [\href{http://arxiv.org/abs/1511.01102}{{\tt arXiv:1511.01102}}].

\bibitem{Sunyaev:1970eu}
R.~A. Sunyaev and {\relax Ya}.~B. Zeldovich, {\it {Small scale fluctuations of
  relic radiation}},  {\em Astrophys. Space Sci.} {\bf 7} (1970) 3--19.

\bibitem{Peebles:1970ag}
P.~J.~E. Peebles and J.~T. Yu, {\it {Primeval adiabatic perturbation in an
  expanding universe}},  {\em Astrophys. J.} {\bf 162} (1970) 815--836.

\bibitem{Eisenstein:2005su}
{\bf SDSS} Collaboration, D.~J. Eisenstein et~al., {\it {Detection of the
  baryon acoustic peak in the large-scale correlation function of SDSS luminous
  red galaxies}},  {\em Astrophys. J.} {\bf 633} (2005) 560--574,
  [\href{http://arxiv.org/abs/astro-ph/0501171}{{\tt astro-ph/0501171}}].

\bibitem{Cole:2005sx}
{\bf 2dFGRS} Collaboration, S.~Cole et~al., {\it {The 2dF Galaxy Redshift
  Survey: Power-spectrum analysis of the final dataset and cosmological
  implications}},  {\em Mon. Not. Roy. Astron. Soc.} {\bf 362} (2005) 505--534,
  [\href{http://arxiv.org/abs/astro-ph/0501174}{{\tt astro-ph/0501174}}].

\bibitem{Padmanabhan:2006cia}
{\bf SDSS} Collaboration, N.~Padmanabhan et~al., {\it {The Clustering of
  Luminous Red Galaxies in the Sloan Digital Sky Survey Imaging Data}},  {\em
  Mon. Not. Roy. Astron. Soc.} {\bf 378} (2007) 852--872,
  [\href{http://arxiv.org/abs/astro-ph/0605302}{{\tt astro-ph/0605302}}].

\bibitem{Percival:2007yw}
W.~J. Percival, S.~Cole, D.~J. Eisenstein, R.~C. Nichol, J.~A. Peacock, A.~C.
  Pope, and A.~S. Szalay, {\it {Measuring the Baryon Acoustic Oscillation scale
  using the SDSS and 2dFGRS}},  {\em Mon. Not. Roy. Astron. Soc.} {\bf 381}
  (2007) 1053--1066, [\href{http://arxiv.org/abs/0705.3323}{{\tt
  arXiv:0705.3323}}].

\bibitem{Blake:2011en}
C.~Blake et~al., {\it {The WiggleZ Dark Energy Survey: mapping the
  distance-redshift relation with baryon acoustic oscillations}},  {\em Mon.
  Not. Roy. Astron. Soc.} {\bf 418} (2011) 1707--1724,
  [\href{http://arxiv.org/abs/1108.2635}{{\tt arXiv:1108.2635}}].

\bibitem{Beutler:2011hx}
F.~Beutler, C.~Blake, M.~Colless, D.~H. Jones, L.~Staveley-Smith, L.~Campbell,
  Q.~Parker, W.~Saunders, and F.~Watson, {\it {The 6dF Galaxy Survey: Baryon
  Acoustic Oscillations and the Local Hubble Constant}},  {\em Mon. Not. Roy.
  Astron. Soc.} {\bf 416} (2011) 3017--3032,
  [\href{http://arxiv.org/abs/1106.3366}{{\tt arXiv:1106.3366}}].

\bibitem{Padmanabhan:2012hf}
N.~Padmanabhan, X.~Xu, D.~J. Eisenstein, R.~Scalzo, A.~J. Cuesta, K.~T. Mehta,
  and E.~Kazin, {\it {A 2 per cent distance to $z$=0.35 by reconstructing
  baryon acoustic oscillations - I. Methods and application to the Sloan
  Digital Sky Survey}},  {\em Mon. Not. Roy. Astron. Soc.} {\bf 427} (2012),
  no.~3 2132--2145, [\href{http://arxiv.org/abs/1202.0090}{{\tt
  arXiv:1202.0090}}].

\bibitem{Anderson:2013zyy}
{\bf BOSS} Collaboration, L.~Anderson et~al., {\it {The clustering of galaxies
  in the SDSS-III Baryon Oscillation Spectroscopic Survey: baryon acoustic
  oscillations in the Data Releases 10 and 11 Galaxy samples}},  {\em Mon. Not.
  Roy. Astron. Soc.} {\bf 441} (2014), no.~1 24--62,
  [\href{http://arxiv.org/abs/1312.4877}{{\tt arXiv:1312.4877}}].

\bibitem{Ross:2014qpa}
A.~J. Ross, L.~Samushia, C.~Howlett, W.~J. Percival, A.~Burden, and M.~Manera,
  {\it {The clustering of the SDSS DR7 main Galaxy sample – I. A 4 per cent
  distance measure at $z = 0.15$}},  {\em Mon. Not. Roy. Astron. Soc.} {\bf
  449} (2015), no.~1 835--847, [\href{http://arxiv.org/abs/1409.3242}{{\tt
  arXiv:1409.3242}}].

\bibitem{Blake03}
C.~{Blake} and K.~{Glazebrook}, {\it {Probing Dark Energy Using Baryonic
  Oscillations in the Galaxy Power Spectrum as a Cosmological Ruler}},  {\em
  \apj} {\bf 594} (Sept., 2003) 665--673,
  [\href{http://arxiv.org/abs/astro-ph/0301632}{{\tt astro-ph/0301632}}].

\bibitem{Hu_rings}
W.~Hu and Z.~Haiman, {\it Redshifting rings of power},  {\em Phys. Rev.} {\bf
  D68} (2003) 063004, [\href{http://arxiv.org/abs/astro-ph/0306053}{{\tt
  astro-ph/0306053}}].

\bibitem{Seo_Eisenstein}
H.-J. Seo and D.~J. Eisenstein, {\it Probing dark energy with baryonic acoustic
  oscillations from future large galaxy redshift surveys},  {\em \apj} {\bf
  598} (2003) 720--740, [\href{http://arxiv.org/abs/astro-ph/0307460}{{\tt
  astro-ph/0307460}}].

\bibitem{Padmanabhan:2009yr}
N.~Padmanabhan and M.~White, {\it {Calibrating the Baryon Oscillation Ruler for
  Matter and Halos}},  {\em Phys. Rev.} {\bf D80} (2009) 063508,
  [\href{http://arxiv.org/abs/0906.1198}{{\tt arXiv:0906.1198}}].

\bibitem{Mehta:2011xf}
K.~T. Mehta, H.-J. Seo, J.~Eckel, D.~J. Eisenstein, M.~Metchnik, P.~Pinto, and
  X.~Xu, {\it {Galaxy Bias and its Effects on the Baryon Acoustic Oscillations
  Measurements}},  {\em Astrophys. J.} {\bf 734} (2011) 94,
  [\href{http://arxiv.org/abs/1104.1178}{{\tt arXiv:1104.1178}}].

\bibitem{Seo:2009fp}
H.-J. Seo, J.~Eckel, D.~J. Eisenstein, K.~Mehta, M.~Metchnik, N.~Padmanabhan,
  P.~Pinto, R.~Takahashi, M.~White, and X.~Xu, {\it {High-precision predictions
  for the acoustic scale in the non-linear regime}},  {\em Astrophys. J.} {\bf
  720} (2010) 1650--1667, [\href{http://arxiv.org/abs/0910.5005}{{\tt
  arXiv:0910.5005}}].

\bibitem{Desjacques:2016bnm}
V.~Desjacques, D.~Jeong, and F.~Schmidt, {\it {Large-Scale Galaxy Bias}},
  \href{http://arxiv.org/abs/1611.09787}{{\tt arXiv:1611.09787}}.

\bibitem{Ellis:2012rn}
{\bf PFS Team} Collaboration, R.~Ellis et~al., {\it {Extragalactic science,
  cosmology, and Galactic archaeology with the Subaru Prime Focus
  Spectrograph}},  {\em Publ. Astron. Soc. Jap.} {\bf 66} (2014), no.~1 R1,
  [\href{http://arxiv.org/abs/1206.0737}{{\tt arXiv:1206.0737}}].

\bibitem{Aghamousa:2016zmz}
{\bf DESI} Collaboration, A.~Aghamousa et~al., {\it {The DESI Experiment Part
  I: Science,Targeting, and Survey Design}},
  \href{http://arxiv.org/abs/1611.00036}{{\tt arXiv:1611.00036}}.

\bibitem{Aghamousa:2016sne}
{\bf DESI} Collaboration, A.~Aghamousa et~al., {\it {The DESI Experiment Part
  II: Instrument Design}},  \href{http://arxiv.org/abs/1611.00037}{{\tt
  arXiv:1611.00037}}.

\bibitem{Busca:2012bu}
N.~G. Busca et~al., {\it {Baryon Acoustic Oscillations in the Ly-$\alpha$
  forest of BOSS quasars}},  {\em Astron. Astrophys.} {\bf 552} (2013) A96,
  [\href{http://arxiv.org/abs/1211.2616}{{\tt arXiv:1211.2616}}].

\bibitem{Slosar:2013fi}
A.~Slosar et~al., {\it {Measurement of Baryon Acoustic Oscillations in the
  Lyman-alpha Forest Fluctuations in BOSS Data Release 9}},  {\em JCAP} {\bf
  1304} (2013) 026, [\href{http://arxiv.org/abs/1301.3459}{{\tt
  arXiv:1301.3459}}].

\bibitem{Font-Ribera:2013wce}
{\bf BOSS} Collaboration, A.~Font-Ribera et~al., {\it {Quasar-Lyman $\alpha$
  Forest Cross-Correlation from BOSS DR11 : Baryon Acoustic Oscillations}},
  {\em JCAP} {\bf 1405} (2014) 027, [\href{http://arxiv.org/abs/1311.1767}{{\tt
  arXiv:1311.1767}}].

\bibitem{Delubac:2014aqe}
{\bf BOSS} Collaboration, T.~Delubac et~al., {\it {Baryon acoustic oscillations
  in the Lyα forest of BOSS DR11 quasars}},  {\em Astron. Astrophys.} {\bf
  574} (2015) A59, [\href{http://arxiv.org/abs/1404.1801}{{\tt
  arXiv:1404.1801}}].

\bibitem{Bautista:2017zgn}
J.~E. Bautista et~al., {\it {Measurement of baryon acoustic oscillation
  correlations at $z=2.3$ with SDSS DR12 Ly$\alpha$-Forests}},  {\em Astron.
  Astrophys.} {\bf 603} (2017) A12,
  [\href{http://arxiv.org/abs/1702.00176}{{\tt arXiv:1702.00176}}].

\bibitem{Betoule:2014frx}
{\bf SDSS} Collaboration, M.~Betoule et~al., {\it {Improved cosmological
  constraints from a joint analysis of the SDSS-II and SNLS supernova
  samples}},  {\em Astron. Astrophys.} {\bf 568} (2014) A22,
  [\href{http://arxiv.org/abs/1401.4064}{{\tt arXiv:1401.4064}}].

\bibitem{1997PhT....50k..32B}
C.~L. {Bennett}, M.~S. {Turner}, and M.~{White}, {\it {The cosmic Rosetta
  stone}},  {\em Physics Today} {\bf 50} (Nov., 1997) 32--38.

\bibitem{Hu_Dodelson}
W.~Hu and S.~Dodelson, {\it Cosmic microwave background anisotropies},  {\em
  Ann. Rev. Astron. Astrophys.} {\bf 40} (2002) 171--216,
  [\href{http://arxiv.org/abs/astro-ph/0110414}{{\tt astro-ph/0110414}}].

\bibitem{Bond:1997wr}
J.~R. Bond, G.~Efstathiou, and M.~Tegmark, {\it {Forecasting cosmic parameter
  errors from microwave background anisotropy experiments}},  {\em Mon. Not.
  Roy. Astron. Soc.} {\bf 291} (1997) L33--L41,
  [\href{http://arxiv.org/abs/astro-ph/9702100}{{\tt astro-ph/9702100}}].

\bibitem{Melchiorri:2002ux}
A.~Melchiorri, L.~Mersini-Houghton, C.~J. Odman, and M.~Trodden, {\it {The
  State of the dark energy equation of state}},  {\em Phys. Rev.} {\bf D68}
  (2003) 043509, [\href{http://arxiv.org/abs/astro-ph/0211522}{{\tt
  astro-ph/0211522}}].

\bibitem{frieman_03}
J.~A. {Frieman}, D.~{Huterer}, E.~V. {Linder}, and M.~S. {Turner}, {\it
  {Probing dark energy with supernovae: Exploiting complementarity with the
  cosmic microwave background}},  {\em Phys. Rev. D} {\bf 67} (Apr., 2003)
  083505, [\href{http://arxiv.org/abs/astro-ph/0208100}{{\tt
  astro-ph/0208100}}].

\bibitem{Sherwin:2011gv}
B.~D. Sherwin et~al., {\it {Evidence for dark energy from the cosmic microwave
  background alone using the Atacama Cosmology Telescope lensing
  measurements}},  {\em Phys. Rev. Lett.} {\bf 107} (2011) 021302,
  [\href{http://arxiv.org/abs/1105.0419}{{\tt arXiv:1105.0419}}].

\bibitem{Jaffe:2000tx}
{\bf Boomerang} Collaboration, A.~H. Jaffe et~al., {\it {Cosmology from
  MAXIMA-1, BOOMERANG and COBE / DMR CMB observations}},  {\em Phys. Rev.
  Lett.} {\bf 86} (2001) 3475--3479,
  [\href{http://arxiv.org/abs/astro-ph/0007333}{{\tt astro-ph/0007333}}].

\bibitem{Sievers:2002tq}
J.~L. Sievers et~al., {\it {Cosmological parameters from Cosmic Background
  Imager observations and comparisons with BOOMERANG, DASI, and MAXIMA}},  {\em
  Astrophys. J.} {\bf 591} (2003) 599--622,
  [\href{http://arxiv.org/abs/astro-ph/0205387}{{\tt astro-ph/0205387}}].

\bibitem{WMAP_1}
{\bf WMAP} Collaboration, D.~N. Spergel et~al., {\it First year wilkinson
  microwave anisotropy probe (wmap) observations: Determination of cosmological
  parameters},  {\em \apjs} {\bf 148} (2003) 175,
  [\href{http://arxiv.org/abs/astro-ph/0302209}{{\tt astro-ph/0302209}}].

\bibitem{WMAP_3}
D.~N. {Spergel}, R.~{Bean}, O.~{Dor{\'e}}, M.~R. {Nolta}, C.~L. {Bennett},
  J.~{Dunkley}, G.~{Hinshaw}, N.~{Jarosik}, E.~{Komatsu}, L.~{Page}, H.~V.
  {Peiris}, L.~{Verde}, M.~{Halpern}, R.~S. {Hill}, A.~{Kogut}, M.~{Limon},
  S.~S. {Meyer}, N.~{Odegard}, G.~S. {Tucker}, J.~L. {Weiland}, E.~{Wollack},
  and E.~L. {Wright}, {\it {Three-Year Wilkinson Microwave Anisotropy Probe
  (WMAP) Observations: Implications for Cosmology}},  {\em \apjs} {\bf 170}
  (June, 2007) 377--408, [\href{http://arxiv.org/abs/astro-ph/0603449}{{\tt
  astro-ph/0603449}}].

\bibitem{Komatsu:2008hk}
{\bf WMAP} Collaboration, E.~Komatsu et~al., {\it {Five-Year Wilkinson
  Microwave Anisotropy Probe (WMAP) Observations: Cosmological
  Interpretation}},  {\em Astrophys. J. Suppl.} {\bf 180} (2009) 330--376,
  [\href{http://arxiv.org/abs/0803.0547}{{\tt arXiv:0803.0547}}].

\bibitem{Komatsu:2010fb}
{\bf WMAP} Collaboration, E.~Komatsu et~al., {\it {Seven-Year Wilkinson
  Microwave Anisotropy Probe (WMAP) Observations: Cosmological
  Interpretation}},  {\em Astrophys. J. Suppl.} {\bf 192} (2011) 18,
  [\href{http://arxiv.org/abs/1001.4538}{{\tt arXiv:1001.4538}}].

\bibitem{Hinshaw:2012aka}
{\bf WMAP} Collaboration, G.~Hinshaw et~al., {\it {Nine-Year Wilkinson
  Microwave Anisotropy Probe (WMAP) Observations: Cosmological Parameter
  Results}},  {\em Astrophys. J. Suppl.} {\bf 208} (2013) 19,
  [\href{http://arxiv.org/abs/1212.5226}{{\tt arXiv:1212.5226}}].

\bibitem{Ade:2013zuv}
{\bf Planck} Collaboration, P.~A.~R. Ade et~al., {\it {Planck 2013 results.
  XVI. Cosmological parameters}},  {\em Astron. Astrophys.} {\bf 571} (2014)
  A16, [\href{http://arxiv.org/abs/1303.5076}{{\tt arXiv:1303.5076}}].

\bibitem{Hou:2012xq}
Z.~Hou et~al., {\it {Constraints on Cosmology from the Cosmic Microwave
  Background Power Spectrum of the 2500 deg$^2$ SPT-SZ Survey}},  {\em
  Astrophys. J.} {\bf 782} (2014) 74,
  [\href{http://arxiv.org/abs/1212.6267}{{\tt arXiv:1212.6267}}].

\bibitem{Sievers:2013ica}
{\bf Atacama Cosmology Telescope} Collaboration, J.~L. Sievers et~al., {\it
  {The Atacama Cosmology Telescope: Cosmological parameters from three seasons
  of data}},  {\em JCAP} {\bf 1310} (2013) 060,
  [\href{http://arxiv.org/abs/1301.0824}{{\tt arXiv:1301.0824}}].

\bibitem{Sachs:1967er}
R.~K. Sachs and A.~M. Wolfe, {\it {Perturbations of a cosmological model and
  angular variations of the microwave background}},  {\em Astrophys. J.} {\bf
  147} (1967) 73--90. [Gen. Rel. Grav.39,1929(2007)].

\bibitem{Hu:1993xh}
W.~Hu and N.~Sugiyama, {\it {The Small scale integrated Sachs-Wolfe effect}},
  {\em Phys. Rev.} {\bf D50} (1994) 627--631,
  [\href{http://arxiv.org/abs/astro-ph/9310046}{{\tt astro-ph/9310046}}].

\bibitem{Hu:2001fb}
W.~Hu, {\it {Dark synergy: Gravitational lensing and the CMB}},  {\em Phys.
  Rev.} {\bf D65} (2002) 023003,
  [\href{http://arxiv.org/abs/astro-ph/0108090}{{\tt astro-ph/0108090}}].

\bibitem{Bean:2003fb}
R.~Bean and O.~Dore, {\it {Probing dark energy perturbations: The Dark energy
  equation of state and speed of sound as measured by WMAP}},  {\em Phys. Rev.}
  {\bf D69} (2004) 083503, [\href{http://arxiv.org/abs/astro-ph/0307100}{{\tt
  astro-ph/0307100}}].

\bibitem{Song:2006jk}
Y.-S. Song, I.~Sawicki, and W.~Hu, {\it {Large-Scale Tests of the DGP Model}},
  {\em Phys. Rev.} {\bf D75} (2007) 064003,
  [\href{http://arxiv.org/abs/astro-ph/0606286}{{\tt astro-ph/0606286}}].

\bibitem{Bartelmann_Schneider}
M.~Bartelmann and P.~Schneider, {\it {Weak Gravitational Lensing}},  {\em Phys.
  Rept.} {\bf 340} (2001) 291--472,
  [\href{http://arxiv.org/abs/astro-ph/9912508}{{\tt astro-ph/9912508}}].

\bibitem{Hoekstra_Jain}
H.~{Hoekstra} and B.~{Jain}, {\it {Weak Gravitational Lensing and Its
  Cosmological Applications}},  {\em Annual Review of Nuclear and Particle
  Science} {\bf 58} (Nov., 2008) 99--123,
  [\href{http://arxiv.org/abs/0805.0139}{{\tt arXiv:0805.0139}}].

\bibitem{Huterer_GRG}
D.~Huterer, {\it {Weak lensing, dark matter and dark energy}},  {\em Gen. Rel.
  Grav.} {\bf 42} (2010) 2177--2195,
  [\href{http://arxiv.org/abs/1001.1758}{{\tt arXiv:1001.1758}}].

\bibitem{Takada_Jain}
M.~Takada and B.~Jain, {\it Cosmological parameters from lensing power spectrum
  and bispectrum tomography},  {\em Mon. Not. Roy. Astron. Soc.} {\bf 348}
  (2004) 897, [\href{http://arxiv.org/abs/astro-ph/0310125}{{\tt
  astro-ph/0310125}}].

\bibitem{Hu:1999ek}
W.~Hu, {\it {Power spectrum tomography with weak lensing}},  {\em Astrophys.
  J.} {\bf 522} (1999) L21--L24,
  [\href{http://arxiv.org/abs/astro-ph/9904153}{{\tt astro-ph/9904153}}].

\bibitem{Jarvis:2005ck}
M.~Jarvis, B.~Jain, G.~Bernstein, and D.~Dolney, {\it {Dark energy constraints
  from the CTIO Lensing Survey}},  {\em Astrophys. J.} {\bf 644} (2006) 71--79,
  [\href{http://arxiv.org/abs/astro-ph/0502243}{{\tt astro-ph/0502243}}].

\bibitem{Massey:2007gh}
R.~Massey et~al., {\it {COSMOS: 3D weak lensing and the growth of structure}},
  {\em Astrophys. J. Suppl.} {\bf 172} (2007) 239--253,
  [\href{http://arxiv.org/abs/astro-ph/0701480}{{\tt astro-ph/0701480}}].

\bibitem{Schrabback:2009ba}
T.~Schrabback et~al., {\it {Evidence for the accelerated expansion of the
  Universe from weak lensing tomography with COSMOS}},  {\em Astron.
  Astrophys.} {\bf 516} (2010) A63, [\href{http://arxiv.org/abs/0911.0053}{{\tt
  arXiv:0911.0053}}].

\bibitem{Lin:2011bc}
{\bf SDSS} Collaboration, H.~Lin, S.~Dodelson, H.-J. Seo, M.~Soares-Santos,
  J.~Annis, J.~Hao, D.~Johnston, J.~M. Kubo, R.~R.~R. Reis, and M.~Simet, {\it
  {The SDSS Coadd: Cosmic Shear Measurement}},  {\em Astrophys. J.} {\bf 761}
  (2012) 15, [\href{http://arxiv.org/abs/1111.6622}{{\tt arXiv:1111.6622}}].

\bibitem{Heymans:2013fya}
C.~Heymans et~al., {\it {CFHTLenS tomographic weak lensing cosmological
  parameter constraints: Mitigating the impact of intrinsic galaxy
  alignments}},  {\em Mon. Not. Roy. Astron. Soc.} {\bf 432} (2013) 2433,
  [\href{http://arxiv.org/abs/1303.1808}{{\tt arXiv:1303.1808}}].

\bibitem{Huff:2011aa}
E.~M. Huff, T.~Eifler, C.~M. Hirata, R.~Mandelbaum, D.~Schlegel, and U.~Seljak,
  {\it {Seeing in the dark. 2. Cosmic shear in the Sloan Digital Sky Survey}},
  {\em Mon. Not. Roy. Astron. Soc.} {\bf 440} (2014), no.~2 1322--1344,
  [\href{http://arxiv.org/abs/1112.3143}{{\tt arXiv:1112.3143}}].

\bibitem{Jee:2015jta}
M.~J. Jee, J.~A. Tyson, S.~Hilbert, M.~D. Schneider, S.~Schmidt, and
  D.~Wittman, {\it {Cosmic Shear Results from the Deep Lens Survey - II: Full
  Cosmological Parameter Constraints from Tomography}},  {\em Astrophys. J.}
  {\bf 824} (2016), no.~2 77, [\href{http://arxiv.org/abs/1510.03962}{{\tt
  arXiv:1510.03962}}].

\bibitem{Hildebrandt:2016iqg}
H.~Hildebrandt et~al., {\it {KiDS-450: Cosmological parameter constraints from
  tomographic weak gravitational lensing}},  {\em Mon. Not. Roy. Astron. Soc.}
  {\bf 465} (2017) 1454, [\href{http://arxiv.org/abs/1606.05338}{{\tt
  arXiv:1606.05338}}].

\bibitem{Troxel:2017xyo}
{\bf DES} Collaboration, M.~A. Troxel et~al., {\it {Dark Energy Survey Year 1
  Results: Cosmological Constraints from Cosmic Shear}},
  \href{http://arxiv.org/abs/1708.01538}{{\tt arXiv:1708.01538}}.

\bibitem{Huterer:2005ez}
D.~Huterer, M.~Takada, G.~Bernstein, and B.~Jain, {\it {Systematic errors in
  future weak lensing surveys: Requirements and prospects for
  self-calibration}},  {\em Mon. Not. Roy. Astron. Soc.} {\bf 366} (2006)
  101--114, [\href{http://arxiv.org/abs/astro-ph/0506030}{{\tt
  astro-ph/0506030}}].

\bibitem{Heymans}
C.~{Heymans}, M.~{White}, A.~{Heavens}, C.~{Vale}, and L.~{van Waerbeke}, {\it
  {Potential sources of contamination to weak lensing measurements: constraints
  from N-body simulations}},  {\em \mnras} {\bf 371} (Sept., 2006) 750--760,
  [\href{http://arxiv.org/abs/astro-ph/0604001}{{\tt astro-ph/0604001}}].

\bibitem{Mandelbaum:2013esa}
R.~Mandelbaum et~al., {\it {The Third Gravitational Lensing Accuracy Testing
  (GREAT3) Challenge Handbook}},  {\em Astrophys. J. Suppl.} {\bf 212} (2014)
  5, [\href{http://arxiv.org/abs/1308.4982}{{\tt arXiv:1308.4982}}].

\bibitem{Ma:2005rc}
Z.-M. Ma, W.~Hu, and D.~Huterer, {\it {Effect of photometric redshift
  uncertainties on weak lensing tomography}},  {\em Astrophys. J.} {\bf 636}
  (2005) 21--29, [\href{http://arxiv.org/abs/astro-ph/0506614}{{\tt
  astro-ph/0506614}}].

\bibitem{Hearin:2010jr}
A.~P. Hearin, A.~R. Zentner, Z.~Ma, and D.~Huterer, {\it {A General Study of
  the Influence of Catastrophic Photometric Redshift Errors on Cosmology with
  Cosmic Shear Tomography}},  {\em Astrophys. J.} {\bf 720} (2010) 1351--1369,
  [\href{http://arxiv.org/abs/1002.3383}{{\tt arXiv:1002.3383}}].

\bibitem{Huterer_Takada}
D.~Huterer and M.~Takada, {\it {Calibrating the Nonlinear Matter Power
  Spectrum: Requirements for Future Weak Lensing Surveys}},  {\em Astropart.
  Phys.} {\bf 23} (2005) 369--376,
  [\href{http://arxiv.org/abs/astro-ph/0412142}{{\tt astro-ph/0412142}}].

\bibitem{Rudd_Zentner_Kravtsov}
D.~H. Rudd, A.~R. Zentner, and A.~V. Kravtsov, {\it {Effects of Baryons and
  Dissipation on the Matter Power Spectrum}},  {\em Astrophys. J.} {\bf 672}
  (2008) 19--32, [\href{http://arxiv.org/abs/astro-ph/0703741}{{\tt
  astro-ph/0703741}}].

\bibitem{Hearin:2009hz}
A.~P. Hearin and A.~R. Zentner, {\it {The Influence of Galaxy Formation Physics
  on Weak Lensing Tests of General Relativity}},  {\em JCAP} {\bf 0904} (2009)
  032, [\href{http://arxiv.org/abs/0904.3334}{{\tt arXiv:0904.3334}}].

\bibitem{Takada:2008fn}
M.~Takada and B.~Jain, {\it {The Impact of Non-Gaussian Errors on Weak Lensing
  Surveys}},  {\em Mon. Not. Roy. Astron. Soc.} {\bf 395} (2009) 2065--2086,
  [\href{http://arxiv.org/abs/0810.4170}{{\tt arXiv:0810.4170}}].

\bibitem{Taylor:2012kz}
A.~Taylor, B.~Joachimi, and T.~Kitching, {\it {Putting the Precision in
  Precision Cosmology: How accurate should your data covariance matrix be?}},
  {\em Mon. Not. Roy. Astron. Soc.} {\bf 432} (2013) 1928,
  [\href{http://arxiv.org/abs/1212.4359}{{\tt arXiv:1212.4359}}].

\bibitem{Dodelson:2013uaa}
S.~Dodelson and M.~D. Schneider, {\it {The Effect of Covariance Estimator Error
  on Cosmological Parameter Constraints}},  {\em Phys. Rev.} {\bf D88} (2013)
  063537, [\href{http://arxiv.org/abs/1304.2593}{{\tt arXiv:1304.2593}}].

\bibitem{Joachimi:2015mma}
B.~Joachimi et~al., {\it {Galaxy alignments: An overview}},  {\em Space Sci.
  Rev.} {\bf 193} (2015), no.~1-4 1--65,
  [\href{http://arxiv.org/abs/1504.05456}{{\tt arXiv:1504.05456}}].

\bibitem{Abbott:2017wau}
{\bf DES} Collaboration, T.~M.~C. Abbott et~al., {\it {Dark Energy Survey Year
  1 Results: Cosmological Constraints from Galaxy Clustering and Weak
  Lensing}},  \href{http://arxiv.org/abs/1708.01530}{{\tt arXiv:1708.01530}}.

\bibitem{Aihara:2017tri}
H.~Aihara et~al., {\it {First Data Release of the Hyper Suprime-Cam Subaru
  Strategic Program}},  \href{http://arxiv.org/abs/1702.08449}{{\tt
  arXiv:1702.08449}}.

\bibitem{Cai:2011wj}
Y.-C. Cai and G.~Bernstein, {\it {Combining weak lensing tomography and
  spectroscopic redshift surveys}},  {\em Mon. Not. Roy. Astron. Soc.} {\bf
  422} (2012) 1045--1056, [\href{http://arxiv.org/abs/1112.4478}{{\tt
  arXiv:1112.4478}}].

\bibitem{Joachimi:2010xb}
B.~Joachimi, R.~Mandelbaum, F.~B. Abdalla, and S.~L. Bridle, {\it {Constraints
  on intrinsic alignment contamination of weak lensing surveys using the
  MegaZ-LRG sample}},  {\em Astron. Astrophys.} {\bf 527} (2011) A26,
  [\href{http://arxiv.org/abs/1008.3491}{{\tt arXiv:1008.3491}}].

\bibitem{Font-Ribera:2013rwa}
A.~Font-Ribera, P.~McDonald, N.~Mostek, B.~A. Reid, H.-J. Seo, and A.~Slosar,
  {\it {DESI and other dark energy experiments in the era of neutrino mass
  measurements}},  {\em JCAP} {\bf 1405} (2014) 023,
  [\href{http://arxiv.org/abs/1308.4164}{{\tt arXiv:1308.4164}}].

\bibitem{Eriksen:2014zua}
M.~Eriksen and E.~Gaztanaga, {\it {Combining Spectroscopic and Photometric
  Surveys: Same or different sky?}},  {\em Mon. Not. Roy. Astron. Soc.} {\bf
  451} (2015), no.~2 1553--1560, [\href{http://arxiv.org/abs/1412.8429}{{\tt
  arXiv:1412.8429}}].

\bibitem{vanUitert:2017ieu}
E.~van Uitert et~al., {\it {KiDS+GAMA: Cosmology constraints from a joint
  analysis of cosmic shear, galaxy-galaxy lensing and angular clustering}},
  \href{http://arxiv.org/abs/1706.05004}{{\tt arXiv:1706.05004}}.

\bibitem{Joudaki:2017zdt}
S.~Joudaki et~al., {\it {KiDS-450 + 2dFLenS: Cosmological parameter constraints
  from weak gravitational lensing tomography and overlapping redshift-space
  galaxy clustering}},  \href{http://arxiv.org/abs/1707.06627}{{\tt
  arXiv:1707.06627}}.

\bibitem{Wittman01}
D.~Wittman, J.~A. Tyson, V.~E. Margoniner, J.~G. Cohen, and I.~P. Dell'Antonio,
  {\it {Discovery of a Galaxy Cluster via Weak Lensing}},  {\em Astrophys. J.}
  {\bf 557} (2001) L89--L92, [\href{http://arxiv.org/abs/astro-ph/0104094}{{\tt
  astro-ph/0104094}}].

\bibitem{Wittman02}
D.~Wittman, V.~E. Margoniner, J.~A. Tyson, J.~G. Cohen, and I.~P. Dell'Antonio,
  {\it {Weak Lensing Discovery and Tomography of a Cluster at z=0.68}},  {\em
  Astrophys. J.} {\bf 597} (2003) 218--224,
  [\href{http://arxiv.org/abs/astro-ph/0210120}{{\tt astro-ph/0210120}}].

\bibitem{Wittman06}
D.~Wittman et~al., {\it {First Results On Shear-Selected Clusters From the Deep
  Lens Survey: Optical Imaging, Spectroscopy, and X-ray Followup}},  {\em
  Astrophys. J.} {\bf 643} (2006) 128,
  [\href{http://arxiv.org/abs/astro-ph/0507606}{{\tt astro-ph/0507606}}].

\bibitem{Schirmer07}
M.~{Schirmer}, T.~{Erben}, M.~{Hetterscheidt}, and P.~{Schneider}, {\it
  {GaBoDS: the Garching-Bonn Deep Survey. IX. A sample of 158 shear-selected
  mass concentration candidates}},  {\em Astron. Astrophys.} {\bf 462} (Feb.,
  2007) 875--887, [\href{http://arxiv.org/abs/astro-ph/0607022}{{\tt
  astro-ph/0607022}}].

\bibitem{Dietrich07}
J.~P. {Dietrich}, T.~{Erben}, G.~{Lamer}, P.~{Schneider}, A.~{Schwope},
  J.~{Hartlap}, and M.~{Maturi}, {\it {BLOX: the Bonn lensing, optical, and
  X-ray selected galaxy clusters. I. Cluster catalog construction}},  {\em
  Astron. Astrophys.} {\bf 470} (Aug., 2007) 821--834,
  [\href{http://arxiv.org/abs/0705.3455}{{\tt arXiv:0705.3455}}].

\bibitem{Miyazaki07}
S.~{Miyazaki}, T.~{Hamana}, R.~S. {Ellis}, N.~{Kashikawa}, R.~J. {Massey},
  J.~{Taylor}, and A.~{Refregier}, {\it {A Subaru Weak-Lensing Survey. I.
  Cluster Candidates and Spectroscopic Verification}},  {\em Astrophys. J.}
  {\bf 669} (Nov., 2007) 714--728, [\href{http://arxiv.org/abs/0707.2249}{{\tt
  arXiv:0707.2249}}].

\bibitem{Clowe_Bullet}
D.~Clowe et~al., {\it {A direct empirical proof of the existence of dark
  matter}},  {\em Astrophys. J.} {\bf 648} (2006) L109--L113,
  [\href{http://arxiv.org/abs/astro-ph/0608407}{{\tt astro-ph/0608407}}].

\bibitem{Allen:2011zs}
S.~W. Allen, A.~E. Evrard, and A.~B. Mantz, {\it {Cosmological Parameters from
  Observations of Galaxy Clusters}},  {\em Ann. Rev. Astron. Astrophys.} {\bf
  49} (2011) 409--470, [\href{http://arxiv.org/abs/1103.4829}{{\tt
  arXiv:1103.4829}}].

\bibitem{Heitmann:2015xma}
K.~Heitmann et~al., {\it {The Mira–Titan Universe: Precision Predictions for
  Dark Energy Surveys}},  {\em Astrophys. J.} {\bf 820} (2016), no.~2 108,
  [\href{http://arxiv.org/abs/1508.02654}{{\tt arXiv:1508.02654}}].

\bibitem{Press_Schechter}
W.~H. Press and P.~Schechter, {\it {Formation of galaxies and clusters of
  galaxies by selfsimilar gravitational condensation}},  {\em Astrophys. J.}
  {\bf 187} (1974) 425--438.

\bibitem{Gunn:1972sv}
J.~E. Gunn and J.~R. Gott, III, {\it {On the Infall of Matter into Clusters of
  Galaxies and Some Effects on Their Evolution}},  {\em Astrophys. J.} {\bf
  176} (1972) 1--19.

\bibitem{Zentner:2006vw}
A.~R. Zentner, {\it {The Excursion Set Theory of Halo Mass Functions, Halo
  Clustering, and Halo Growth}},  {\em Int. J. Mod. Phys.} {\bf D16} (2007)
  763--816, [\href{http://arxiv.org/abs/astro-ph/0611454}{{\tt
  astro-ph/0611454}}].

\bibitem{Tinker}
J.~L. Tinker et~al., {\it {Toward a halo mass function for precision cosmology:
  the limits of universality}},  {\em Astrophys. J.} {\bf 688} (2008) 709--728,
  [\href{http://arxiv.org/abs/0803.2706}{{\tt arXiv:0803.2706}}].

\bibitem{Linder:2003dr}
E.~V. Linder and A.~Jenkins, {\it {Cosmic structure and dark energy}},  {\em
  Mon. Not. Roy. Astron. Soc.} {\bf 346} (2003) 573,
  [\href{http://arxiv.org/abs/astro-ph/0305286}{{\tt astro-ph/0305286}}].

\bibitem{Baldi:2012ky}
M.~Baldi, {\it {Dark Energy Simulations}},  {\em Phys. Dark Univ.} {\bf 1}
  (2012) 162--193, [\href{http://arxiv.org/abs/1210.6650}{{\tt
  arXiv:1210.6650}}].

\bibitem{Vikhlinin:2008ym}
A.~Vikhlinin et~al., {\it {Chandra Cluster Cosmology Project III: Cosmological
  Parameter Constraints}},  {\em Astrophys. J.} {\bf 692} (2009) 1060--1074,
  [\href{http://arxiv.org/abs/0812.2720}{{\tt arXiv:0812.2720}}].

\bibitem{Evrard:2014cea}
A.~E. Evrard, P.~Arnault, D.~Huterer, and A.~Farahi, {\it {A model for
  multiproperty galaxy cluster statistics}},  {\em Mon. Not. Roy. Astron. Soc.}
  {\bf 441} (2014), no.~4 3562--3569,
  [\href{http://arxiv.org/abs/1403.1456}{{\tt arXiv:1403.1456}}].

\bibitem{Rykoff:2011xi}
E.~S. Rykoff, B.~P. Koester, E.~Rozo, J.~Annis, A.~E. Evrard, S.~M. Hansen,
  J.~Hao, D.~E. Johnston, T.~A. McKay, and R.~H. Wechsler, {\it {Robust Optical
  Richness Estimation with Reduced Scatter}},  {\em Astrophys. J.} {\bf 746}
  (2012) 178, [\href{http://arxiv.org/abs/1104.2089}{{\tt arXiv:1104.2089}}].

\bibitem{Rykoff:2013ovv}
{\bf SDSS} Collaboration, E.~S. Rykoff et~al., {\it {redMaPPer I: Algorithm and
  SDSS DR8 Catalog}},  {\em Astrophys. J.} {\bf 785} (2014) 104,
  [\href{http://arxiv.org/abs/1303.3562}{{\tt arXiv:1303.3562}}].

\bibitem{Oguri:2010vi}
M.~Oguri and M.~Takada, {\it {Combining cluster observables and stacked weak
  lensing to probe dark energy: Self-calibration of systematic uncertainties}},
   {\em Phys. Rev.} {\bf D83} (2011) 023008,
  [\href{http://arxiv.org/abs/1010.0744}{{\tt arXiv:1010.0744}}].

\bibitem{Hu_Keeton}
W.~Hu and C.~R. Keeton, {\it {Three-dimensional mapping of dark matter}},  {\em
  Phys. Rev.} {\bf D66} (2002) 063506,
  [\href{http://arxiv.org/abs/astro-ph/0205412}{{\tt astro-ph/0205412}}].

\bibitem{Marian_Smith_Bernstein}
L.~Marian, R.~E. Smith, and G.~M. Bernstein, {\it {The cosmology dependence of
  weak lensing cluster counts}},  {\em Astrophys. J.} {\bf 698} (2009)
  L33--L36, [\href{http://arxiv.org/abs/0811.1991}{{\tt arXiv:0811.1991}}].

\bibitem{Dietrich_Hartlap}
J.~P. Dietrich and J.~Hartlap, {\it {Cosmology with the shear-peak
  statistics}},  {\em Mon. Not. Roy. Astron. Soc.} {\bf in press} (2010)
  [\href{http://arxiv.org/abs/0906.3512}{{\tt arXiv:0906.3512}}].

\bibitem{Rozo:2009jj}
{\bf DSDD} Collaboration, E.~Rozo et~al., {\it {Cosmological Constraints from
  the SDSS maxBCG Cluster Catalog}},  {\em Astrophys. J.} {\bf 708} (2010)
  645--660, [\href{http://arxiv.org/abs/0902.3702}{{\tt arXiv:0902.3702}}].

\bibitem{Mantz:2009fw}
A.~Mantz, S.~W. Allen, D.~Rapetti, and H.~Ebeling, {\it {The Observed Growth of
  Massive Galaxy Clusters I: Statistical Methods and Cosmological
  Constraints}},  {\em Mon. Not. Roy. Astron. Soc.} {\bf 406} (2010)
  1759--1772, [\href{http://arxiv.org/abs/0909.3098}{{\tt arXiv:0909.3098}}].

\bibitem{Tinker:2011pv}
J.~L. Tinker et~al., {\it {Cosmological Constraints from Galaxy Clustering and
  the Mass-to-Number Ratio of Galaxy Clusters}},  {\em Astrophys. J.} {\bf 745}
  (2012) 16, [\href{http://arxiv.org/abs/1104.1635}{{\tt arXiv:1104.1635}}].

\bibitem{Mantz:2014xba}
A.~B. Mantz, S.~W. Allen, R.~G. Morris, D.~A. Rapetti, D.~E. Applegate, P.~L.
  Kelly, A.~von~der Linden, and R.~W. Schmidt, {\it {Cosmology and astrophysics
  from relaxed galaxy clusters – II. Cosmological constraints}},  {\em Mon.
  Not. Roy. Astron. Soc.} {\bf 440} (2014), no.~3 2077--2098,
  [\href{http://arxiv.org/abs/1402.6212}{{\tt arXiv:1402.6212}}].

\bibitem{deHaan:2016qvy}
{\bf SPT} Collaboration, T.~de~Haan et~al., {\it {Cosmological Constraints from
  Galaxy Clusters in the 2500 square-degree SPT-SZ Survey}},  {\em Astrophys.
  J.} {\bf 832} (2016), no.~1 95, [\href{http://arxiv.org/abs/1603.06522}{{\tt
  arXiv:1603.06522}}].

\bibitem{Schmidt:2010jr}
F.~Schmidt, {\it {Dynamical Masses in Modified Gravity}},  {\em Phys. Rev.}
  {\bf D81} (2010) 103002, [\href{http://arxiv.org/abs/1003.0409}{{\tt
  arXiv:1003.0409}}].

\bibitem{Becker:2010xj}
M.~R. Becker and A.~V. Kravtsov, {\it {On the Accuracy of Weak Lensing Cluster
  Mass Reconstructions}},  {\em Astrophys. J.} {\bf 740} (2011) 25,
  [\href{http://arxiv.org/abs/1011.1681}{{\tt arXiv:1011.1681}}].

\bibitem{Rasia:2012jz}
E.~Rasia et~al., {\it {Lensing and X-ray mass estimates of clusters
  (SIMULATION)}},  {\em New J. Phys.} {\bf 14} (2012) 055018,
  [\href{http://arxiv.org/abs/1201.1569}{{\tt arXiv:1201.1569}}].

\bibitem{vonderLinden:2012kh}
A.~von~der Linden et~al., {\it {Weighing the Giants – I. Weak-lensing masses
  for 51 massive galaxy clusters: project overview, data analysis methods and
  cluster images}},  {\em Mon. Not. Roy. Astron. Soc.} {\bf 439} (2014), no.~1
  2--27, [\href{http://arxiv.org/abs/1208.0597}{{\tt arXiv:1208.0597}}].

\bibitem{Rykoff:2016trm}
{\bf DES} Collaboration, E.~S. Rykoff et~al., {\it {The redMaPPer Galaxy
  Cluster Catalog From DES Science Verification Data}},  {\em Astrophys. J.
  Suppl.} {\bf 224} (2016), no.~1 1,
  [\href{http://arxiv.org/abs/1601.00621}{{\tt arXiv:1601.00621}}].

\bibitem{Huterer:2004rf}
D.~Huterer, A.~Kim, L.~M. Krauss, and T.~Broderick, {\it {Redshift accuracy
  requirements for future supernova and number count surveys}},  {\em
  Astrophys. J.} {\bf 615} (2004) 595,
  [\href{http://arxiv.org/abs/astro-ph/0402002}{{\tt astro-ph/0402002}}].

\bibitem{Lima:2007kx}
M.~Lima and W.~Hu, {\it {Photometric Redshift Requirements for Self-Calibration
  of Cluster Dark Energy Studies}},  {\em Phys. Rev.} {\bf D76} (2007) 123013,
  [\href{http://arxiv.org/abs/0709.2871}{{\tt arXiv:0709.2871}}].

\bibitem{Levine:2002uq}
E.~S. Levine, A.~E. Schulz, and M.~J. White, {\it {Future galaxy cluster
  surveys: The Effect of theory uncertainty on constraining cosmological
  parameters}},  {\em Astrophys. J.} {\bf 577} (2002) 569--578,
  [\href{http://arxiv.org/abs/astro-ph/0204273}{{\tt astro-ph/0204273}}].

\bibitem{Majumdar:2003mw}
S.~Majumdar and J.~J. Mohr, {\it {Self calibration in cluster studies of dark
  energy: Combining the cluster redshift distribution, the power spectrum and
  mass measurements}},  {\em Astrophys. J.} {\bf 613} (2004) 41--50,
  [\href{http://arxiv.org/abs/astro-ph/0305341}{{\tt astro-ph/0305341}}].

\bibitem{Lima:2005tt}
M.~Lima and W.~Hu, {\it {Self-calibration of cluster dark energy studies:
  Observable-mass distribution}},  {\em Phys. Rev.} {\bf D72} (2005) 043006,
  [\href{http://arxiv.org/abs/astro-ph/0503363}{{\tt astro-ph/0503363}}].

\bibitem{Rozo:2012xa}
E.~Rozo, J.~G. Bartlett, A.~E. Evrard, and E.~S. Rykoff, {\it {Closing the
  loop: a self-consistent model of optical, X-ray and Sunyaev–Zel'dovich
  scaling relations for clusters of Galaxies}},  {\em Mon. Not. Roy. Astron.
  Soc.} {\bf 438} (2014), no.~1 78--96,
  [\href{http://arxiv.org/abs/1204.6305}{{\tt arXiv:1204.6305}}].

\bibitem{Brainerd95}
T.~G. Brainerd, R.~D. Blandford, and I.~Smail, {\it {MEASURING GALAXY MASSES
  USING GALAXY-GALAXY GRAVITATIONAL LENSING}},  {\em Astrophys. J.} {\bf 466}
  (1996) 623, [\href{http://arxiv.org/abs/astro-ph/9503073}{{\tt
  astro-ph/9503073}}].

\bibitem{Fischer}
{\bf SDSS} Collaboration, P.~Fischer et~al., {\it {Weak lensing with SDSS
  commissioning data: The Galaxy mass correlation function to $1h^{-1}$ Mpc}},
  {\em Astron. J.} {\bf 120} (2000) 1198--1208,
  [\href{http://arxiv.org/abs/astro-ph/9912119}{{\tt astro-ph/9912119}}].

\bibitem{Hoek_Yee_Glad}
H.~Hoekstra, H.~K.~C. Yee, and M.~D. Gladders, {\it {Properties of galaxy dark
  matter halos from weak lensing}},  {\em Astrophys. J.} {\bf 606} (2004)
  67--77, [\href{http://arxiv.org/abs/astro-ph/0306515}{{\tt
  astro-ph/0306515}}].

\bibitem{Sheldon:2003xj}
{\bf SDSS} Collaboration, E.~S. Sheldon et~al., {\it {The Galaxy - mass
  correlation function measured from weak lensing in the SDSS}},  {\em Astron.
  J.} {\bf 127} (2004) 2544--2564,
  [\href{http://arxiv.org/abs/astro-ph/0312036}{{\tt astro-ph/0312036}}].

\bibitem{Mandelbaum:2005nx}
R.~Mandelbaum, U.~Seljak, G.~Kauffmann, C.~M. Hirata, and J.~Brinkmann, {\it
  {Galaxy halo masses and satellite fractions from galaxy-galaxy lensing in the
  sdss: stellar mass, luminosity, morphology, and environment dependencies}},
  {\em Mon. Not. Roy. Astron. Soc.} {\bf 368} (2006) 715,
  [\href{http://arxiv.org/abs/astro-ph/0511164}{{\tt astro-ph/0511164}}].

\bibitem{Johnston:2007uc}
{\bf SDSS} Collaboration, D.~E. Johnston, E.~S. Sheldon, R.~H. Wechsler,
  E.~Rozo, B.~P. Koester, J.~A. Frieman, T.~A. McKay, A.~E. Evrard, M.~R.
  Becker, and J.~Annis, {\it {Cross-correlation Weak Lensing of SDSS galaxy
  Clusters II: Cluster Density Profiles and the Mass--Richness Relation}},
  \href{http://arxiv.org/abs/0709.1159}{{\tt arXiv:0709.1159}}.

\bibitem{Choi:2012kf}
A.~Choi, J.~A. Tyson, C.~B. Morrison, M.~J. Jee, S.~J. Schmidt, V.~E.
  Margoniner, and D.~M. Wittman, {\it {Galaxy-Mass Correlations on 10 Mpc
  Scales in the Deep Lens Survey}},  {\em Astrophys. J.} {\bf 759} (2012) 101,
  [\href{http://arxiv.org/abs/1208.3904}{{\tt arXiv:1208.3904}}].

\bibitem{Velander:2013jga}
M.~Velander et~al., {\it {CFHTLenS: The relation between galaxy dark matter
  haloes and baryons from weak gravitational lensing}},  {\em Mon. Not. Roy.
  Astron. Soc.} {\bf 437} (2014), no.~3 2111--2136,
  [\href{http://arxiv.org/abs/1304.4265}{{\tt arXiv:1304.4265}}].

\bibitem{Leauthaud:2016jdb}
A.~Leauthaud et~al., {\it {Lensing is Low: Cosmology, Galaxy Formation, or New
  Physics?}},  {\em Mon. Not. Roy. Astron. Soc.} {\bf 467} (2017) 3024,
  [\href{http://arxiv.org/abs/1611.08606}{{\tt arXiv:1611.08606}}].

\bibitem{Prat:2017goa}
{\bf DES} Collaboration, J.~Prat et~al., {\it {Dark Energy Survey Year 1
  Results: Galaxy-Galaxy Lensing}},
  \href{http://arxiv.org/abs/1708.01537}{{\tt arXiv:1708.01537}}.

\bibitem{Miralda-Escude91}
J.~{Miralda-Escude}, {\it {The correlation function of galaxy ellipticities
  produced by gravitational lensing}},  {\em Astrophys. J.} {\bf 380} (Oct.,
  1991) 1--8.

\bibitem{Wilson:2001}
G.~{Wilson}, N.~{Kaiser}, G.~A. {Luppino}, and L.~L. {Cowie}, {\it {Galaxy Halo
  Masses from Galaxy-Galaxy Lensing}},  {\em Astrophys. J.} {\bf 555} (July,
  2001) 572--584, [\href{http://arxiv.org/abs/astro-ph/0008504}{{\tt
  astro-ph/0008504}}].

\bibitem{Kleinheinrich04}
M.~{Kleinheinrich}, P.~{Schneider}, H.~{Rix}, T.~{Erben}, C.~{Wolf},
  M.~{Schirmer}, K.~{Meisenheimer}, A.~{Borch}, S.~{Dye}, Z.~{Kovacs}, and
  L.~{Wisotzki}, {\it {Weak lensing measurements of dark matter halos of
  galaxies from COMBO-17}},  {\em Astron. Astrophys.} {\bf 455} (Aug., 2006)
  441--451.

\bibitem{Mandelbaum_profiles}
R.~Mandelbaum et~al., {\it {Density profiles of galaxy groups and clusters from
  SDSS galaxy-galaxy weak lensing}},  {\em Mon. Not. Roy. Astron. Soc.} {\bf
  372} (2006) 758--776, [\href{http://arxiv.org/abs/astro-ph/0605476}{{\tt
  astro-ph/0605476}}].

\bibitem{Johnston07}
{\bf SDSS} Collaboration, D.~E. Johnston et~al., {\it {Cross-correlation Weak
  Lensing of SDSS galaxy Clusters II: Cluster Density Profiles and the
  Mass--Richness Relation}},  {\em arXiv:0709.1159} (2007)
  [\href{http://arxiv.org/abs/0709.1159}{{\tt arXiv:0709.1159}}].

\bibitem{Leauthaud_ML}
A.~Leauthaud et~al., {\it {A Weak Lensing Study of X-ray Groups in the COSMOS
  survey: Form and Evolution of the Mass-Luminosity Relation}},  {\em
  Astrophys. J.} {\bf 709} (2010) 97--114,
  [\href{http://arxiv.org/abs/0910.5219}{{\tt arXiv:0910.5219}}].

\bibitem{Sheldon09_ML}
{\bf SDSS} Collaboration, E.~S. Sheldon et~al., {\it {Cross-correlation Weak
  Lensing of SDSS Galaxy Clusters III: Mass-to-light Ratios}},  {\em Astrophys.
  J.} {\bf 703} (2009) 2232--2248, [\href{http://arxiv.org/abs/0709.1162}{{\tt
  arXiv:0709.1162}}].

\bibitem{Hu_Jain}
W.~Hu and B.~Jain, {\it {Joint Galaxy-Lensing Observables and the Dark
  Energy}},  {\em Phys. Rev.} {\bf D70} (2004) 043009,
  [\href{http://arxiv.org/abs/astro-ph/0312395}{{\tt astro-ph/0312395}}].

\bibitem{Schmidt_MGWL}
F.~Schmidt, {\it {Weak Lensing Probes of Modified Gravity}},  {\em Phys. Rev.}
  {\bf D78} (2008) 043002, [\href{http://arxiv.org/abs/0805.4812}{{\tt
  arXiv:0805.4812}}].

\bibitem{Kochanek:1995ap}
C.~S. Kochanek, {\it {Is there a cosmological constant?}},  {\em Astrophys. J.}
  {\bf 466} (1996) 638, [\href{http://arxiv.org/abs/astro-ph/9510077}{{\tt
  astro-ph/9510077}}].

\bibitem{Huterer:2003uf}
D.~Huterer and C.-P. Ma, {\it {Constraints on the Inner Cluster Mass Profile
  and the Power Spectrum Normalization from Strong Lensing Statistics}},  {\em
  Astrophys. J.} {\bf 600} (2004) L7,
  [\href{http://arxiv.org/abs/astro-ph/0307301}{{\tt astro-ph/0307301}}].

\bibitem{Chae:2002mb}
K.~H. Chae et~al., {\it {Constraints on cosmological parameters from the
  analysis of the Cosmic Lens All sky Survey radio - selected gravitational
  lens statistics}},  {\em Phys. Rev. Lett.} {\bf 89} (2002) 151301,
  [\href{http://arxiv.org/abs/astro-ph/0209602}{{\tt astro-ph/0209602}}].

\bibitem{Refsdal:1964nw}
S.~Refsdal, {\it {On the possibility of determining Hubble's parameter and the
  masses of galaxies from the gravitational lens effect}},  {\em Mon. Not. Roy.
  Astron. Soc.} {\bf 128} (1964) 307.

\bibitem{Linder:2011dr}
E.~V. Linder, {\it {Lensing Time Delays and Cosmological Complementarity}},
  {\em Phys. Rev.} {\bf D84} (2011) 123529,
  [\href{http://arxiv.org/abs/1109.2592}{{\tt arXiv:1109.2592}}].

\bibitem{Suyu:2012aa}
S.~H. Suyu et~al., {\it {Two accurate time-delay distances from strong lensing:
  Implications for cosmology}},  {\em Astrophys. J.} {\bf 766} (2013) 70,
  [\href{http://arxiv.org/abs/1208.6010}{{\tt arXiv:1208.6010}}].

\bibitem{Suyu:2009by}
S.~H. Suyu, P.~J. Marshall, M.~W. Auger, S.~Hilbert, R.~D. Blandford, L.~V.~E.
  Koopmans, C.~D. Fassnacht, and T.~Treu, {\it {Dissecting the Gravitational
  Lens B1608+656. II. Precision Measurements of the Hubble Constant, Spatial
  Curvature, and the Dark Energy Equation of State}},  {\em Astrophys. J.} {\bf
  711} (2010) 201--221, [\href{http://arxiv.org/abs/0910.2773}{{\tt
  arXiv:0910.2773}}].

\bibitem{Linder:2004hx}
E.~V. Linder, {\it {Strong gravitational lensing and dark energy
  complementarity}},  {\em Phys. Rev.} {\bf D70} (2004) 043534,
  [\href{http://arxiv.org/abs/astro-ph/0401433}{{\tt astro-ph/0401433}}].

\bibitem{Paraficz:2009xj}
D.~Paraficz and J.~Hjorth, {\it {Gravitational lenses as cosmic rulers: density
  of dark matter and dark energy from time delays and velocity dispersions}},
  {\em Astron. Astrophys.} {\bf 507} (2009) L49,
  [\href{http://arxiv.org/abs/0910.5823}{{\tt arXiv:0910.5823}}].

\bibitem{Jee:2014uxa}
I.~Jee, E.~Komatsu, and S.~H. Suyu, {\it {Measuring angular diameter distances
  of strong gravitational lenses}},  {\em JCAP} {\bf 1511} (2015), no.~11 033,
  [\href{http://arxiv.org/abs/1410.7770}{{\tt arXiv:1410.7770}}].

\bibitem{Jee:2015yra}
I.~Jee, E.~Komatsu, S.~H. Suyu, and D.~Huterer, {\it {Time-delay Cosmography:
  Increased Leverage with Angular Diameter Distances}},  {\em JCAP} {\bf 1604}
  (2016), no.~04 031, [\href{http://arxiv.org/abs/1509.03310}{{\tt
  arXiv:1509.03310}}].

\bibitem{Treu:2016ljm}
T.~Treu and P.~J. Marshall, {\it {Time Delay Cosmography}},  {\em Astron.
  Astrophys. Rev.} {\bf 24} (2016), no.~1 11,
  [\href{http://arxiv.org/abs/1605.05333}{{\tt arXiv:1605.05333}}].

\bibitem{Kaiser:1987qv}
N.~Kaiser, {\it {Clustering in real space and in redshift space}},  {\em Mon.
  Not. Roy. Astron. Soc.} {\bf 227} (1987) 1--27.

\bibitem{Song:2008qt}
Y.-S. Song and W.~J. Percival, {\it {Reconstructing the history of structure
  formation using Redshift Distortions}},  {\em JCAP} {\bf 0910} (2009) 004,
  [\href{http://arxiv.org/abs/0807.0810}{{\tt arXiv:0807.0810}}].

\bibitem{Percival:2008sh}
W.~J. Percival and M.~White, {\it {Testing cosmological structure formation
  using redshift-space distortions}},  {\em Mon. Not. Roy. Astron. Soc.} {\bf
  393} (2009) 297, [\href{http://arxiv.org/abs/0808.0003}{{\tt
  arXiv:0808.0003}}].

\bibitem{Linder:2007nu}
E.~V. Linder, {\it {Redshift Distortions as a Probe of Gravity}},  {\em
  Astropart. Phys.} {\bf 29} (2008) 336--339,
  [\href{http://arxiv.org/abs/0709.1113}{{\tt arXiv:0709.1113}}].

\bibitem{delaTorre:2013rpa}
S.~de~la Torre et~al., {\it {The VIMOS Public Extragalactic Redshift Survey
  (VIPERS). Galaxy clustering and redshift-space distortions at z=0.8 in the
  first data release}},  {\em Astron. Astrophys.} {\bf 557} (2013) A54,
  [\href{http://arxiv.org/abs/1303.2622}{{\tt arXiv:1303.2622}}].

\bibitem{Beutler:2013yhm}
{\bf BOSS} Collaboration, F.~Beutler et~al., {\it {The clustering of galaxies
  in the SDSS-III Baryon Oscillation Spectroscopic Survey: Testing gravity with
  redshift-space distortions using the power spectrum multipoles}},  {\em Mon.
  Not. Roy. Astron. Soc.} {\bf 443} (2014), no.~2 1065--1089,
  [\href{http://arxiv.org/abs/1312.4611}{{\tt arXiv:1312.4611}}].

\bibitem{Johnson:2014kaa}
A.~Johnson et~al., {\it {The 6dF Galaxy Velocity Survey: Cosmological
  constraints from the velocity power spectrum}},  {\em Mon. Not. Roy. Astron.
  Soc.} {\bf 444} (2014) 3926, [\href{http://arxiv.org/abs/1404.3799}{{\tt
  arXiv:1404.3799}}].

\bibitem{Huterer:2016uyq}
D.~Huterer, D.~Shafer, D.~Scolnic, and F.~Schmidt, {\it {Testing $\Lambda$CDM
  at the lowest redshifts with SN Ia and galaxy velocities}},  {\em JCAP} {\bf
  1705} (2017), no.~05 015, [\href{http://arxiv.org/abs/1611.09862}{{\tt
  arXiv:1611.09862}}].

\bibitem{Turnbull:2011ty}
S.~J. Turnbull, M.~J. Hudson, H.~A. Feldman, M.~Hicken, R.~P. Kirshner, and
  R.~Watkins, {\it {Cosmic flows in the nearby universe from Type Ia
  Supernovae}},  {\em Mon. Not. Roy. Astron. Soc.} {\bf 420} (2012) 447--454,
  [\href{http://arxiv.org/abs/1111.0631}{{\tt arXiv:1111.0631}}].

\bibitem{Beutler:2012px}
F.~Beutler, C.~Blake, M.~Colless, D.~H. Jones, L.~Staveley-Smith, G.~B. Poole,
  L.~Campbell, Q.~Parker, W.~Saunders, and F.~Watson, {\it {The 6dF Galaxy
  Survey: $z \approx 0$ measurement of the growth rate and $\sigma_8$}},  {\em
  Mon. Not. Roy. Astron. Soc.} {\bf 423} (2012) 3430--3444,
  [\href{http://arxiv.org/abs/1204.4725}{{\tt arXiv:1204.4725}}].

\bibitem{Blake:2013nif}
C.~Blake et~al., {\it {Galaxy And Mass Assembly (GAMA): improved cosmic growth
  measurements using multiple tracers of large-scale structure}},  {\em Mon.
  Not. Roy. Astron. Soc.} {\bf 436} (2013) 3089,
  [\href{http://arxiv.org/abs/1309.5556}{{\tt arXiv:1309.5556}}].

\bibitem{Blake:2011rj}
C.~Blake et~al., {\it {The WiggleZ Dark Energy Survey: the growth rate of
  cosmic structure since redshift z=0.9}},  {\em Mon. Not. Roy. Astron. Soc.}
  {\bf 415} (2011) 2876, [\href{http://arxiv.org/abs/1104.2948}{{\tt
  arXiv:1104.2948}}].

\bibitem{Beutler:2016arn}
{\bf BOSS} Collaboration, F.~Beutler et~al., {\it {The clustering of galaxies
  in the completed SDSS-III Baryon Oscillation Spectroscopic Survey:
  Anisotropic galaxy clustering in Fourier-space}},  {\em Submitted to: Mon.
  Not. Roy. Astron. Soc.} (2016) [\href{http://arxiv.org/abs/1607.03150}{{\tt
  arXiv:1607.03150}}].

\bibitem{Kaiser:1989kb}
N.~Kaiser, {\it {Theoretical implications of deviations from Hubble flow}},
  {\em Mon. Not. Roy. Astron. Soc.} {\bf 231} (1989) 149.

\bibitem{Gorski_etal}
K.~M. {Gorski}, M.~{Davis}, M.~A. {Strauss}, S.~D.~M. {White}, and A.~{Yahil},
  {\it {Cosmological velocity correlations - Observations and model
  predictions}},  {\em Astrophys. J.} {\bf 344} (1989) 1--19.

\bibitem{Haugboelle:2006uc}
T.~Haugboelle, S.~Hannestad, B.~Thomsen, J.~Fynbo, J.~Sollerman, and S.~Jha,
  {\it {The Velocity Field of the Local Universe from Measurements of Type Ia
  Supernovae}},  {\em Astrophys. J.} {\bf 661} (2007) 650--659,
  [\href{http://arxiv.org/abs/astro-ph/0612137}{{\tt astro-ph/0612137}}].

\bibitem{Gordon:2007zw}
C.~Gordon, K.~Land, and A.~Slosar, {\it {Cosmological Constraints from Type Ia
  Supernovae Peculiar Velocity Measurements}},  {\em Phys. Rev. Lett.} {\bf 99}
  (2007) 081301, [\href{http://arxiv.org/abs/0705.1718}{{\tt
  arXiv:0705.1718}}].

\bibitem{Ma:2010ps}
Y.-Z. Ma, C.~Gordon, and H.~A. Feldman, {\it {The peculiar velocity field:
  constraining the tilt of the Universe}},  {\em Phys. Rev.} {\bf D83} (2011)
  103002, [\href{http://arxiv.org/abs/1010.4276}{{\tt arXiv:1010.4276}}].

\bibitem{Dai:2011xm}
D.-C. Dai, W.~H. Kinney, and D.~Stojkovic, {\it {Measuring the cosmological
  bulk flow using the peculiar velocities of supernovae}},  {\em JCAP} {\bf
  1104} (2011) 015, [\href{http://arxiv.org/abs/1102.0800}{{\tt
  arXiv:1102.0800}}].

\bibitem{Nusser:2011tu}
A.~Nusser and M.~Davis, {\it {The cosmological bulk flow: consistency with
  $\Lambda$CDM and $z\approx 0$ constraints on $\sigma_8$ and $\gamma$}},  {\em
  Astrophys. J.} {\bf 736} (2011) 93,
  [\href{http://arxiv.org/abs/1101.1650}{{\tt arXiv:1101.1650}}].

\bibitem{Weyant:2011hs}
A.~Weyant, M.~Wood-Vasey, L.~Wasserman, and P.~Freeman, {\it {An Unbiased
  Method of Modeling the Local Peculiar Velocity Field with Type Ia
  Supernovae}},  {\em Astrophys. J.} {\bf 732} (2011) 65,
  [\href{http://arxiv.org/abs/1103.1603}{{\tt arXiv:1103.1603}}].

\bibitem{Ma:2012tt}
Y.-Z. Ma and D.~Scott, {\it {Cosmic bulk flows on $50 {h}^{-1}$Mpc scales: A
  Bayesian hyper-parameter method and multi-shells likelihood analysis}},  {\em
  Mon. Not. Roy. Astron. Soc.} {\bf 428} (2013) 2017,
  [\href{http://arxiv.org/abs/1208.2028}{{\tt arXiv:1208.2028}}].

\bibitem{Rathaus:2013ut}
B.~Rathaus, E.~D. Kovetz, and N.~Itzhaki, {\it {Studying the Peculiar Velocity
  Bulk Flow in a Sparse Survey of Type-Ia SNe}},  {\em Mon. Not. Roy. Astron.
  Soc.} {\bf 431} (2013) 3678, [\href{http://arxiv.org/abs/1301.7710}{{\tt
  arXiv:1301.7710}}].

\bibitem{Feindt:2013pma}
U.~Feindt et~al., {\it {Measuring cosmic bulk flows with Type Ia Supernovae
  from the Nearby Supernova Factory}},  {\em Astron. Astrophys.} {\bf 560}
  (2013) A90, [\href{http://arxiv.org/abs/1310.4184}{{\tt arXiv:1310.4184}}].

\bibitem{Ma:2013oja}
Y.-Z. Ma and J.~Pan, {\it {An estimation of local bulk flow with the
  maximum-likelihood method}},  {\em Mon. Not. Roy. Astron. Soc.} {\bf 437}
  (2014), no.~2 1996--2004, [\href{http://arxiv.org/abs/1311.6888}{{\tt
  arXiv:1311.6888}}].

\bibitem{Carrick:2015xza}
J.~Carrick, S.~J. Turnbull, G.~Lavaux, and M.~J. Hudson, {\it {Cosmological
  parameters from the comparison of peculiar velocities with predictions from
  the 2M++ density field}},  {\em Mon. Not. Roy. Astron. Soc.} {\bf 450}
  (2015), no.~1 317--332, [\href{http://arxiv.org/abs/1504.04627}{{\tt
  arXiv:1504.04627}}].

\bibitem{Riess:2016jrr}
A.~G. Riess et~al., {\it {A 2.4\% Determination of the Local Value of the
  Hubble Constant}},  {\em Astrophys. J.} {\bf 826} (2016), no.~1 56,
  [\href{http://arxiv.org/abs/1604.01424}{{\tt arXiv:1604.01424}}].

\bibitem{Bernal:2016gxb}
J.~L. Bernal, L.~Verde, and A.~G. Riess, {\it {The trouble with $H_0$}},  {\em
  JCAP} {\bf 1610} (2016), no.~10 019,
  [\href{http://arxiv.org/abs/1607.05617}{{\tt arXiv:1607.05617}}].

\bibitem{Hu:2004kn}
W.~Hu, {\it {Dark energy probes in light of the CMB}},  {\em ASP Conf. Ser.}
  {\bf 339} (2005) 215, [\href{http://arxiv.org/abs/astro-ph/0407158}{{\tt
  astro-ph/0407158}}].

\bibitem{Lee:2007kq}
J.~Lee and D.~Park, {\it {Constraining Dark Energy Equation of State with
  Cosmic Voids}},  {\em Astrophys. J.} {\bf 696} (2009) L10--L12,
  [\href{http://arxiv.org/abs/0704.0881}{{\tt arXiv:0704.0881}}].

\bibitem{Lavaux:2011yh}
G.~Lavaux and B.~D. Wandelt, {\it {Precision cosmography with stacked voids}},
  {\em Astrophys. J.} {\bf 754} (2012) 109,
  [\href{http://arxiv.org/abs/1110.0345}{{\tt arXiv:1110.0345}}].

\bibitem{Sutter:2012wh}
P.~M. Sutter, G.~Lavaux, B.~D. Wandelt, and D.~H. Weinberg, {\it {A public void
  catalog from the SDSS DR7 Galaxy Redshift Surveys based on the watershed
  transform}},  {\em Astrophys. J.} {\bf 761} (2012) 44,
  [\href{http://arxiv.org/abs/1207.2524}{{\tt arXiv:1207.2524}}].

\bibitem{Nadathur:2013bba}
S.~Nadathur and S.~Hotchkiss, {\it {A robust public catalogue of voids and
  superclusters in the SDSS Data Release 7 galaxy surveys}},  {\em Mon. Not.
  Roy. Astron. Soc.} {\bf 440} (2014), no.~2 1248--1262,
  [\href{http://arxiv.org/abs/1310.2791}{{\tt arXiv:1310.2791}}].

\bibitem{Leclercq:2014pga}
F.~Leclercq, J.~Jasche, P.~M. Sutter, N.~Hamaus, and B.~Wandelt, {\it {Dark
  matter voids in the SDSS galaxy survey}},  {\em JCAP} {\bf 1503} (2015),
  no.~03 047, [\href{http://arxiv.org/abs/1410.0355}{{\tt arXiv:1410.0355}}].

\bibitem{Hamaus:2016wka}
N.~Hamaus, A.~Pisani, P.~M. Sutter, G.~Lavaux, S.~Escoffier, B.~D. Wandelt, and
  J.~Weller, {\it {Constraints on Cosmology and Gravity from the Dynamics of
  Voids}},  {\em Phys. Rev. Lett.} {\bf 117} (2016), no.~9 091302,
  [\href{http://arxiv.org/abs/1602.01784}{{\tt arXiv:1602.01784}}].

\bibitem{Sutter:2013yda}
P.~M. Sutter, G.~Lavaux, B.~D. Wandelt, and D.~H. Weinberg, {\it {A response to
  arXiv:1310.2791: A self-consistent public catalogue of voids and
  superclusters in the SDSS Data Release 7 galaxy surveys}},
  \href{http://arxiv.org/abs/1310.5067}{{\tt arXiv:1310.5067}}.

\bibitem{Jain:1999nu}
B.~Jain and L.~V. Van~Waerbeke, {\it {Statistics of dark matter halos from
  gravitational lensing}},  {\em Astrophys. J.} {\bf 530} (2000) L1,
  [\href{http://arxiv.org/abs/astro-ph/9910459}{{\tt astro-ph/9910459}}].

\bibitem{Hamana:2003ts}
T.~Hamana, M.~Takada, and N.~Yoshida, {\it {Searching for massive clusters in
  weak lensing surveys}},  {\em Mon. Not. Roy. Astron. Soc.} {\bf 350} (2004)
  893, [\href{http://arxiv.org/abs/astro-ph/0310607}{{\tt astro-ph/0310607}}].

\bibitem{Hennawi:2004ai}
J.~F. Hennawi and D.~N. Spergel, {\it {Mass selected cluster cosmology. 1:
  Tomography and optimal filtering}},  {\em Astrophys. J.} {\bf 624} (2005) 59,
  [\href{http://arxiv.org/abs/astro-ph/0404349}{{\tt astro-ph/0404349}}].

\bibitem{Marian:2006zp}
L.~Marian and G.~M. Bernstein, {\it {Dark energy constraints from
  lensing-detected galaxy clusters}},  {\em Phys. Rev.} {\bf D73} (2006)
  123525, [\href{http://arxiv.org/abs/astro-ph/0605746}{{\tt
  astro-ph/0605746}}].

\bibitem{Dietrich:2009jq}
J.~P. Dietrich and J.~Hartlap, {\it {Cosmology with the shear-peak
  statistics}},  {\em Mon. Not. Roy. Astron. Soc.} {\bf 402} (2010) 1049,
  [\href{http://arxiv.org/abs/0906.3512}{{\tt arXiv:0906.3512}}].

\bibitem{Kratochvil:2009wh}
J.~M. Kratochvil, Z.~Haiman, and M.~May, {\it {Probing Cosmology with Weak
  Lensing Peak Counts}},  {\em Phys. Rev.} {\bf D81} (2010) 043519,
  [\href{http://arxiv.org/abs/0907.0486}{{\tt arXiv:0907.0486}}].

\bibitem{Liu:2014fzc}
J.~Liu, A.~Petri, Z.~Haiman, L.~Hui, J.~M. Kratochvil, and M.~May, {\it
  {Cosmology constraints from the weak lensing peak counts and the power
  spectrum in CFHTLenS data}},  {\em Phys. Rev.} {\bf D91} (2015), no.~6
  063507, [\href{http://arxiv.org/abs/1412.0757}{{\tt arXiv:1412.0757}}].

\bibitem{Hamana:2015bwa}
T.~Hamana, J.~Sakurai, M.~Koike, and L.~Miller, {\it {Cosmological constraints
  from Subaru weak lensing cluster counts}},  {\em Publ. Astron. Soc. Jap.}
  {\bf 67} (2015), no.~3 34, [\href{http://arxiv.org/abs/1503.01851}{{\tt
  arXiv:1503.01851}}].

\bibitem{Kacprzak:2016vir}
{\bf DES} Collaboration, T.~Kacprzak et~al., {\it {Cosmology constraints from
  shear peak statistics in Dark Energy Survey Science Verification data}},
  {\em Mon. Not. Roy. Astron. Soc.} {\bf 463} (2016), no.~4 3653--3673,
  [\href{http://arxiv.org/abs/1603.05040}{{\tt arXiv:1603.05040}}].

\bibitem{Liu:2016xjb}
J.~Liu and Z.~Haiman, {\it {Origin of weak lensing convergence peaks}},  {\em
  Phys. Rev.} {\bf D94} (2016), no.~4 043533,
  [\href{http://arxiv.org/abs/1606.01318}{{\tt arXiv:1606.01318}}].

\bibitem{Marian:2009wi}
L.~Marian, R.~E. Smith, and G.~M. Bernstein, {\it {The impact of correlated
  projections on weak lensing cluster counts}},  {\em Astrophys. J.} {\bf 709}
  (2010) 286--300, [\href{http://arxiv.org/abs/0912.0261}{{\tt
  arXiv:0912.0261}}].

\bibitem{Lin:2014dua}
C.-A. Lin and M.~Kilbinger, {\it {A new model to predict weak-lensing peak
  counts I. Comparison with $N$-body Simulations}},  {\em Astron. Astrophys.}
  {\bf 576} (2015) A24, [\href{http://arxiv.org/abs/1410.6955}{{\tt
  arXiv:1410.6955}}].

\bibitem{Schmidt:2010ex}
F.~Schmidt and E.~Rozo, {\it {Weak Lensing Peak Finding: Estimators, Filters,
  and Biases}},  {\em Astrophys. J.} {\bf 735} (2011) 119,
  [\href{http://arxiv.org/abs/1009.0757}{{\tt arXiv:1009.0757}}].

\bibitem{Jimenez_Loeb}
R.~Jimenez and A.~Loeb, {\it Constraining cosmological parameters based on
  relative galaxy ages},  {\em \apj} {\bf 573} (2002) 37--42,
  [\href{http://arxiv.org/abs/astro-ph/0106145}{{\tt astro-ph/0106145}}].

\bibitem{Stern:2009ep}
D.~Stern, R.~Jimenez, L.~Verde, M.~Kamionkowski, and S.~A. Stanford, {\it
  {Cosmic Chronometers: Constraining the Equation of State of Dark Energy. I:
  H(z) Measurements}},  {\em JCAP} {\bf 1002} (2010) 008,
  [\href{http://arxiv.org/abs/0907.3149}{{\tt arXiv:0907.3149}}].

\bibitem{Moresco:2012jh}
M.~Moresco et~al., {\it {Improved constraints on the expansion rate of the
  Universe up to z~1.1 from the spectroscopic evolution of cosmic
  chronometers}},  {\em JCAP} {\bf 1208} (2012) 006,
  [\href{http://arxiv.org/abs/1201.3609}{{\tt arXiv:1201.3609}}].

\bibitem{Moresco:2016mzx}
M.~Moresco, L.~Pozzetti, A.~Cimatti, R.~Jimenez, C.~Maraston, L.~Verde,
  D.~Thomas, A.~Citro, R.~Tojeiro, and D.~Wilkinson, {\it {A 6\% measurement of
  the Hubble parameter at $z\sim0.45$: direct evidence of the epoch of cosmic
  re-acceleration}},  {\em JCAP} {\bf 1605} (2016), no.~05 014,
  [\href{http://arxiv.org/abs/1601.01701}{{\tt arXiv:1601.01701}}].

\bibitem{Schutz:1986gp}
B.~F. Schutz, {\it {Determining the Hubble Constant from Gravitational Wave
  Observations}},  {\em Nature} {\bf 323} (1986) 310--311.

\bibitem{Holz_Hughes_05}
D.~E. Holz and S.~A. Hughes, {\it Using gravitational-wave standard sirens},
  {\em \apj} {\bf 629} (2005) 15--22,
  [\href{http://arxiv.org/abs/astro-ph/0504616}{{\tt astro-ph/0504616}}].

\bibitem{Dalal06}
N.~{Dalal}, D.~E. {Holz}, S.~A. {Hughes}, and B.~{Jain}, {\it {Short GRB and
  binary black hole standard sirens as a probe of dark energy}},  {\em Phys.
  Rev. D} {\bf 74} (Sept., 2006) 063006,
  [\href{http://arxiv.org/abs/astro-ph/0601275}{{\tt astro-ph/0601275}}].

\bibitem{Cutler:2009qv}
C.~Cutler and D.~E. Holz, {\it {Ultra-high precision cosmology from
  gravitational waves}},  {\em Phys. Rev.} {\bf D80} (2009) 104009,
  [\href{http://arxiv.org/abs/0906.3752}{{\tt arXiv:0906.3752}}].

\bibitem{Chen:2016tys}
H.-Y. Chen and D.~E. Holz, {\it {Finding the One: Identifying the Host Galaxies
  of Gravitational-Wave Sources}},  \href{http://arxiv.org/abs/1612.01471}{{\tt
  arXiv:1612.01471}}.

\bibitem{Sandage62}
A.~{Sandage}, {\it {The Change of Redshift and Apparent Luminosity of Galaxies
  due to the Deceleration of Selected Expanding Universes.}},  {\em \apj} {\bf
  136} (Sept., 1962) 319--333.

\bibitem{1997fpc..book.....L}
E.~V. {Linder}, {\em {First Principles of Cosmology}}.
\newblock 1997.

\bibitem{Loeb}
A.~{Loeb}, {\it {Direct Measurement of Cosmological Parameters from the Cosmic
  Deceleration of Extragalactic Objects}},  {\em \apjl} {\bf 499} (June, 1998)
  L111, [\href{http://arxiv.org/abs/astro-ph/9802122}{{\tt astro-ph/9802122}}].

\bibitem{Liske:2008ph}
J.~Liske et~al., {\it {Cosmic dynamics in the era of Extremely Large
  Telescopes}},  {\em Mon. Not. Roy. Astron. Soc.} {\bf 386} (2008) 1192--1218,
  [\href{http://arxiv.org/abs/0802.1532}{{\tt arXiv:0802.1532}}].

\bibitem{Cor_Hut_Mel}
P.-S. Corasaniti, D.~Huterer, and A.~Melchiorri, {\it Exploring the dark energy
  redshift desert with the sandage-loeb test},  {\em Phys. Rev.} {\bf D75}
  (2007) 062001, [\href{http://arxiv.org/abs/astro-ph/0701433}{{\tt
  astro-ph/0701433}}].

\bibitem{Quercellini:2010zr}
C.~Quercellini, L.~Amendola, A.~Balbi, P.~Cabella, and M.~Quartin, {\it
  {Real-time Cosmology}},  {\em Phys. Rept.} {\bf 521} (2012) 95--134,
  [\href{http://arxiv.org/abs/1011.2646}{{\tt arXiv:1011.2646}}].

\bibitem{Kim:2014uha}
A.~G. Kim, E.~V. Linder, J.~Edelstein, and D.~Erskine, {\it {Giving Cosmic
  Redshift Drift a Whirl}},  {\em Astropart. Phys.} {\bf 62} (2015) 195--205,
  [\href{http://arxiv.org/abs/1402.6614}{{\tt arXiv:1402.6614}}].

\bibitem{Yu:2013bia}
H.-R. Yu, T.-J. Zhang, and U.-L. Pen, {\it {Method for Direct Measurement of
  Cosmic Acceleration by 21-cm Absorption Systems}},  {\em Phys. Rev. Lett.}
  {\bf 113} (2014) 041303, [\href{http://arxiv.org/abs/1311.2363}{{\tt
  arXiv:1311.2363}}].

\bibitem{Klockner:2015rqa}
H.-R. Kl{\"o}ckner, D.~Obreschkow, C.~Martins, A.~Raccanelli, D.~Champion,
  A.~L. Roy, A.~Lobanov, J.~Wagner, and R.~Keller, {\it {Real time cosmology -
  A direct measure of the expansion rate of the Universe with the SKA}},  {\em
  PoS} {\bf AASKA14} (2015) 027, [\href{http://arxiv.org/abs/1501.03822}{{\tt
  arXiv:1501.03822}}].

\bibitem{1994ApJ...426...38D}
R.~A. {Daly}, {\it {Cosmology with powerful extended radio sources}},  {\em
  \apj} {\bf 426} (May, 1994) 38--50.

\bibitem{Daly:2002kn}
R.~A. Daly and E.~J. Guerra, {\it {Quintessence, cosmology, and FRIIb radio
  galaxies}},  {\em Astron. J.} {\bf 124} (2002) 1831,
  [\href{http://arxiv.org/abs/astro-ph/0209503}{{\tt astro-ph/0209503}}].

\bibitem{Risaliti:2015zla}
G.~Risaliti and E.~Lusso, {\it {A Hubble Diagram for Quasars}},  {\em
  Astrophys. J.} {\bf 815} (2015) 33,
  [\href{http://arxiv.org/abs/1505.07118}{{\tt arXiv:1505.07118}}].

\bibitem{schaefer03}
B.~E. {Schaefer}, {\it {Gamma-Ray Burst Hubble Diagram to z=4.5}},  {\em \apjl}
  {\bf 583} (Feb., 2003) L67--L70,
  [\href{http://arxiv.org/abs/astro-ph/0212445}{{\tt astro-ph/0212445}}].

\bibitem{Amati:2002ny}
L.~Amati et~al., {\it {Intrinsic spectra and energetics of BeppoSAX gamma-ray
  bursts with known redshifts}},  {\em Astron. Astrophys.} {\bf 390} (2002) 81,
  [\href{http://arxiv.org/abs/astro-ph/0205230}{{\tt astro-ph/0205230}}].

\bibitem{Amati:2006ky}
L.~Amati, {\it {The E(p,i) - E(iso) correlation in grbs: updated observational
  status, re-analysis and main implications}},  {\em Mon. Not. Roy. Astron.
  Soc.} {\bf 372} (2006) 233--245,
  [\href{http://arxiv.org/abs/astro-ph/0601553}{{\tt astro-ph/0601553}}].

\bibitem{Schaefer:2006pa}
B.~E. Schaefer, {\it {The Hubble Diagram to Redshift $>6$ from 69 Gamma-Ray
  Bursts}},  {\em Astrophys. J.} {\bf 660} (2007) 16--46,
  [\href{http://arxiv.org/abs/astro-ph/0612285}{{\tt astro-ph/0612285}}].

\bibitem{Holz:2010ck}
D.~E. Holz and S.~Perlmutter, {\it {The most massive objects in the Universe}},
   {\em Astrophys. J.} {\bf 755} (2012) L36,
  [\href{http://arxiv.org/abs/1004.5349}{{\tt arXiv:1004.5349}}].

\bibitem{Mortonson:2010mj}
M.~J. Mortonson, W.~Hu, and D.~Huterer, {\it {Simultaneous Falsification of
  $\Lambda$CDM and Quintessence with Massive, Distant Clusters}},  {\em Phys.
  Rev.} {\bf D83} (2011) 023015, [\href{http://arxiv.org/abs/1011.0004}{{\tt
  arXiv:1011.0004}}].

\bibitem{Hotchkiss:2011ms}
S.~Hotchkiss, {\it {Quantifying the rareness of extreme galaxy clusters}},
  {\em JCAP} {\bf 1107} (2011) 004, [\href{http://arxiv.org/abs/1105.3630}{{\tt
  arXiv:1105.3630}}].

\end{thebibliography}\endgroup

\end{document}